\documentclass[aps,prc,twocolumn,
amsfonts,superscriptaddress,showpacs,showkeys,nofootinbib,floatfix]{revtex4-1}

\usepackage{url}
\usepackage{color}
\usepackage{amssymb}
\usepackage{amsmath}
\usepackage{graphics,epsfig}
\usepackage{verbatim}
\usepackage{amstext}
\usepackage{mathrsfs}
\usepackage{latexsym}
\usepackage{subfigure}
\usepackage{wasysym}
\usepackage{fancyhdr}
\usepackage{epstopdf}
\usepackage{bm}
\usepackage{hyperref}

\begin{document}
\title{Measurement of the structure function of the nearly free neutron using spectator tagging in inelastic  $^2$H$(e,e' p_s)X$ scattering with CLAS}

\newcommand*{\ANL}{Argonne National Laboratory, Argonne, Illinois 60439}
\newcommand*{\ANLindex}{1}
\affiliation{\ANL}
\newcommand*{\ASU}{Arizona State University, Tempe, Arizona 85287-1504}
\newcommand*{\ASUindex}{2}
\affiliation{\ASU}
\newcommand*{\CSUDH}{California State University, Dominguez Hills, Carson, CA 90747}
\newcommand*{\CSUDHindex}{3}
\affiliation{\CSUDH}
\newcommand*{\CANISIUS}{Canisius College, Buffalo, NY}
\newcommand*{\CANISIUSindex}{4}
\affiliation{\CANISIUS}
\newcommand*{\CMU}{Carnegie Mellon University, Pittsburgh, Pennsylvania 15213}
\newcommand*{\CMUindex}{5}
\affiliation{\CMU}
\newcommand*{\CUA}{Catholic University of America, Washington, D.C. 20064}
\newcommand*{\CUAindex}{6}
\affiliation{\CUA}
\newcommand*{\SACLAY}{CEA, Centre de Saclay, Irfu/Service de Physique Nucl\'eaire, 91191 Gif-sur-Yvette, France}
\newcommand*{\SACLAYindex}{7}
\affiliation{\SACLAY}
\newcommand*{\CNU}{Christopher Newport University, Newport News, Virginia 23606}
\newcommand*{\CNUindex}{8}
\affiliation{\CNU}
\newcommand*{\UCONN}{University of Connecticut, Storrs, Connecticut 06269}
\newcommand*{\UCONNindex}{9}
\affiliation{\UCONN}
\newcommand*{\EDINBURGH}{Edinburgh University, Edinburgh EH9 3JZ, United Kingdom}
\newcommand*{\EDINBURGHindex}{10}
\affiliation{\EDINBURGH}
\newcommand*{\FU}{Fairfield University, Fairfield CT 06824}
\newcommand*{\FUindex}{11}
\affiliation{\FU}
\newcommand*{\FIU}{Florida International University, Miami, Florida 33199}
\newcommand*{\FIUindex}{12}
\affiliation{\FIU}
\newcommand*{\FSU}{Florida State University, Tallahassee, Florida 32306}
\newcommand*{\FSUindex}{13}
\affiliation{\FSU}
\newcommand*{\GWUI}{The George Washington University, Washington, DC 20052}
\newcommand*{\GWUIindex}{14}
\affiliation{\GWUI}
\newcommand*{\HAMP}{Hampton University, Hampton, Virginia 23668}
\newcommand*{\HAMPindex}{15}
\affiliation{\HAMP}
\newcommand*{\HOUS}{University of Houston,  Houston, TX 77004}
\newcommand*{\HOUSindex}{16}
\affiliation{\HOUS}
\newcommand*{\ISU}{Idaho State University, Pocatello, Idaho 83209}
\newcommand*{\ISUindex}{17}
\affiliation{\ISU}
\newcommand*{\INFNFE}{INFN, Sezione di Ferrara, 44100 Ferrara, Italy}
\newcommand*{\INFNFEindex}{18}
\affiliation{\INFNFE}
\newcommand*{\INFNFR}{INFN, Laboratori Nazionali di Frascati, 00044 Frascati, Italy}
\newcommand*{\INFNFRindex}{19}
\affiliation{\INFNFR}
\newcommand*{\INFNGE}{INFN, Sezione di Genova, 16146 Genova, Italy}
\newcommand*{\INFNGEindex}{20}
\affiliation{\INFNGE}
\newcommand*{\INFNRO}{INFN, Sezione di Roma Tor Vergata, 00133 Rome, Italy}
\newcommand*{\INFNROindex}{21}
\affiliation{\INFNRO}
\newcommand*{\ORSAY}{Institut de Physique Nucl\'eaire ORSAY, Orsay, France}
\newcommand*{\ORSAYindex}{22}
\affiliation{\ORSAY}
\newcommand*{\ITEP}{Institute of Theoretical and Experimental Physics, Moscow, 117259, Russia}
\newcommand*{\ITEPindex}{23}
\affiliation{\ITEP}
\newcommand*{\JMU}{James Madison University, Harrisonburg, Virginia 22807}
\newcommand*{\JMUindex}{24}
\affiliation{\JMU}
\newcommand*{\KNU}{Kyungpook National University, Daegu 702-701, Republic of Korea}
\newcommand*{\KNUindex}{25}
\affiliation{\KNU}
\newcommand*{\LPSC}{LPSC, Universite Joseph Fourier, CNRS/IN2P3, INPG, Grenoble, France}
\newcommand*{\LPSCindex}{26}
\affiliation{\LPSC}
\newcommand*{\MSSU}{Mississippi State University, Mississippi State, MS 39762}
\newcommand*{\MSSUindex}{27}
\affiliation{\MSSU}
\newcommand*{\UNH}{University of New Hampshire, Durham, New Hampshire 03824-3568}
\newcommand*{\UNHindex}{28}
\affiliation{\UNH}
\newcommand*{\NSU}{Norfolk State University, Norfolk, Virginia 23504}
\newcommand*{\NSUindex}{29}
\affiliation{\NSU}
\newcommand*{\OHIOU}{Ohio University, Athens, Ohio  45701}
\newcommand*{\OHIOUindex}{30}
\affiliation{\OHIOU}
\newcommand*{\ODU}{Old Dominion University, Norfolk, Virginia 23529}
\newcommand*{\ODUindex}{31}
\affiliation{\ODU}
\newcommand*{\RPI}{Rensselaer Polytechnic Institute, Troy, New York 12180-3590}
\newcommand*{\RPIindex}{32}
\affiliation{\RPI}
\newcommand*{\ROMAII}{Universita' di Roma Tor Vergata, 00133 Rome Italy}
\newcommand*{\ROMAIIindex}{33}
\affiliation{\ROMAII}
\newcommand*{\MSU}{Skobeltsyn Institute of Nuclear Physics, Lomonosov Moscow State University, 119234 Moscow, Russia}
\newcommand*{\MSUindex}{34}
\affiliation{\MSU}
\newcommand*{\SCAROLINA}{University of South Carolina, Columbia, South Carolina 29208}
\newcommand*{\SCAROLINAindex}{35}
\affiliation{\SCAROLINA}
\newcommand*{\JLAB}{Thomas Jefferson National Accelerator Facility, Newport News, Virginia 23606}
\newcommand*{\JLABindex}{36}
\affiliation{\JLAB}
\newcommand*{\GLASGOW}{University of Glasgow, Glasgow G12 8QQ, United Kingdom}
\newcommand*{\GLASGOWindex}{37}
\affiliation{\GLASGOW}
\newcommand*{\URICH}{University of Richmond, Richmond, Virginia 23173}
\newcommand*{\URICHindex}{38}
\affiliation{\URICH}
\newcommand*{\UTFSM}{Universidad T\'{e}cnica Federico Santa Mar\'{i}a, Casilla 110-V Valpara\'{i}so, Chile}
\newcommand*{\UTFSMindex}{39}
\affiliation{\UTFSM}
\newcommand*{\VIRGINIA}{University of Virginia, Charlottesville, Virginia 22901}
\newcommand*{\VIRGINIAindex}{40}
\affiliation{\VIRGINIA}
\newcommand*{\WM}{College of William and Mary, Williamsburg, Virginia 23187-8795}
\newcommand*{\WMindex}{41}
\affiliation{\WM}
\newcommand*{\YEREVAN}{Yerevan Physics Institute, 375036 Yerevan, Armenia}
\newcommand*{\YEREVANindex}{42}
\affiliation{\YEREVAN}
 
\newcommand*{\NOWUVA}{ University of Virginia, Charlottesville, Virginia 22901}
\newcommand*{\NOWORSAY} {Institut de Physique Nucl\'eaire ORSAY, Orsay, France}
\newcommand*{\NOWUTFSM}{ Universidad T\'{e}cnica Federico Santa Mar\'{i}a, Casilla 110-V Valpara\'{i}so, Chile}
\newcommand*{\NOWJLAB}{ Thomas Jefferson National Accelerator Facility, Newport News, Virginia 23606}
\newcommand*{\NOWROMAII}{ Universita' di Roma Tor Vergata, 00133 Rome Italy}
\newcommand*{\NOWODU}{Old Dominion University, Norfolk, Virginia 23529}

\author{S.~Tkachenko}
\altaffiliation[Current address: ]{\NOWUVA}
\affiliation{\ODU}
\author{N.~Baillie}
\affiliation{\WM}
\affiliation{\HAMP}
\author{S.E.~Kuhn} 
     \email{skuhn@odu.edu}
     \thanks{Corresponding author.}
\affiliation{\ODU}
\author{J.~Zhang} 
\altaffiliation[Current address: ]{\NOWUVA}
\affiliation{\ODU}
\affiliation{\JLAB}
\author{J.~Arrington}
\affiliation{\ANL}
\author{P.~Bosted}
\affiliation{\WM}
\affiliation{\JLAB}
\author{S. B\"ultmann}
\affiliation{\ODU}
\author{M.E.~Christy}
\affiliation{\HAMP}
\author{D. Dutta}
\affiliation{\MSSU}
\author{R.~Ent}
\affiliation{\JLAB}
% ...

\collaboration{The CLAS Collaboration}
\noaffiliation

 \date{\today}

\begin{abstract}
\begin{description}
\item[Background]
Much less is known about neutron structure than that of the proton due
to the absence of free neutron targets.  Neutron information is usually
extracted from data on nuclear targets such as deuterium, requiring
corrections for nuclear binding and nucleon off-shell effects.
These corrections are model dependent and have significant uncertainties,
especially for large values of the Bjorken scaling variable $x$.
As a consequence, the same data can lead to different conclusions,
for example, about the behavior of the $d$ quark distribution in the
proton at large $x$.
\item[Purpose]
The Barely Off-shell Nucleon Structure (BONuS) experiment at Jefferson
Lab measured the inelastic electron--deuteron scattering cross section,
tagging spectator protons in coincidence with the scattered electrons.
This method reduces nuclear binding uncertainties significantly and has
allowed for the first time a (nearly) model-independent extraction of
the neutron structure function $F_2(x,Q^2)$ in the resonance and
deep-inelastic regions.
\item[Method]
A novel compact radial time projection chamber was built to detect
protons with momentum between 70 and 150~MeV/$c$ and over a nearly
$4\pi$ angular range.  For the extraction of the free-neutron
structure function $F_2^n$, spectator protons at backward angles
($>100^\circ$ relative to the momentum transfer) and with momenta
below 100~MeV/$c$ were selected, ensuring that the scattering took
place on a nearly free neutron.  The scattered electrons were detected
with Jefferson Lab's CLAS spectrometer, with data taken at beam
energies near 2, 4 and 5~GeV. 
\item[Results]
The extracted neutron structure function $F_2^n$
and its ratio to the inclusive deuteron structure function $F_2^d$
are presented in both 
the resonance and deep-inelastic regions for momentum transfer squared
$Q^2$ between 0.7 and 5~GeV$^2/c^2$, invariant mass $W$ between
1 and 2.7~GeV/$c^2$, and Bjorken $x$ between 0.25 and 0.6 (in the DIS
region).
The dependence of the semi-inclusive cross section on the spectator
proton momentum and angle is investigated, and tests of the spectator
mechanism for different kinematics are performed.  
\item[Conclusions]
Our data
set on the structure function ratio $F_2^n/F_2^d$
can be used
to study neutron
resonance excitations, test quark-hadron duality in the neutron,
develop more precise parametrizations of structure functions,
as well as investigate binding effects (including possible mechanisms
for the nuclear EMC effect) and provide a first glimpse of the
asymptotic behavior of $d/u$ at $x \to 1$.
\end{description}
\end{abstract}


\maketitle{}

\section{Introduction}
\label{sec:intro}

The advent of high-luminosity beams at modern accelerator facilities
such as CEBAF (Continuous Electron Beam Accelerator Facility) at
Jefferson Lab has opened the way for dedicated programs of nucleon
structure measurements with unprecedented precision.
The data have allowed phenomena such as quark-hadron duality and the
transition to scaling in transverse and longitudinal nucleon structure
functions to be accurately verified, as well as precision studies
to be conducted of the flavor and spin structure of the proton
in kinematic regions previously inaccessible (see, {\it e.g.},
Refs.~\cite{Christy, Chen:2011zzp, Melnitchouk05} and references
therein).

In particular, the region of large parton (quark) momentum fraction
($x \gtrsim 0.5$), which is experimentally challenging because of the small cross
sections involved, has seen a resurgence of interest in recent years
\cite{Holt10, JMO13}, especially at Jefferson Lab with its unique
access to large $x$.
Part of this interest has been the promise to resolve decades-long
questions about parton distribution functions (PDFs) at large $x$,
such as the behavior of the unpolarized $d/u$ or polarized
$\Delta d/d$ ratios in the $x \to 1$ limit.  At large four-momentum
transfer squared, $Q^2 \gg 1$ GeV$^2/c^2$, these offer relatively clean probes of the
strong interaction dynamics of valence quarks in the nucleon.
To access  information on $d$ quarks, and in particular these ratios,
one needs electron scattering data from both proton and neutron targets.
However, while experiments have been able to map out in great detail
the characteristics of the proton at large $x$, determining the
corresponding structure of the neutron has proved to be much more
difficult.

At lower values of $Q^2$ (of order 1  GeV$^2/c^2$), 
the large-$x$ region is dominated by nucleon
resonances, among which the $\Delta (1232)$ is the lowest-mass
 excitation.  A fundamental question here is whether the
ratio $\sigma_n/\sigma_p$ of neutron to proton 
inclusive electron scattering cross sections for
the $N \to \Delta(1232)$ transition is unity, as would be expected
for a pure isovector transition ($\Delta I = 1$).  Existing deuteron
electroproduction data \cite{Bleckwenn, Kobberling, Stuart} indicate
that the isotensor ($\Delta I = 2$) contribution is small but
non-negligible.  
Similarly, comparing inclusive cross sections on the neutron with those on the proton for the
higher-lying (overlapping) resonance excitations can provide constraints on
the isospin structure of the resonant and non-resonant contributions to the 
total cross section.
Finally, neutron structure functions in the resonance region are
needed to conclusively test Bloom-Gilman duality \cite{Bloom69}
in the neutron.

The absence of free neutron targets has meant that in practice light
nuclei such as the deuteron and $^3$He are routinely used as effective
neutron targets.  In regions of kinematics where most of the neutron's
momentum is carried by a single valence quark, or where the spectrum
is dominated by resonances, different choices for models of nuclear
corrections can lead to significant uncertainties in the neutron
cross sections \cite{slacdata1, Melnitchouk96, Arrington09, CJ_accardi,
Arrington:2011qt, Owens:2012bv}.
Consequently our ability to determine unambiguously the isospin
structure of the nucleon PDFs, as well as the spectrum of the
excited states of the nucleon, has been severely limited.
For example, in the nucleon resonance region there are large
uncertainties in the neutron to $N^*$ transition helicity amplitudes
extracted from deuteron measurements, while in the deep-inelastic
scattering (DIS) region the $d$-quark PDF is poorly determined
beyond $x \sim 0.6$.
Aside from the intrinsic value of such knowledge, a practical
ramification is that the large-$x$ PDF uncertainties can in some
cases propagate to influence production rates of particles,
including those predicted beyond the Standard Model, at high-energy
colliders such as the Large Hadron Collider \cite{Kuhlmann00,
Brady:2011hb}.

To move beyond this impasse, it has been suggested \cite{frstr,
cda2_2, mel_sarg_strik, sargsian} that one can minimize the nuclear
model uncertainties by selecting (or ``tagging'') final states in the
electron--deuteron scattering process in which the proton is produced
with small momentum in the backward hemisphere relative to the
momentum transfer.  This minimizes the probability of rescattering
of the ``spectator'' proton with the rest of the hadronic debris,
thereby ensuring that the reaction took place on a neutron close
to its mass shell \cite{cda4, Cosyn11}.

The first direct extraction of 
inclusive scattering data on a nearly free neutron
using this spectator tagging technique was performed with the BONuS
(Barely Off-shell Nucleon Structure) experiment at Jefferson Lab,
which ran in 2005 in Hall~B using CLAS and a novel Radial Time
Projection Chamber (RTPC) capable of detecting protons with momenta
down to 70~MeV/$c$.
In a first report \cite{prl}, a representative sample of the BONuS
neutron spectra was presented, allowing a first glimpse into the
inclusive neutron excited mass spectrum and the neutron $F_2^n$
structure function at large $x$, essentially free of nuclear
correction uncertainties.  In this paper we present the full BONuS
data sample.  These data cover a large kinematic range, from the
quasielastic peak to the region of final-state hadron masses
$W \approx 2.7$~GeV/$c^2$, and $Q^2$ from 0.7 to 5~GeV$^2$/$c^2$.

In Sec.~\ref{sec_physics} we review the basic formulas for describing
spectator proton tagging in semi-inclusive scattering from the deuteron
within the impulse approximation (IA), and discuss various corrections
to the IA due to final-state interactions, nucleon off-shellness and
other effects.  An overview of the experimental setup is presented in
Sec.~\ref{sec_setup}, where we outline the novel features of the BONuS
RTPC.  Details of the data analysis are given in Sec.~\ref{analres},
which describes the event selection and background subtraction,
and two different methods of analysis.
The results of the experiment are presented in Sec.~\ref{results}.
We present results both for the ``spectator limit'' (slow, backward
protons), which can be used to constrain models of neutron structure
with minimal nuclear binding uncertainties, and for kinematics
in which nuclear and final-state interaction effects are enhanced
(forward and higher-momentum protons).  Our analysis allows us to
identify kinematic regions in which the spectator approximation can
be used for extracting the free neutron structure function.
Finally, in Sec.~\ref{sec_summary} we summarize our findings and
discuss future extensions of the spectator tagging technique planned
at the energy-upgraded 12~GeV Jefferson Lab facility.


\section{Physics overview}
\label{sec_physics}

In this section we review the physics motivation for the BONuS
experiment and the formalism employed to analyze semi-inclusive
scattering from the deuteron with a tagged spectator proton.
We discuss the accuracy of the nuclear impulse approximation used
to extract the neutron structure function from the semi-inclusive
cross section, and examine various corrections to the IA from
final-state interactions and nucleon off-shell effects.

\subsection{Motivation}

There are a number of reasons why knowledge of the free neutron
structure functions is vital for our understanding of the quark
structure of the nucleon, and nonperturbative QCD more generally.
In the nucleon resonance region, an accurate determination of
neutron structure functions is needed for the extraction of
the full isospin dependence of the resonant and nonresonant
contributions to the inclusive neutron cross section.
Knowledge of the neutron resonance structure is also needed
for the model-independent verification of Bloom-Gilman duality
in the neutron \cite{Bloom69, Niculescu:2000tk, Arrington:2003nt,
Melnitchouk05}, and for understanding the transition between
the resonance and deep-inelastic regions.  While existing
model-dependent studies \cite{Malace1} suggest a common origin
of duality for the neutron and proton, proof of this requires
neutron resonance data that are free of nuclear model assumptions.

Unfortunately, the absence of high-density, free neutron targets
has usually forced neutron structure to be extracted from inclusive
scattering experiments on nuclear targets, such as the deuteron.
Such extractions, however, necessarily involve model-dependent methods
to account for nuclear effects in the deuteron \cite{Malace1}.
The extraction of the neutron structure function in the resonance
region from inclusive nuclear data is particularly challenging because
of Fermi smearing, which acts to reduce the distinctiveness of the
resonance peaks from the nonresonant background \cite{Kahn:2008nq}.

Of course, definitive tests of quark-hadron duality must involve
data from both the resonance and DIS regions.  For the latter,
the parton model allows the structure of the nucleon to be
characterized in terms of the nucleon's valence $u$- and $d$-quark
momentum distributions.  Following many years of DIS and other
high-energy scattering experiments, a detailed picture has emerged
of the structure of the nucleon at intermediate and small values
of Bjorken $x$.
%
%
The abundance of high-precision proton structure function ($F_2^p$)
data has, due to the preferential coupling of the photon to $u$
quarks compared with $d$ quarks in the proton, allowed an accurate
determination of the $u$-quark PDF at both small and large values
of $x$.

\begin{figure}
\begin{center}
\includegraphics[width=0.45\textwidth]{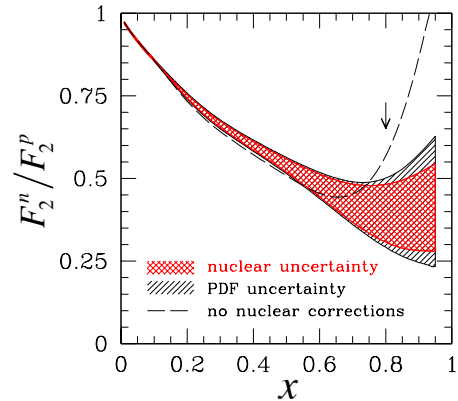}
\end{center}
\vspace*{-1.cm}
\caption{(Color online)
	Ratio of inclusive neutron to proton structure functions
	$F_2^n/F_2^p$ from the CJ global PDF analysis \cite{CJ_accardi}.
	The shaded bands illustrate the range of possible values for the ratio
	from nuclear corrections and experimental uncertainties. The vertical arrow indicates the edge of the region
in $x$ where the ratio is constrained by data ($x \lesssim 0.8$).}
\label{fig_CJnp_ratio}
\end{figure}

The corresponding $d$-quark distribution could be similarly
constrained by neutron structure function ($F_2^n$) data,
and the $d/u$ ratio extracted, at leading order in the strong
coupling constant and for $x \gtrsim 0.5$, via
\begin{eqnarray}
\frac{d}{u} \approx \frac{4F_2^n/F_2^p - 1}{4 - F_2^n/F_2^p},
\label{eq_doveru}
\end{eqnarray}
where the approximation neglects strange and heavier quarks.
At high values of $x$ (where large nucleon momenta contribute
significantly in nuclei) the uncertainties associated with the
nuclear corrections propagate to the extracted neutron structure
functions, and hence to the $F_2^n/F_2^p$ ratio \cite{slacdata1,
Melnitchouk96, Arrington09, Arrington:2011qt, CJ_accardi,
Owens:2012bv}.
The results for $F_2^n/F_2^p$ from a recent global fit by the
CTEQ-Jefferson Lab (CJ) Collaboration \cite{CJ_accardi} are
illustrated in Fig.~\ref{fig_CJnp_ratio}, showing both the
uncertainties from nuclear corrections and experiment.
Beyond $x \approx 0.5$ the current data not only prevent us from
understanding the basic nonperturbative dynamics responsible for
the behavior of $d/u$ in the $x \to 1$ limit, for which predictions
range from 0 to $\approx 0.5$ \cite{Melnitchouk96, Holt10}, but
can also impact our ability to reliably compute QCD cross sections
in high-energy collider experiments which have sensitivity to the
$d$-quark PDF \cite{Brady:2011hb}.

Measurement of the free neutron structure function would also allow
for a model-independent determination of the size of the nuclear
correction in the deuteron through the construction of the
$F_2^d/(F_2^p + F_2^n)$ ratio. 
This would provide data that could discriminate between various
detailed models of nuclear effects in the deuteron \cite{frstr,
meln_schr_thom, Kulagin06, West:1972qj, Frankfurt:1988nt,
Kaptar:1991hx}, thereby solving the decades-long question
about the magnitude of the nuclear EMC effect in the deuteron.
Finally, reliable parametrizations for $F_2^n$ are needed to extract
ratios of nuclear to nucleon structure functions from inclusive
measurements on nuclear targets, and on spin structure functions
from polarization asymmetries in inclusive scattering.

\subsection{Spectator tagging}

Since the deuteron is a weakly bound system with binding energy
$\epsilon_d = -2.2$~MeV (only about 0.1\% of the deuteron mass),
on average the deuteron structure function may be reasonably well
approximated by a sum of free proton and neutron structure functions.
At large values of $x$, however, the deuteron structure functions
receive increasingly greater contributions from nucleons carrying
a larger fraction of the deuteron's momentum.
These contributions are sensitive to the details of the high-momentum
tails of the deuteron wave function, which are not as well constrained
by nucleon--nucleon scattering data as the low-momentum components.
Consequently, in the high-$x$ region there is a more significant
dependence on the model for the smearing of the nucleon structure
due to binding and Fermi motion effects, as well as to possible
modifications of nucleon structure when the nucleon is off its
mass shell.

The nuclear model uncertainties in the extraction of the neutron
structure function from inclusive electron--deuteron scattering
data can be significantly reduced by detecting low-momentum protons
produced at backward kinematics, relative to the momentum transfer,
in coincidence with the scattered electron,
\begin{equation}
  e + d \to e + p_s + X.
\label{eq_specreac}
\end{equation}
The restriction to low momenta ensures that the scattering takes place
on a nearly on-shell neutron \cite{mel_sarg_strik, cda2_2, sargsian},
while tagging backward-moving spectator protons ($p_s$) minimizes
final-state interaction effects \cite{cda4, Cosyn11}.

The cross section for the semi-inclusive electroproduction of a proton
with four-momentum $p_s^\mu = (E_s, \bm{p}_s)$ can be written in the 
deuteron rest frame as \cite{sargsian, Cosyn11}
\begin{eqnarray}
\frac{d\sigma}{dx dQ^2 d^3\bm{p}_s / E_s}
&=& \frac{4\pi \alpha^2_{\rm em}}{xQ^4}
    \left( 1-y-\frac{x^2 y^2 M^2}{Q^2} \right)	\nonumber\\
& & \hspace*{-3cm}
\times
\left[
  F_L^d\
+\ \left( \frac{Q^2}{2\bm{q}^2} + \tan^2\frac{\theta}{2} \right)
   \frac{\nu}{M} F_T^d 				
\right.						\nonumber\\
& & \hspace*{-2.5cm}
+\ \sqrt{ \frac{Q^2}{2\bm{q}^2} + \tan^2\frac{\theta}{2} }\,
   \cos\phi\, F_{TL}^d\
+\ \cos2\phi\, F_{TT}^d
\Big],
\label{spec_crosssec}
\end{eqnarray}
where $\alpha_{\rm em}$ is the electromagnetic fine structure constant,
and
  $E_s = \sqrt{M^2 + \bm{p}_s^2}$ and $M$ are the energy and mass,
respectively, of the spectator proton produced at an azimuthal angle
$\phi$ around the $z$ axis (defined along the $\bm{q}$ direction).
The four-momentum transfer to the deuteron is given by
  $q^\mu = (\nu, \bm{q})$,
with
  $Q^2 \equiv -q^2$ and
  $x = Q^2/2M\nu$ the usual Bjorken scaling variable evaluated in the
target rest frame.
The variable
  $y = \nu/E_e$ denotes the fractional loss of the electron energy $E_e$,
and $\theta$ is the electron scattering angle.

The semi-inclusive deuteron structure functions $F^d_L$, $F^d_T$,
$F^d_{TL}$ and $F^d_{TT}$ depend on the variables $x$, $Q^2$,
the light-cone momentum fraction of the spectator proton
$\alpha_s = (E_s - p_s^z)/M$, and the spectator proton transverse
momentum $p_s^{\perp}$.  In terms of the angle between the outgoing
spectator proton and the direction of $\bm{q}$, the longitudinal
and transverse spectator momenta are given by
      $p_s^z = |\bm{p}_s| \cos \theta_{pq}$ and
$p_s^{\perp} = |\bm{p}_s| \sin \theta_{pq}$, respectively.
Integrating over the azimuthal angle $\phi$, the terms proportional
to $F^d_{TL}$ and $F^d_{TT}$ vanish, and the cross section of
Eq.~\eqref{spec_crosssec} becomes proportional to the familiar
combination of semi-inclusive (SI) structure functions
  $(2\nu/M) \tan^2(\theta/2)\, F_1^{d\, {\rm (SI)}}
   + F_2^{d\, {\rm (SI)}}$,
where
\begin{subequations}
\begin{align}
F_1^{d\, {\rm (SI)}}
&= {1\over 2} F_T^d,				\\
F_2^{d\, {\rm (SI)}}
&= F_L^d + \frac{x}{\rho^2} F_T^d,
\end{align}
\label{eq_sistrfun}%
\end{subequations}%
with $\rho^2 = 1 + 4M^2x^2/Q^2$.
%
%
The semi-inclusive structure functions $F_{1,2}^{d\, {\rm (SI)}}$
are then related to the inclusive deuteron structure functions
$F_{1,2}^d$ simply by integrating over the spectator proton momentum
$\bm{p}_s$.

\begin{figure}
\begin{center}
\includegraphics[width=0.45\textwidth]{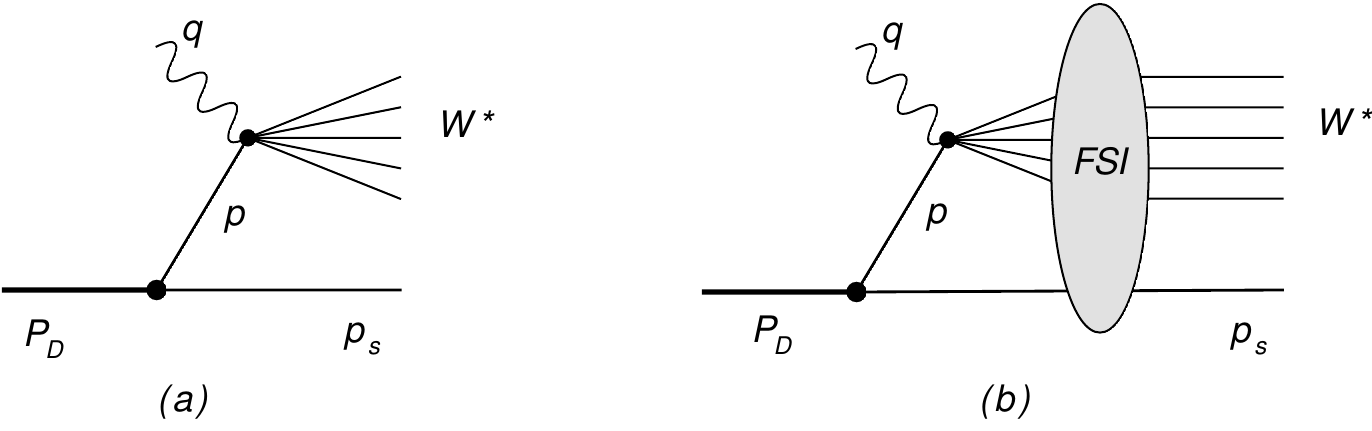}
\end{center}
\caption{Semi-inclusive scattering from a deuteron with detection
	of a spectator proton, $p_s$, within the framework of
	(a) the nuclear impulse approximation, and
	(b) including the effects of final-state interactions.}
\label{fig_spec_ia_fsi}
\end{figure}

In the nuclear impulse approximation, illustrated in
Fig.~\ref{fig_spec_ia_fsi}(a), the virtual photon scatters
incoherently from the bound neutron with four-momentum $p^\mu$,
where
  $p^\mu + p_s^\mu = P_d^\mu = (M_d, \bm{0})$ in the deuteron
rest frame, with $M_d$ the deuteron mass.
In this case the semi-inclusive deuteron structure functions can be
written as products of the structure functions of the bound neutron
and the nuclear spectral function $S(\alpha_s,p_s^\perp)$
\cite{sargsian, Cosyn11},
\begin{subequations}
\begin{eqnarray}
F_1^{d\, {\rm (SI)}}(x,Q^2,\alpha_s,p_s^\perp)
&\approx& S(\alpha_s,p_s^\perp)				\nonumber\\
& & \hspace*{-3.5cm}
\times
\left[ 
  F_1^{n, \rm eff}(x^*,Q^2,\alpha,p^\perp)
+ \frac{p^{\perp 2}}{2p \cdot q}\,
  F_2^{n, \rm eff}(x^*,Q^2,\alpha,p^\perp)
\right]							\nonumber\\
& &							\\
F_2^{d\, {\rm (SI)}}(x,Q^2,\alpha_s,p_{\perp s})
&\approx& S(\alpha_s,p_s^\perp)\, \frac{M\nu}{p\cdot q}	\nonumber\\
& & \hspace*{-3.5cm}
\times
\left[  \Big( 1+\sqrt{1-\frac{Q^2}{2\bm{q}^2}} \Big)^2
	\Big( \alpha + \frac{2 p \cdot q}{(\nu+|\bm{q}|) M_d} \Big)^2
      + \frac{Q^2}{2\bm{q}^2}\, \frac{p_s^{\perp 2}}{M^2} 
\right]							\nonumber\\
& & \hspace*{-3.5cm}
\times\ \, F_2^{n, \rm eff}(x^*,Q^2,\alpha,p^\perp),
\end{eqnarray}
\label{eq_eff_strfun}%
\end{subequations}%
where $F_{1,2}^{n, \rm eff}$ are the bound or ``effective''
neutron structure functions.  In the on-shell limit, the bound
neutron structure functions reduce to the free neutron structure functions,
  $F_{1,2}^{n, \rm eff} \to F_{1,2}^n$,
but in general are functions of the off-shell neutron's invariant
Bjorken variable
\begin{equation}
x^* = {Q^2 \over 2 p \cdot q} \approx {x \over \alpha},
\end{equation}
the struck neutron's light-cone momentum fraction
  $\alpha = 2-\alpha_s$,
and its transverse momentum
  $\bm{p}^\perp = -\bm{p}_s^\perp$.
Alternatively, one can also express $F_{1,2}^{n, \rm eff}$
as a function of the final-state invariant mass squared
\begin{equation}
W^{*2} = (p + q)^2 = p^2 + {Q^2 (1-x^*) \over x^*},
\end{equation}
where $p^2 = (M_d-E_s)^2 - \bm{p}^2$ is the invariant mass squared
of the off-shell nucleon.  Note that in the on-shell limit,
the struck nucleon's Bjorken variable $x^* \to x$,
while $W^{*2}$ reduces to the invariant mass squared
  $W^2 = M^2 + Q^2 (1-x)/x$
for a free nucleon at rest.

The nuclear spectral function $S$ describes the probability of finding
an off-shell neutron in the deuteron with momentum $(\alpha, p^{\perp})$
and an on-shell proton with momentum $(\alpha_s, p_s^\perp)$.  
It is determined by the square of the deuteron wave function
$|\psi_d(p)|^2$ and kinematic factors that depend on the framework
used to compute the nuclear structure function.  These factors
coincide in the limit where both nucleons are on-shell, but differ
in the off-shell behavior \cite{mel_sarg_strik, sargsian}.
The expressions in Eqs.~\eqref{eq_eff_strfun} for the semi-inclusive structure
functions can be used to extract the free neutron $F_{1,2}^n$ structure
functions in the limit
  $\alpha_s \to 1$ and
  $p_s^\perp \to 0$.
Of course, the experimental data will only be available for some minimum
value of $p_s^\perp$, which will introduce some uncertainty into the
on-shell extrapolation, as discussed in the following sections.

While uncertainties in the nucleon--nucleon interaction at short
distances lead to significant dependence of the inclusive deuteron
structure function on the deuteron wave function for $x \gtrsim 0.6$
\cite{CJ_accardi}, restricting the spectator proton momenta to
$|\bm{p}_s| \lesssim 100$~MeV/$c$ renders these negligible.
Furthermore, comparisons of spectral functions computed within the
instant-form and light-front formulations suggest \cite{mel_sarg_strik}
that at these momenta and $\alpha \lesssim 1.1$ the model dependence
of the spectral function is at the few percent level.

\subsection{Beyond the impulse approximation}
\label{beyond}

\subsubsection{Final state interactions}

Although Eqs.~\eqref{eq_eff_strfun} describe semi-inclusive proton
production in the nuclear impulse approximation, interactions between
the recoil proton and the hadronic debris of the scattered neutron,
illustrated in Fig.~\ref{fig_spec_ia_fsi}(b), can in principle distort
the momentum distribution of the detected protons.
Microscopic calculations of the final state interaction (FSI) effects
within hadronization models and the distorted wave impulse approximation
suggest strong suppression of FSIs at backward spectator proton angles
$\theta_{pq}$ relative to the photon direction \cite{cda4, Cosyn11,
Frankfurt94}.

The main uncertainty in estimating the role of FSIs is the spectator
proton--hadronic debris ($X$) scattering cross section $\sigma_{p X}$.
Frankfurt {\it et al.} \cite{Frankfurt94} estimated this from the
$^2$H$(e,e'p)n$ break-up reaction at high energies using data on soft
neutron production in muon DIS from heavy nuclei \cite{Adams95}.
At backward angles FSIs were found to contribute less than 5\% to the
cross section for $p_s^\perp < 100$~MeV/$c$ and $\alpha_s < 1.5$.

In the hadronization model of Ciofi~degli~Atti {\it et al.} \cite{cda4}
the rescattering cross section $\sigma_{p X}$ was derived from a color
flux tube picture, and found to grow logarithmically with time.
Including the effects of color string breaking and gluon bremsstrahlung,
the resulting FSI corrections were again small in the backward
hemisphere, amounting to $\lesssim 5\%$ for spectator angles
  $\theta_{pq} > 120^\circ$ and $|\bm{p}_s| \lesssim 100$~MeV/$c$.
For larger momenta, 
  $|\bm{p}_s| \approx 200$~MeV/$c$,
FSIs enhance the spectral function by $\approx 20\%$ at backward angles.
FSI are most pronounced in perpendicular kinematics,
  $\theta_{pq} \sim 90^\circ$,
where they can be used as a tool to study the process of hadronization
in nuclei.  Models such as that of Ciofi~degli~Atti {\it et al.}
\cite{cda4} predict that in this angular region, FSI can lead to
either a suppression (for $|\bm{p}_s| \le 200$~MeV/$c$) or a significant
enhancement (for $|\bm{p}_s| \ge 400$~MeV/$c$) of the cross section.
In all existing models, however, it is clear that FSIs can be minimized
to $\lesssim 5\%$ by restricting proton momenta to
  $|\bm{p}_s| \lesssim 100$~MeV/$c$ and spectator angles to
  $\theta_{pq} \gtrsim 100^\circ$,
which serves as a guide for the kinematic cuts utilized in the BONuS
experiment.

\subsubsection{Target fragmentation}

Backward kinematics also suppresses hadronization of low-momentum
protons produced from the debris of the struck neutron \cite{cda2_1,
cda2_2, Bosveld94}.  Although a potentially important contribution
in the forward hemisphere (current fragmentation region), direct
fragmentation into protons was found by Simula \cite{cda2_2} to be
negligible for $\theta_{pq} \gtrsim 90^\circ$ even for large momenta
$p_s$.

\subsubsection{Nucleon off-shell effects}

The dependence of the bound neutron structure functions on the neutron's
off-shell mass squared $p^2 \approx M^2 + 2M\epsilon_d - 2 \bm{p}_s^2$
can introduce additional deviations of the extracted structure
functions in Eqs.~\eqref{eq_eff_strfun} from their on-shell values.
On the other hand, the restriction to low-momentum protons
guarantees that the neutron's virtuality $M^2-p^2$ does not exceed
$\approx 13$~MeV$^2/c^2$ for $p_s = 100$~MeV/$c$, and
$\approx  7$~MeV$^2/c^2$ for $p_s =  70$~MeV/$c$,
the lower acceptance limit of the BONuS detector.

Determining the effect of the nucleon's virtuality on its structure
from first principles is extremely challenging, and in fact cannot
be rigorously defined independently of the nucleon's environment.
The off-shell effects have been estimated within several models
of the nucleon, including dynamical quark--diquark models
\cite{meln_schr_thom, Kulagin06} and effective models in which the
bound nucleon structure functions are evaluated at shifted kinematics
\cite{Heller90, gross_liuti1}.

In the covariant quark--(spectator) diquark model of Melnitchouk
{\it et al.} \cite{meln_schr_thom}, scattering from a bound nucleon is
described in terms of relativistic vertex functions that parametrize
the nucleon--quark--(spectator) diquark interaction, with the vertex
functions constrained by inclusive $F_2^p$ and $F_2^d$ data.
The off-shell effects at low $p_s$ are small as expected, and
increase at higher momenta.  For $p_s < 100$~MeV/$c$, the correction
is essentially zero at $x \approx 0.3$, and does not exceed
$\approx 1\%$ at larger $x$.

A similar model introduced by Gross and Liuti \cite{gross_liuti1}
describes scattering from an off-shell nucleon in terms of a
relativistic quark spectral function, with the bound nucleon structure
function evaluated at a shifted value of $x$ that depends on the mass
of the diquark, the bound nucleon momentum, and the binding energy.
The effects are again small at low spectator proton momenta,
$\lesssim 2\%$ for $p_s < 100$~MeV/$c$, increasing to around 5\%
for $p_s = 200$~MeV/$c$.

Simply on the basis of kinematics, Heller and Thomas \cite{Heller90}
also estimated the role of nucleon off-shellness within an instant form
approach, in which the bound nucleon structure function was evaluated
at a shifted energy transfer that is correlated with the degree to
which the nucleon is off its energy shell.  The off-shell modifications
here were found to be $\lesssim 1\%$ for low spectator momenta
$p_s \approx 100$~MeV/$c$.

In all cases considered, therefore, the effects of the neutron's
off-shellness play only a very minor role as long as spectator
proton momenta are restricted to values $p_s < 100$~MeV/$c$.
At larger $p_s$ the off-shell effects can be studied in conjunction
with data from earlier experiments \cite{Klimenko06}, which measured
spectator proton spectra over the range $280 < p_s < 700$~MeV/$c$,
as a means of probing the medium modifications of the nucleon's
quark structure.

\section{Experimental setup}
\label{sec_setup}
The BONuS experiment was conducted in Hall B of the Thomas Jefferson National Accelerator Facility (TJNAF or Jefferson Lab).
Electrons from the CEBAF beam were scattered off a deuteron target and detected by CLAS. The spectator protons were detected with an RTPC designed specially for this experiment.

CEBAF  is a superconducting radio frequency accelerator facility capable of delivering
continuous polarized electron beams with
energies up to 6 GeV. (It is presently being upgraded for up to 12 GeV beam energy.)
During the BONuS experiment, beam energies of approximately 1.1, 2.14, 4.23, and
5.27 GeV  with beam currents from 2 nA up to 55 nA were employed.

\subsection{CLAS}

The Hall B end station houses CLAS, the 	``CEBAF Large Acceptance Spectrometer''.
 CLAS can detect particles for $\theta$ angles $8^\circ$ -- $142^\circ$ and for approximately 80$\%$
of $2\pi$ in $\phi$. It employs a toroidal magnetic field of up to 2 T produced by 6 superconducting coils.
CLAS consists of several layers of particle detectors, each separated into 6 azimuthal sectors by the torus
magnet coils: 
\begin{enumerate}
\item Drift chambers (DC), which determine charged particle trajectories. They are capable of a momentum  resolution of $\delta p/p \leq 0.5\%$ and angular track resolution of $\delta \theta \leq 1$ mrad, $\delta \phi \leq 5$ mrad for 1 GeV/$c$ 
particles~\cite{Mestayer:2000we}.
\item Cherenkov counters (CC) for electron-pion separation (used in the trigger). CLAS Cherenkov counters are capable of distinguishing pions and electrons up to momenta of approximately 2.8 GeV/$c$~\cite{Adams:2001kk}.
\item Scintillation counters (SC) for time-of-flight (TOF) measurements. 
The counters cover the $\theta$ range between $8^\circ$ and $142^\circ$ and the entire active range in $\phi$ (for a total area of 206 $\rm m^2$) \cite{essmith}. The time resolution of the system is between 70 ps (for the shortest counters) and 165 ps (for the longest counters).
\item Electromagnetic calorimeters (EC) to identify electrons and to detect neutral particles like photons and neutrons. 
The EC are used to trigger on
electrons at energies above 0.5 GeV. 
The sampling fraction is approximately 0.3 for electrons of 3 GeV and greater, and for smaller energies, there is a monotonic decrease to about 0.25 for electrons of 0.5 GeV~ \cite{ec}. The average rms resolution is 2.3 cm for electron showers with more than 0.5 GeV  of energy deposited in the scintillator. The timing resolution of the EC for electrons averages to 200 ps over the entire detector.
\end{enumerate}

All detectors listed above are standard CLAS equipment and have been in Hall B for over a decade. CLAS is described in
detail in Ref.~ \cite{mecking}. They were complemented by a dedicated RTPC utilizing Gas
Electron Multipliers (GEMs) that was built specifically for this experiment (see below). It was designed to detect heavily ionizing, slow moving protons that can not travel far from the target.

\subsection{Radial Time Projection Chamber}

\begin{figure*}[htb!]
  \includegraphics[width=0.8\textwidth]{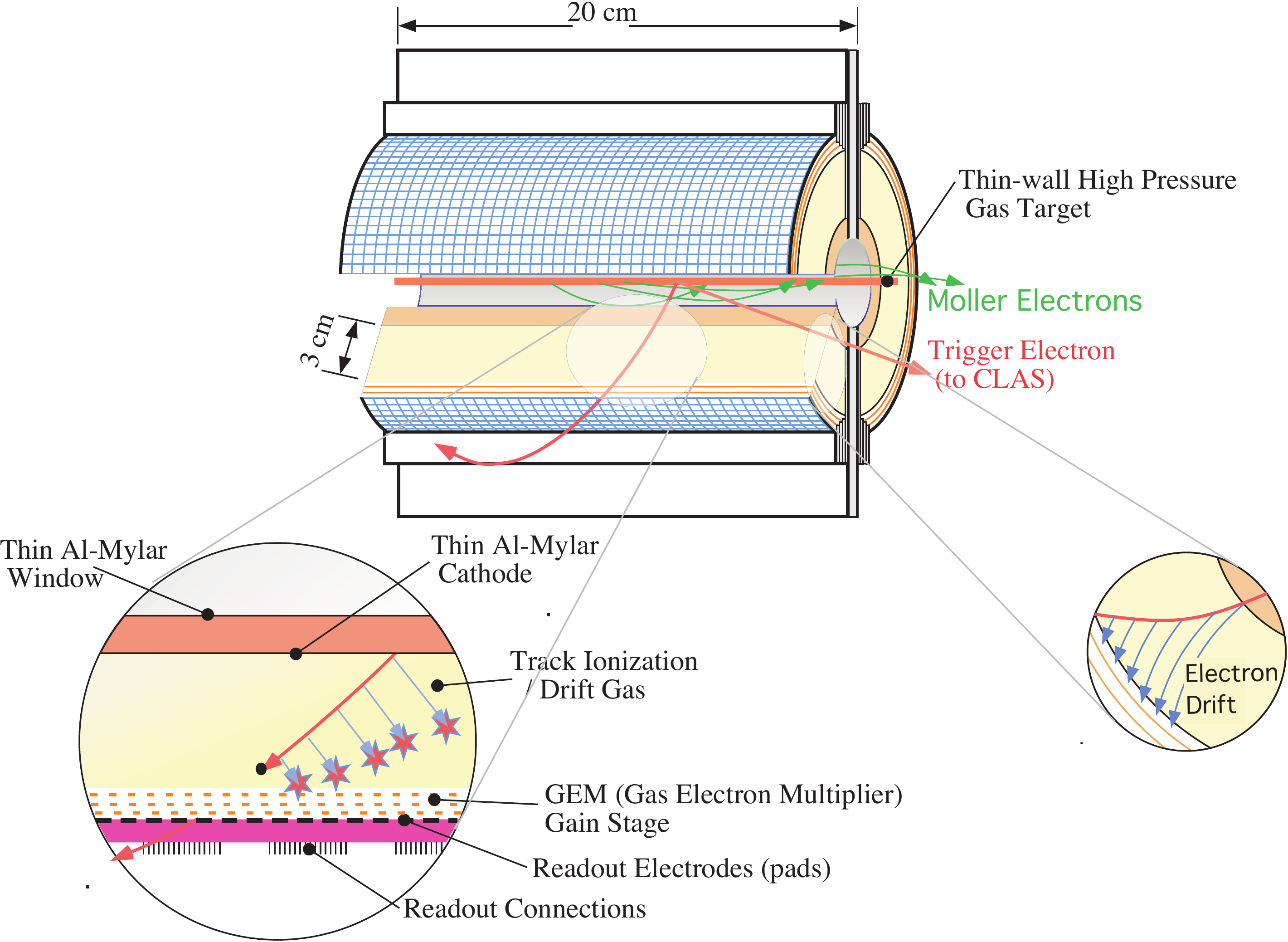}
\caption{(Color Online) Schematics of the BONuS RTPC. See text for details.
\label{rtpc}}
\end{figure*}

To identify events in which a proton is a
mere ``spectator'' to the electron-neutron collision, we needed to select
events in which the detected proton is  moving backwards  with
 low momentum (around or below 0.1~GeV/$c$). To register such protons, we needed a detector that
provides good coverage in the backward hemisphere  (with respect to the direction
of the electron beam), and is close enough to the target to
be able to detect these heavily ionizing low energy protons before they get
stopped. An RTPC~\cite{bonusnim} utilizing GEMs was 
constructed for this experiment to fulfill these requirements
(see Fig.~\ref{rtpc}). The RTPC was surrounded by a solenoid magnet, run at 3.5~T and 4.7~T, that served to analyze proton momenta and, in addition, to deflect Moeller electron trajectories, making them stay clear of all sensitive detector volumes. 

The capability of time projection chambers (TPCs)  to provide a complete 3D picture of particle trajectories in
the detector volume, as well as particle identification through
specific energy loss, $dE/dx$, combined with the low mass density of this kind of detector, made it a natural choice for 
our purposes. The BONuS RTPC utilizes gas for its sensitive volume to reduce the mass density the protons have to traverse.
The more common axial TPC would not have been a good choice for the following reasons:
\begin{itemize}
\item The solenoid magnet length is less than its diameter, and so it does not have magnetic field lines parallel to each other over a reasonable length.
\item Detecting forward moving high-momentum particles with CLAS requires minimizing the end cap density, the region where a lot of equipment is normally situated in axial TPCs. 
\item The RTPC configuration made it easier to stay clear of the Moeller electrons.
\end{itemize}
RTPCs, in which electrons drift radially outwards from the cylindrical central cathode to the anode located on a concentric cylinder, have been previously used, {\it e.g.}, by the STAR \cite{star} and CERES \cite{ceres} collaborations. In this configuration, the electric and magnetic fields are no longer parallel, which leads to complex electron drift trajectories. In addition, curved readout pad planes are required. For these reasons RTPCs have a more complex structure.

\begin{figure}[hbt]
  \includegraphics[width=0.45\textwidth]{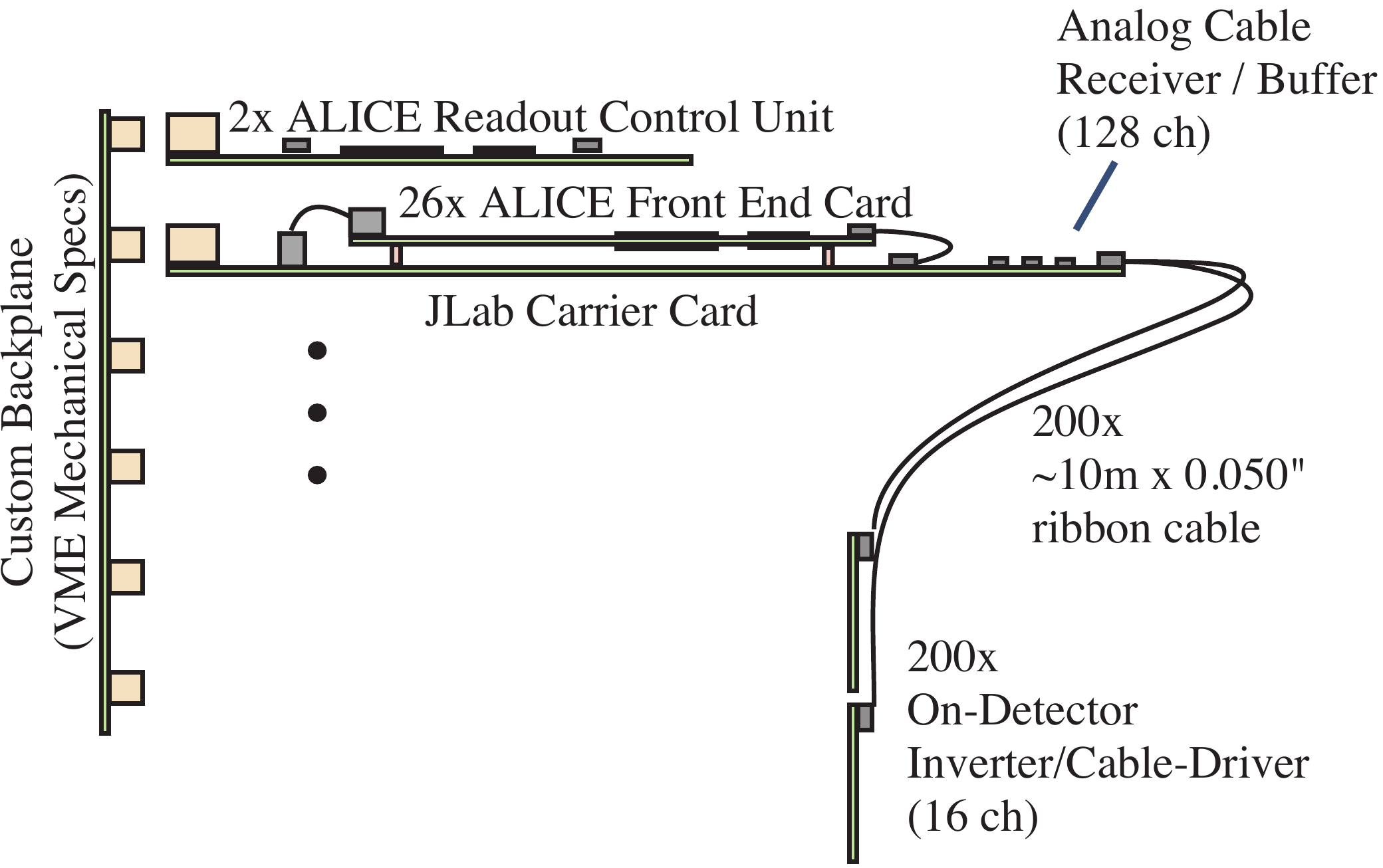}
\caption{(Color Online) BONuS data readout scheme.\label{fig_readout}}
\end{figure}

Since the charge collected at the readout pads is proportional to the
energy loss of the particle, the signal amplitude at the pads as a function of time 
provides information on the specific energy loss of the
particle. A particle's momentum and charge can be found from the
curvature of its trajectory in the magnetic field, hence the particle can
be identified. This requires a quasi-continuous readout of amplitude information
from the pads, generating a potentially large data flow. We designed the BONuS RTPC around
custom integrated
circuits built for the
large TPC used in the ALICE heavy-ion experiment at CERN~\cite{ALTRO,ALICE} (see Fig.~\ref{fig_readout}).

Figure \ref{rtpc} shows the BONuS RTPC with the integrated 7 atm deuterium gas target on its axis.
The target has a fiducial length of 17 cm (visible by the RTPC) and inner diameter of 0.6 cm with 50-$\mu$m Kapton
walls.
The detector surrounds the target at close distance with the center of the RTPC moved 25 mm with respect to the
 target center for better coverage of the backwards hemisphere, where spectator protons are expected. Upon exiting the target and traversing a buffer volume filled with 1 atm helium gas (providing a low mass density region for  Moeller electrons to escape in the forward direction),  protons pass a ground plane  
located at a radius of 2 cm and then the cathode surface at a radius of 3 cm. Upon traversing the cathode, the protons enter the sensitive ionization volume (covering radial distances from 3 cm to 6 cm), filled with an approximately 80$\%$ He/20$\%$ dimethyl ether (DME) mixture. 
Helium as the main component of the mixture provides the necessary low density, which minimizes the energy loss of  slow protons. When traversing the sensitive volume, the spectator ionizes the gas and the released electrons drift towards the amplification and readout stages (see below). The drift region
voltage of the RTPC was kept at 1500 V for all runs.
The resulting electric field produces a sufficiently short clearing time in the drift region without making the cathode voltage so high that a breakdown could occur. 

The BONuS RTPC uses Gaseous Electron Multipliers (GEMs)~\cite{gem_sauli} to amplify the signal 
from the drift electrons. GEM foils are mechanically flexible, robust, and relatively low cost
structures, which can be used in a variety of gases 
and can be placed very close to readout pads, thus decreasing the effects of
charge diffusion. An additional advantage is that they can be formed into non-planar shapes -- the BONuS RTPC was
the first detector to use cylindrically curved GEM foils. 
A total of 3 GEM layers yielded an overall amplification factor of over 1000 during the experimental run.
The GEM gain was limited by the requirement that  non-linearities (saturation) for slow spectator protons had to be avoided. This made the RTPC fairly insensitive to minimum ionizing particles (\textit{i.e.} electrons). 
  The first GEM layer is at 6 cm radius, followed by two more GEM layers at 6.3 and 6.6 cm radius and
the readout pad board at 6.9 cm radius.
The space outside the pad board, within the bore of the solenoidal magnet,
was reserved for preamplifiers and cables. The front and rear caps of the drift region are made of printed-circuit boards patterned with metal traces forming the field cage necessary to make the drift field between the concentric cylinders as close as possible to that between two infinite concentric cylinders. The overall length of the active volume is about 20 cm.

\begin{figure}
  \includegraphics[width=0.49\textwidth]{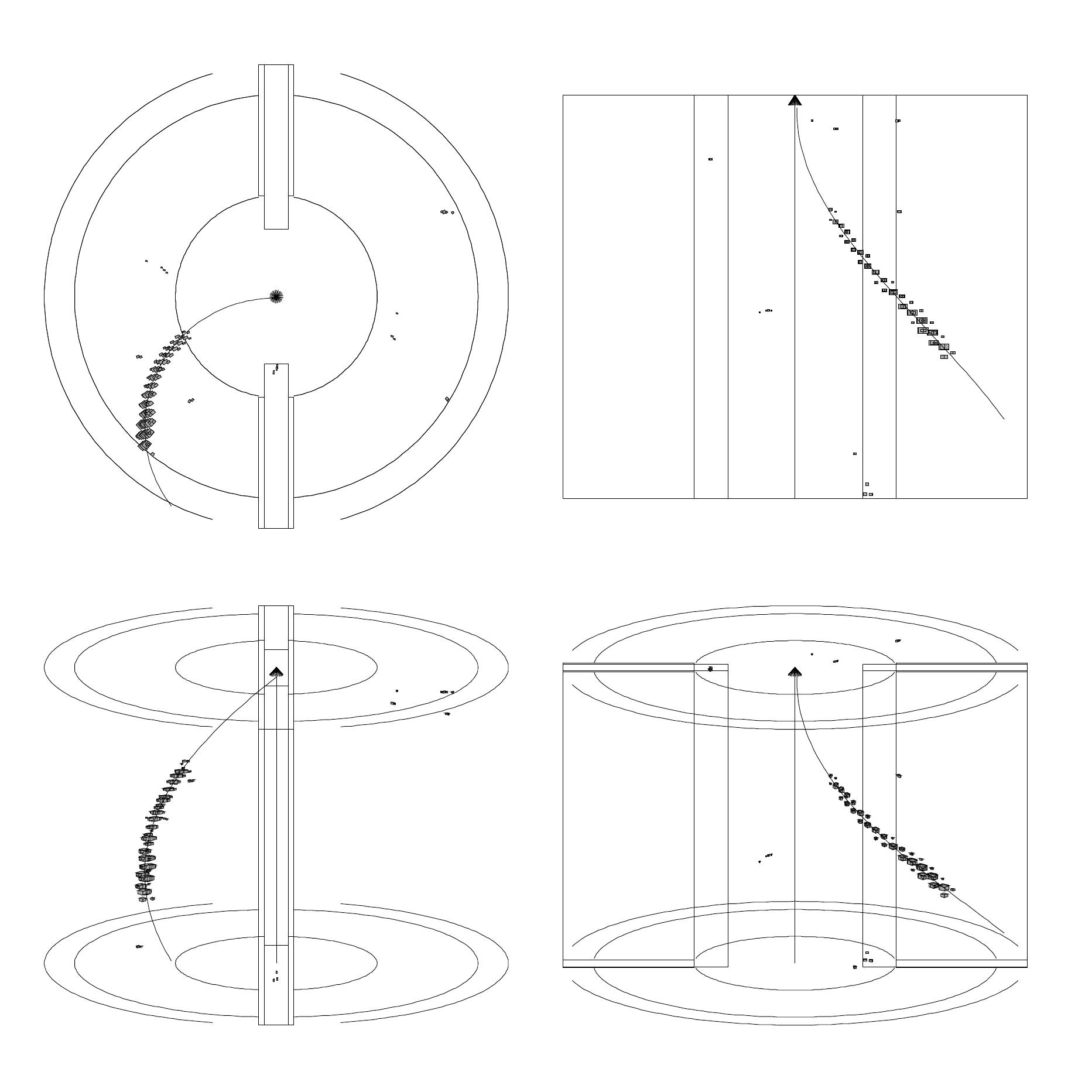}
  \caption{An RTPC event in several views (top row: 2-dimensional projections on end cap and center plane; bottom row: two different 3-dimensional views, the second rotated by $90^\circ$) . Black blobs indicate ionized charge traced back to the spot of the ionization, solid lines going through them indicate fitted tracks. An outline of the RTPC is overlaid.\label{rtpc_event}}
\end{figure}

The RTPC is segmented into two semi-circular halves, each covering an azimuthal angle of around 150$^{\circ}$. 
The readout pads have dimensions of 0.5 cm $\times$ 0.45 cm, thus covering approximately 3.5$^{\circ}$ in azimuthal angle and 0.45 cm along the axis of the cylinder each. Pad rows along the axis of the RTPC are shifted with respect to each other to minimize the probability of a whole track being contained in the same row of pads, thus improving the track resolution. The RTPC is capable of detecting spectator protons with momenta from 0.07 to 0.15~GeV/$c$. Below this range, protons are stopped too soon to leave a substantial track in the RTPC, and above that range, protons are too fast, so that the radius of curvature of their trajectories is too large to confidently reconstruct their momenta (often, they are seen as infinite momentum particles).
Figure \ref{rtpc_event} shows a reconstructed RTPC event. A candidate track curved by the solenoid field is shown. 
The sizes of the symbols indicate the amount of charge collected on a pad. The signal was further amplified, processed by the ALICE readout system, sent to VME crates, and then to Readout Controllers  within the standard CLAS data acquisition system. This system allowed us to read out approximately 1-kB events at a rate of about 500 Hz.

The BONuS event readout was initiated by the standard CLAS electron trigger system selecting interactions with a high probability of having an electron track in CLAS. The data recorded for each event is composed of the time slices (in 114 ns increments) and amplitudes (10 bits) of all RTPC pad signals above threshold for a time period extending from 1.7 $\rm \mu s$ before to 9.7 $\rm \mu s$ after a trigger. This interval is about 1.5 times the maximum drift time in the RTPC.
See Ref.~\cite{bonusnim} for a detailed discussion of the BONuS RTPC.

\section{Analysis}
\label{analres}
\subsection{First pass analysis}
The analysis of the data proceeded in several steps.
As a first step, all detector elements of CLAS and the RTPC were calibrated. After this, all raw digitizations
written to tape were converted into reconstructed events with momentum four-vectors assigned to each
identified particle. Finally, corrections to improve the tracking resolution, including effects like ionization energy
loss of all charged particles, were applied. Most of these steps are part of a standard CLAS analysis
(see, \textit{e.g}.,  \cite{mythesis} for a more detailed description), with the exception of the work related to the RTPC, which
was first used in this experiment.

\subsubsection{RTPC calibration}

Two kinds of calibrations are needed for the RTPC:
\begin{itemize}
\item Drift velocity calibration -- finding time-to-distance correspondence for drifting electrons.
\item Pad gain calibration -- finding the correspondence between registered charge and ionization energy loss.
\end{itemize}

For the drift velocity calibration, ionization electron paths were generated using the MAGBOLTZ program \cite{magboltz}. The result is a function converting any pad signal (given by the pad coordinates and the arrival time
$T_{sig}$) to a spatial point~\cite{bonusnim}:
\begin{equation}
  (x,y,z) = f_{xyz}(j,T_{sig};V_{cathode},V_{GEM},R_{gas}, B_{sol}),
\end{equation}
where $j$ is the pad number and $T_{sig}$ is the time difference between the start time (given by the electron trigger) and
the time when the signal was recorded at the pad. The function $f_{xyz}$ depends on the cathode voltage, $V_{cathode}$,  the GEM voltage, $V_{GEM}$, the solenoidal magnetic field $B_{sol}$, and the fraction $R_{gas}$ of helium in the He/DME drift gas  mixture. 

\begin{figure} [ht]
  \includegraphics[width=0.4\textwidth]{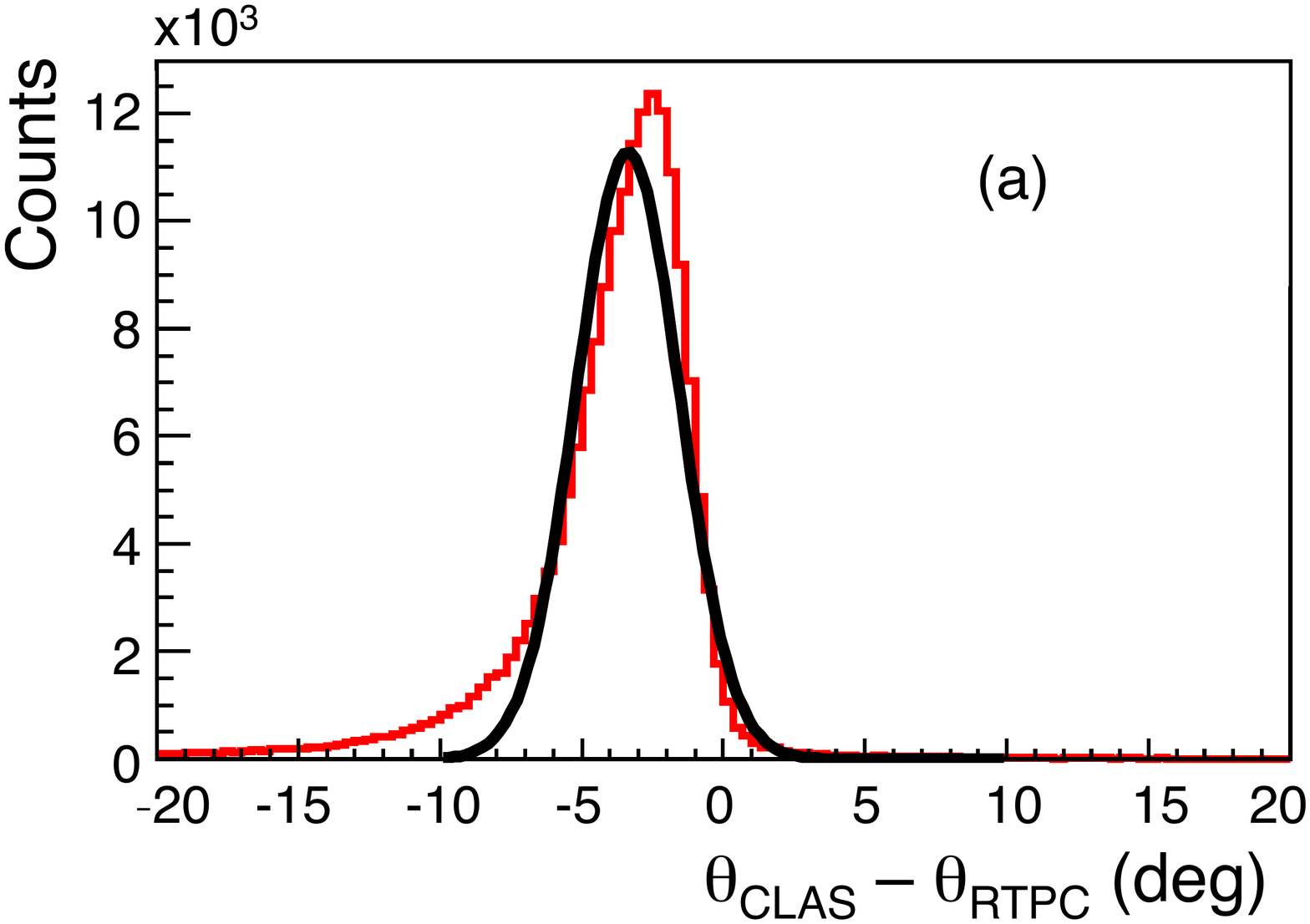}
  \includegraphics[width=0.4\textwidth]{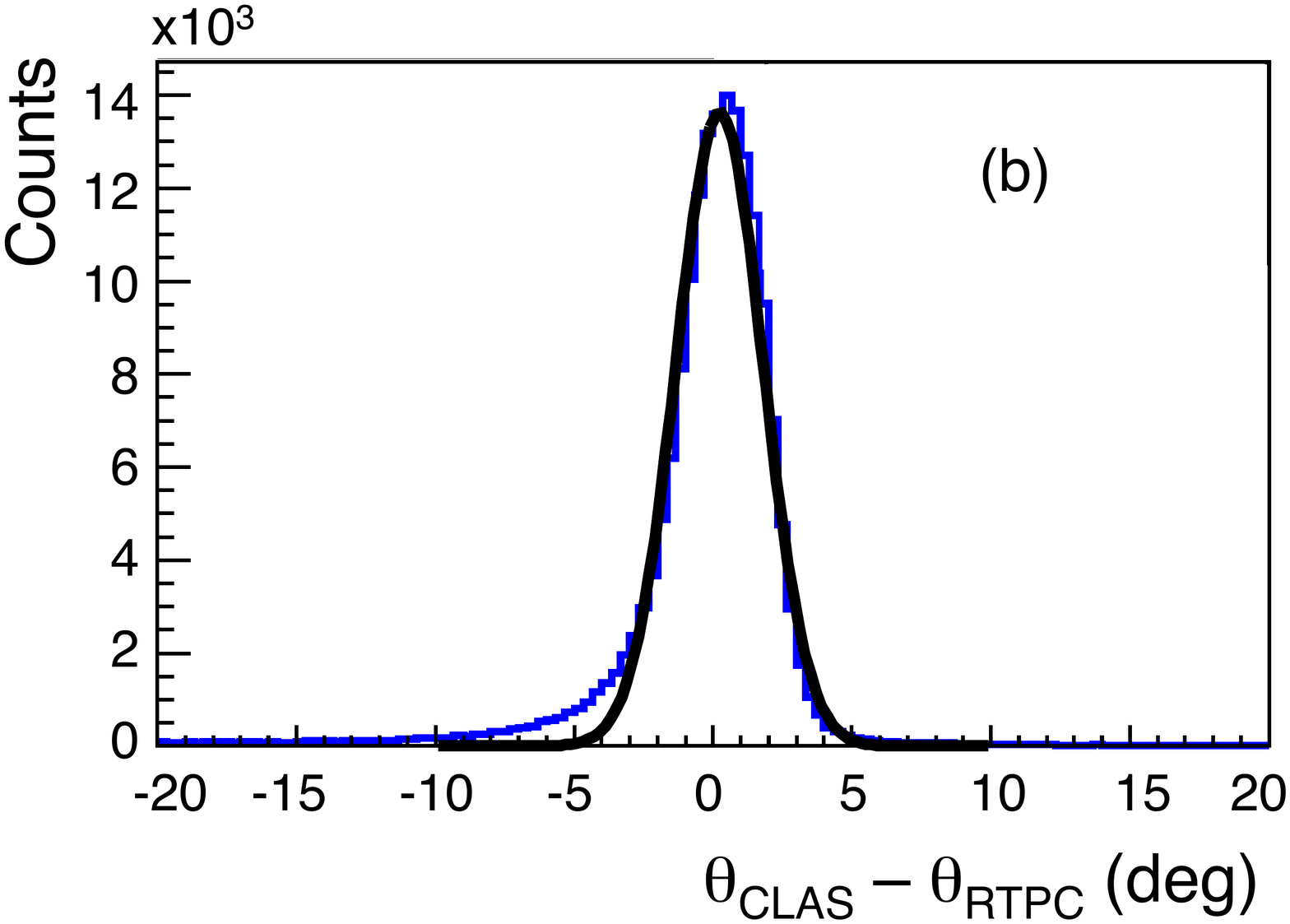}
\caption{(Color online) Comparison of electron scattering angles, as reported by the RTPC and CLAS, before (a) and after (b) calibration. The comparison is shown for the left half of the RTPC; the right half results
are similar. Both experimental distributions (thin colored lines) and Gaussian fits to  them (thick lines) are shown. \label{fig_dth}}
\end{figure}

To correct for our imperfect knowledge of the magnetic field and gas mixture as well as the start time offset, this function was fine-tuned using information from the CLAS detector. A special run with an increased RTPC voltage was conducted so that electrons registered in CLAS were also visible in the RTPC. Cross-checking information from the two detectors allowed us to find optimal parameters for the function $f_{xyz}$. 
Figure~\ref{fig_dth} demonstrates this comparison of track scattering angles between the RTPC and CLAS and shows much better agreement of the angles after the final calibration of the RTPC (bottom). A similar improvement was seen in the reconstructed
$z$ vertex agreement. Some minor discrepancies can still be seen in the CLAS~--~RTPC comparison. Those were taken care of by means of the RTPC and CLAS momentum corrections (see below). 

By comparing average signal sizes from readout pads, we found that the effective detector gain varied considerably across the surface of the RTPC \cite{bonusnim}, most likely due to non-uniformities in the GEM foils or their distance from each other. Therefore, we had to accurately determine the relative responses of all 3200 pads before useful $dE/dx$ information could be extracted from the data.
After the drift velocity/trajectory calibration described above, each track momentum was determined. Using the momentum, the average $dE/dx$ expected for a proton was calculated for the track using the Bethe-Bloch formula (see, for example, \cite{leo}). Using the drift paths obtained in the drift velocity/trajectory calibration, the number of ionization electrons expected to drift to each pad $j$ was determined. Given the measured charge on that pad, we calibrated
its gain $G(j)$ in an iterative procedure.

\begin{figure}
\includegraphics[width=0.4\textwidth]{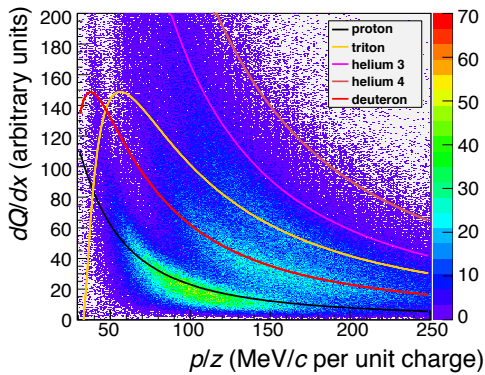}
\caption[(Color online) The ionization density distribution of particles registered by the RTPC after the RTPC gain calibration.]{(Color online) The ionization density distribution of particles registered by the RTPC after the RTPC gain calibration. The solid curves are calculated based on the Bethe-Bloch formula for $dE/dx$ for various particles, in order from bottom to top: proton, deuteron, triton, helion ($^3$He), alpha ($^4$He).
The target was filled with $^4$He gas and the electron energy was 2 GeV for this measurement. \label{fig_dedxafter}}
\end{figure}

The obtained gain-normalization factors were used to scale the raw pulse heights. The same procedure was repeated  excluding tracks whose measured $dE/dx$ after the first iteration was inconsistent with that of protons. The second pass gain-normalization factors were retained and used for the final analysis.
Figure~\ref{fig_dedxafter} shows the  extracted ionization density distributions 
after gain calibration versus measured momentum,  with
the expected functional correlation (from the Bethe-Bloch formula for energy loss $dE/dx$ which should be 
proportional to ionization per unit length) overlaid. One can clearly distinguish several bands belonging to
final state protons, deuterons and heavier nuclei (for these data, the target was temporarily filled with $^4$He gas).

\subsubsection{RTPC momentum corrections}
\label{sec_rtpc_corr}
To determine spectator proton momenta at the vertex from the measured track curvature within the annulus
of the sensitive drift region (ranging from 3 to 6 cm from the beam axis), two additional corrections were applied:
\begin{enumerate}
  \item The track curvature itself was corrected for possible biases in fitting a helical track to the observed
  ionization pattern, as well as for finite position resolution, magnetic field
  inhomogeneities and possible deviation of the ideal (simulated) drift paths and drift velocities from the actual ones.
  \item The corrected curvatures were then converted to momenta at the vertex, after accounting for energy loss in the target gas 
  and the intervening material before reaching the sensitive drift volume.
\end{enumerate}

The mapping between measured curvature and vertex momentum was based on a GEANT4 simulation \cite{geant4}. A
large number of events was generated over  the full range of target $z$ (coordinate along the beam axis) and spectator proton momenta and angles, $p_s,\theta, \phi$. They were subsequently run through a full simulation of the RTPC including signal
conversion and track reconstruction. By comparing the results of the simulation (in terms of the reconstructed
 radius of curvature and angle $\theta$ of the tracks)  with the thrown momenta, we extracted a one-to-one correspondence between the measured radius of curvature and the vertex spectator momentum, accounting for energy loss (see \cite{jzthesis} for more details).

To improve the accuracy of the momentum reconstruction,
we used fully exclusive $^2$H$(e,e'p\pi^-p)$ events, where the first three particles were detected with CLAS and the
last proton with the RTPC. We compared the missing momentum
from the electron, pion and proton measured in CLAS with the reconstructed momentum of the proton detected
in the RTPC. The average agreement of these two quantities was optimized by adjusting the
six parameters of the following correction formulas:
\begin{subequations}
  \begin{align}
    R_{new}      &= R_{old}/(1 + p_1 \cdot R_{old} + p_2)  \\
    \theta_{new} &= (1 + p_3) \cdot \theta_{old} + p_4  \\
    \phi_{new}   &=(1 + p_5) \cdot \phi_{old} + p_6,
  \end{align}
\end{subequations}
where $R_{new}$  and $R_{old}$ are the corrected and reconstructed radius of curvature, respectively, $\theta_{new}$  and $\theta_{old}$ are the corrected and reconstructed polar angle, respectively, and $\phi_{new}$  and $\phi_{old}$ are the corrected and reconstructed azimuthal angle, respectively. $p_1$ $\ldots$ $p_6$ are the fit parameters. All parameters
turned out to be small, leading to corrections of order 2\% on $R$ and less than 1 mrad on $ \theta$ and
$\phi$.

The RTPC--measured momentum distribution of coincident protons
 after these two corrections was similar to the one expected from the pure
spectator picture (given by the deuteron wave function in momentum space), although the measured spectrum falls off 
somewhat faster than predicted. This can be attributed to the RTPC reconstruction efficiency which falls off for higher spectator momenta (due to insufficient charge and track curvature for a reliable track reconstruction). 
We were able to partially correct this efficiency fall-off using the ratio of the number of fully exclusive
$^2$H$(e,e'p\pi^-p)$ to $^2$H$(e,e'p\pi^-)X$  events, where the first three particles in either case were detected with CLAS and we looked for the inferred proton in the RTPC.

\subsubsection{CLAS momentum corrections}
Momenta of particles reconstructed with CLAS were also corrected for minor imperfections (wire misalignments, torus
and solenoid
magnetic field deviations from the ideal field maps used in the reconstruction, beam offset from the ideal center line) and effects like multiple scattering and energy loss. These corrections
have been applied and studied in previous experiments~\cite{Klimenko06,nguler}. We determined correction parameters using a fit to fully 
exclusive BONuS data ($ep \to ep$ and $ep \to ep\pi^+\pi^-$ reactions), following the method described in~\cite{Klimenko06}. After applying all corrections, both the centroid and the widths
of the proton missing-mass peaks were well within the established CLAS resolution and accuracy.
   
\subsection{Event selection and background subtraction}
\label{sec_cuts}

\subsubsection{Particle ID cuts}
For the selection of semi-inclusive D$(e,e'p_s)X$ events, we developed criteria to identify  scattered electrons, $e'$, detected by  CLAS, and spectator protons, $p_s$, detected by the RTPC.

Trigger particles were identified as electrons if they passed the following selection cuts:
\begin{itemize}
      \item Track curvature consistent with a negative charge.
      \item Cherenkov counter signal above the equivalent of 2 photo-electrons for momenta below
        3.0 GeV/$c$. Above this limit, pions can emit Cherenkov radiation and the CC becomes inefficient
        for pion discrimination. (We still required a signal above the equivalent of 1 photo-electron in this case,
        to discriminate against heavier particles like kaons and protons).
        In addition,    geometrical and temporal matching between the CC signal and the measured track was required to eliminate coincidences between CC noise and charged particle tracks, which can result in pions masquerading as electrons \cite{OsipenkoCut}.
      \item Total energy deposited in the EC  above a momentum-dependent threshold consistent with the EC shower sampling
      fraction of $\approx 0.25 - 0.3$. 
      \item At least 0.06 GeV visible energy in the first (front) layers of the EC, which is significantly higher than that 
      expected for minimum-ionizing
      particles like pions.
      \item Track within the fiducial volume (part of the detector with high detection efficiency and no physical obstructions).
\end{itemize}
In addition, the momentum of the trigger electron was required to be larger than 20$\%$ of the beam energy to avoid the kinematic region where radiative corrections and backgrounds become fairly large.

Spectator protons were defined by the following selection cuts
\begin{itemize}
\item Reliable fit of the track in the RTPC ($\chi^2/$d.o.f of the fit less than 4).
\item Positively charged particle.
\item More than 5 pads register above-threshold charge.
\item Energy loss $dE/dx$ consistent with that expected for protons (see Fig.~\ref{fig_dedxafter}; particles with energy loss
more than 2 standard deviations above or less than 3 standard deviations below the measured proton $dE/dx$ distribution
 were rejected). 
\item Beginning and endpoint of the ionization trail reconstructed by the RTPC  within 0.5 cm of the corresponding physical chamber boundary (this is basically a timing cut, since out-of-time tracks will be reconstructed at the wrong radial
positions).
\item $z$ coordinate of the vertex is inside the fiducial target region (between $-6$ cm and $+10$ cm of the RTPC center).
\end{itemize}
In addition, for good electron-proton coincidence events we required that the difference between the $z$ coordinate of the electron
vertex, $z_e$, as reconstructed by CLAS, and the $z$ coordinate of the proton vertex, $z_p$, as reconstructed by the RTPC, be no larger than 1.5 cm
(to exclude accidental coincidences, see below).

Coincident events that passed all cuts were registered in 4-dimensional bins in the kinematic variables $x^*$ or $W^*$, 
$Q^2$, $p_s$ and $\cos \theta_{pq}$. In addition, all  electron events 
from inclusive D$(e,e')X$ that pass the electron cuts above were accumulated
in bins of scattered-electron energy, $E'$, and angle, $\theta_e$.

\subsubsection{\label{sec_acc_bg}Accidental background subtraction}
While the cut on the distance between electron and proton vertices (see above) removes most of the accidental coincidences, the
remainder (when the trigger electron and an unrelated RTPC proton happen to originate within 1.5 cm from
each other) must be quantified and subtracted.

\begin{figure}
\includegraphics[width=0.45\textwidth]{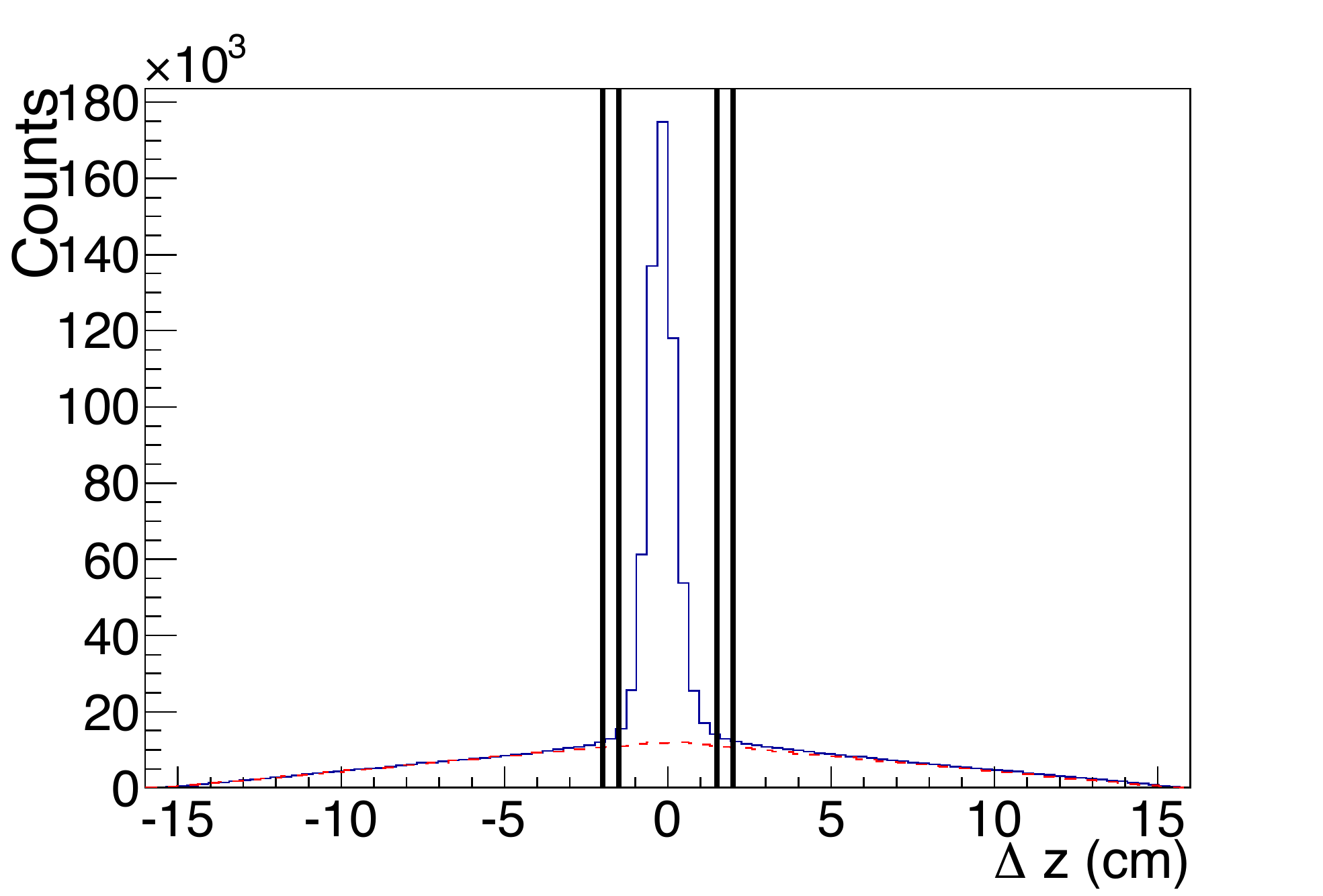}
\caption{(Color online) Representative plot of the
$\Delta z$ distribution for coincidences (solid histogram)
between electrons (in CLAS) and
spectator protons (in the RTPC) from the 5 GeV data set.
Inner vertical lines indicate the region selected for data analysis ($-1.5  \ldots +1.5$ cm). The dashed line indicates 
the corresponding distribution for 
accidental coincidences 
obtained by matching tracks from different events (see text),
cross-normalized to the data outside the outer vertical lines ($-2.0 \ldots +2.0$ cm).
The good agreement in the ``wings'' outside those lines indicates that the shape of the background
is well represented by this method.
See text for more details.
 \label{fig_randz}}
\end{figure}

Such random coincidences can be simulated by taking the trigger electron from one event
(without requiring a matching proton) and the RTPC proton from another event. 
Since spectator protons are distributed rather uniformly in angle (see Section~\ref{results}),
such pairs provide  very good proxies for  true random coincidences.
Using kinematic information from the chosen electron-proton random pair, all quantities in which real data are binned, $Q^2$, $W^*$, $x^*$, $p_s$ and $\cos \theta_{pq}$, are calculated, and the coincidence assigned to the corresponding bin. If the distance between the vertices of the electron and the proton, $\Delta z = z_e - z_p$, is less than 1.5 cm, the event would emulate a random coincidence under the signal. If $\Delta z$ is larger than 2 cm, we consider it a ``wing'' event. Then, after going over all the events within a bin, we  form a scaling ratio, $R_{acc}$, of the number of coincidences under the signal divided by the number of ``wing'' events, separately for each of our 
kinematic bins. 

All same-event experimental coincidences between electrons and RTPC protons are separated into the same categories, ``wing'' events (those with $|\Delta z|>2$ cm) and ``signal'' (peak) events (those with $|\Delta z| < 1.5$ cm).
Then, the number of observed
``wing'' events is scaled by the ratio $R_{acc}$ to yield the number of random coincidences under the peak. The resulting accidental background events are subtracted from the
events within the peak for each kinematic bin.

A sample of the distribution of both same-event and scaled random coincidences is shown in 
Fig.~\ref{fig_randz}; the solid histogram shows the distribution of coincident events from the same ``beam
bucket'' while the dashed line shows the simulated random distribution,
normalized to the wings (outside $\pm 2$ cm). One can clearly see that our method leads to an excellent
approximation of the accidental background in the wings.
After subtracting the accidental distribution from the data, the remaining distribution
is well described by a Gaussian with a resolution of about 0.7 cm ($1 \sigma$).

\subsubsection{Pair symmetric and pion contamination}\label{bgnd}
Electron scattering experiments typically have to account for contamination of the electron sample by 
$e^+/e^-$ pair symmetric contributions as well as the
possible contribution from negative pions misidentified as electrons. 

Pair symmetric
background comes from Dalitz decays ($\pi^0 \rightarrow \gamma e^+ e^-$) and photons converting to $e^+/e^-$ pairs  inside the target enclosure. The decay electron can then be misinterpreted as a scattered beam electron.
The rate of this background (at most a few percent of the electron rate) has been extensively studied in previous CLAS experiments~\cite{nguler} for the case of inclusive electron scattering off isoscalar targets (like deuteron)
and can be parametrized with a simple exponential in both electron and pion momentum and angle.  
This parametrization was applied as a correction to the inclusive D$(e,e')X$ data (between 0\% and 3\%, with
an  average  of about 1\%). For the tagged data,
the correction should be even smaller since it is proportional to the rate of $\pi^0$ and photon production off the
neutron in deuteron (all other channels are automatically subtracted in our treatment of accidental
backgrounds). We therefore did not correct the tagged data and instead included
an overall systematic uncertainty of 1\% due to pair symmetric backgrounds.

Negative pions can be misidentified as electrons if they pass all cuts. 
The size of this contamination was studied in great detail for similar kinematics in an earlier experiment~\cite{nguler}, and it was found to be at most 1\% -- 2\%  for the same set of electron cuts we applied in this work. Since this correction is small compared to other possible systematic effects, it was not applied
to the data but included in the total systematic uncertainty budget.

\subsection{Monte-Carlo based analysis}

To extract quantities of interest from the background-corrected yields, we used two different analysis methods. The first one
uses a full Monte-Carlo
simulation of the experiment to correct for acceptance effects (``Monte Carlo method''),
 while the second one is based on ratios of measured quantities only (``Ratio method''). The Ratio method was used 
for the extraction of the free neutron structure function $F_2^n$ reported by Baillie \textit{et al.}~\cite{prl} and
in this paper;
 it is summarized
 in Section~\ref{altanal}. Some additional results reported below cover a larger range in spectator momenta
and angles of the spectator proton relative to the momentum transfer vector ${\mathbf q}$ and were obtained using the
Monte Carlo method, which is described in detail in the following.
We show a comparison of the results obtained with both methods in 
Section~\ref{compareNS}.

\subsubsection{Event generator}

For the Monte-Carlo based analysis, we simulated both tagged D$(e,e' p_s)X$ events
(where $p_s$ is the spectator proton) and fully inclusive D$(e,e')X$ events (to determine
empirical detector inefficiencies not accounted for by our simulation).
For both processes, we used the same event generator
to (at least partially) cancel model dependencies. We included two basic processes in the generator:
\begin{enumerate}
\item Elastic scattering off deuteron, D$(e,e')$D. We used the well-known deuteron form factors~\cite{HallCdFF}
 and the
prescription by Mo and Tsai~\cite{motsai} to estimate the radiative tail contribution from this process to D$(e,e')X$, which turned out
to be a very small correction to the inclusive cross section in our region of interest. (Obviously, it does not contribute
at all to the tagged cross section).
\item Quasi-free scattering off either a proton or a neutron inside deuteron, within a simple plane wave spectator approximation. This process was further subdivided into quasi-elastic scattering (where the struck nucleon stays
intact) and inelastic scattering off one nucleon (with the other being a spectator). These two processes are described
in more detail below.
Our generator did not contain 
additional processes  like coherent pion production, final-state interactions and other two-nucleon effects; 
therefore, the ratio of measured to simulated tagged data can be interpreted as a direct test of the spectator
picture. On the other hand, these processes do not affect the overall strength of the inclusive cross section
significantly except perhaps in the dip region between the quasi-elastic and the Delta resonance peak.
\end{enumerate}


To simulate scattering off a bound nucleon inside deuteron, we used a simple spectator formalism
where one nucleon is considered to be on-shell and does not participate in the reaction while the other
one is off the mass shell. 
In this picture, the energy and momentum of the off-shell bound nucleon
$p^\mu = (E, {\mathbf p})$  are related to the spectator nucleon momentum ${\mathbf p}_s$ as 
\begin{subequations}
  \begin{align}
    E &= M_d - \sqrt{M^2+p_s^2} \\
    {\mathbf p} &= -{\mathbf p}_s
  \end{align}
\end{subequations}
with $M_d$ the deuteron mass (see Sec.~\ref{sec_physics}). The off-shell mass of the struck nucleon is
\begin{equation}
  M^* = \sqrt{E^2-p_s^2}.
\end{equation}

The initial momentum of the struck nucleon is generated at random with weight
\begin{equation}
  P({\mathbf p}) = |\psi({\mathbf p})|^2,
\end{equation}
where $\psi({\mathbf p})$ is the Paris deuteron wavefunction \cite{pariswf} rescaled using the light-cone formalism \cite{brodsky_lightcone} within the approach by Frankfurt and Strikman~\cite{Frankfurt:1988nt}.

The scattered electron kinematics are generated in the rest frame of the struck nucleon. The scattered electrons are 
distributed according to the
radiated cross section on a nucleon at rest. The distributions are kinematically corrected for the nucleon off-shell mass.
The (quasi)elastic scattering cross section is given by the Rosenbluth formula:
\begin{equation}
  \frac{d\sigma}{d\Omega} = \left(\frac{d\sigma}{d\Omega}\right)_{Point}\frac{1}{\epsilon} \left( \tau G_M^2(Q^2)
  + \epsilon G_E^2(Q^2) \right) \frac{1}{1+\tau},
\label{eq_rosen}
\end{equation}
where $\epsilon = 1/[1+2(1+\tau)\tan^2(\theta_e/2)]$ is the linear polarization of the virtual photon, $G_E$ and $G_M$ are Sachs form factors, and $\tau = Q^2/(4M^2)$. 
We used the parametrization of the proton form factors by Arrington~\cite{Arrington09} and the 
parametrization of Kubon {\it et al.}~\cite{Kubon:2001rj} for $G_{Mn}$ and the Galster~\cite{Galster:1971kv}
 parametrization
for $G_{En}$.
Higher order QED effects and the elastic radiative tail are calculated using the full prescription of Mo and Tsai \cite{motsai}. 

Inelastic events off protons and neutrons in deuteron are generated similarly to the quasi-elastic ones. The cross section is evaluated
 using
\begin{equation}
  \frac{d\sigma}{dE'\,d\Omega} = \left( \frac{d\sigma}{d\Omega} \right)_{Point}
  \frac{2MxF_2(x,Q^2)}{\epsilon Q^2} \frac{1+\epsilon R(x,Q^2)}{1+R(x,Q^2)},
\label{eq_dis_cs_gener}
\end{equation}
where 
$$R=\frac{\sigma_L}{\sigma_T} = \frac{F_2}{2 x F_1} \left(1 +\frac{Q^2}{\nu^2} \right) -1,$$
$\sigma_L$ and $\sigma_T$ being the longitudinal and transverse virtual photo-absorption cross sections. 
The proton and neutron structure functions are taken from Bosted and Christy \cite{bosted_christy}.
Radiative effects are simulated using the code ``RCSLACPOL''  \cite{rcllacpol} which is based on the 
prescription by Mo and Tsai.
The event generator also simulates the (rather small) external radiative energy loss \textit{before} scattering, due to  exit and entrance windows and gas in the beam path, while external radiative and other energy losses after the scattering are included in the detector
simulation (see below). 

The fully inclusive sample is formed by generating quasi-elastic and inelastic events from both the neutron and the proton (integrated over all spectator momenta), plus the radiative elastic tail from $^2$H$(e,e')^2$H. The simulated tagged sample contains only quasi-elastic and inelastic scattering events off bound neutrons, with information on the generated spectator proton being 
kept in addition to that on the scattered electron.

\subsubsection{Detector simulation}
\label{SimulRate}
The generated events are then run through a Monte-Carlo simulation of the experimental set up which includes external radiation and ionization losses after the scattering. The target and RTPC parts of the setup are simulated in detail using the same GEANT4-based simulation package that was used for the RTPC momentum corrections, described in Section \ref{sec_rtpc_corr}. The standard CLAS part of the setup is simulated using the existing GEANT3-based \cite{geant3} package called GSIM. After particle paths through the RTPC are simulated in GEANT4, the output information at the boundary is written to files which serve as input for the GSIM package. To simulate inefficiencies of the CLAS detector, the GSIM Post Processing package (GPP) is run after GSIM.  It makes the GSIM output look more like real data by accounting for dead scintillators and wires and adding some Gaussian smearing to the data to match the
measured detector resolution.

After the generated events are tracked  through the simulated detectors, one obtains files with simulated detector responses for the generated events. Finally, these files are processed by the usual data processing program (RECSIS), the same one used for processing experimental events. After applying the same fiducial and kinematic
 cuts as for the experimental data, we separately
accumulate simulated data from quasi-elastic as well as inelastic scattering off a neutron inside deuteron.
These data are binned
in the same kinematic bins as the experimental tagged data.

Then all events from the elastic, quasi-elastic
and inelastic simulations are combined, after passing inclusive electron cuts, to simulate the inclusive
electron rate. Pair symmetric and pion contamination corrections (see Section~\ref{bgnd}) are applied to these simulated data.
Since the inclusive D$(e,e')$ cross section is well known, the ratio of the inclusive data to
the simulation can be used to extract  remaining inefficiencies of the trigger and of detector elements
like the CC and the EC that were not fully implemented in our simulation. For this purpose this ratio is calculated,
for each beam energy,
in bins of the final electron energy and scattering angle, $E'$ and $\theta_e$. 
The tabulated ratio is used as a weighting factor for each simulated tagged event, depending
on its electron kinematics. This factor turned out to be around 0.85 on average, with a 
standard deviation of 0.072 around this mean.
We used this standard deviation to estimate the point-to-point systematic uncertainty of this correction
as 8.5\%.

\subsubsection{Final data set}
\label{dataextract}

\begin{figure}[h!]
  \includegraphics[width=0.5\textwidth]{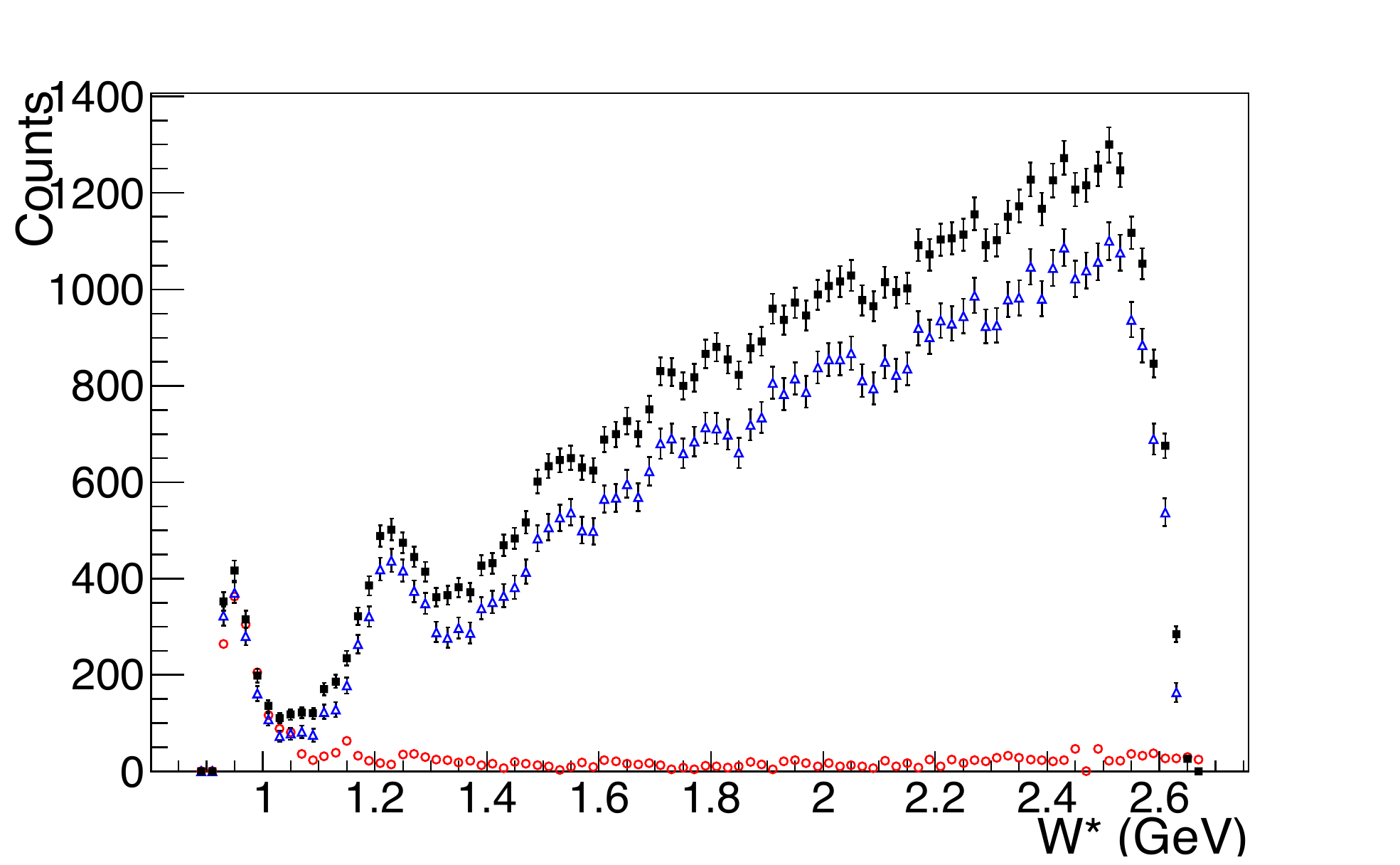}
  \caption{(Color online) $W^*$ distributions  (for 1.10 GeV$^2/c^2 < Q^2 <$2.23 GeV$^2/c^2$) of measured counts for 
the  5.3 GeV beam energy with spectator protons  detected at angles greater than about 100$^\circ$ and momenta between 70 and 85~MeV/$c$.
The data are shown before (top, black squares) and after subtraction of accidental coincidences and other backgrounds (lower blue triangles). Also shown are the normalized simulated counts for elastic scattering off a neutron inside
the deuteron, including the radiative tail (open red circles, bottom). Note the good agreement between this
simulation and the data in the quasi-elastic region, $W^* < 1$ GeV.
\label{fig_simW5gev}}
\end{figure}

The remaining steps of the Monte Carlo method require us to subtract the quasi-elastic radiative tail from the tagged
neutron data, and to normalize our results to account for any remaining RTPC inefficiency not captured by the
GEANT4 simulation.
So, as the next step, we normalize
the  simulated quasi-elastic events (including radiative tail) on the bound neutron to  the measured quasi-elastic strength,
integrated over the region 
0.88 GeV/$c^2$ $< W^* <$ 1 GeV/$c^2$, for each bin in $Q^2$ and spectator kinematics. Figure~\ref{fig_simW5gev} shows
the resulting simulated spectrum as function of $W^*$ for a specific bin in $Q^2$, spectator kinematics and
beam energy, together with the data before and after subtracting experimental backgrounds. The shapes
of the simulated and measured spectra agree well in the region $W^* < 1$ GeV/$c^2$, giving us
confidence that the radiative tail is reasonably well represented by this procedure.
We then subtract this
normalized simulated spectrum from the measured one over the whole $W^*$ range to remove the
 (quasi-) elastic radiative tail  from the measured spectrum. 

The remaining experimental spectrum is due only to inelastic $^2$H$(e,e'p_s)X$ events and can be 
compared to the simulated inelastic spectrum. 
However, the latter must still be normalized to account for the overall efficiency of the RTPC.
In particular, we find that the simulation of the RTPC response did not fully capture the experimentally 
observed RTPC track reconstruction efficiency within cuts, and that this
efficiency varies as a function of proton momentum (from about 0.6 at the lowest $p_s$
down to 0.23 at the upper limit of our $p_s$ range). For this reason, we derive a
normalization factor $N(p_s, E_b)$ for each of our 4 bins in spectator momentum $p_s$.
This factor is also allowed to vary between the different time spans corresponding to each of the  beam energy
settings
used in our experiment (indicated by the dependence on the
 variable $E_b$). We determine this factor using events
in the range $-1 < \cos\theta_{pq} < -0.2$ (backward kinematics). 
According to theoretical expectations and our own data (see next Section), the  spectator picture
works best in this kinematic region.
We match the measured spectrum to the simulated one
in a kinematic region  where the ratio between the two is found to be flat:
$W^*=$
 2.0 -- 2.2 GeV/$c^2$ for both the 4 and 5 GeV data,
within the lowest fully accepted $Q^2$ bin for each energy. The resulting agreement between data
and simulation can be seen in Fig.~\ref{fig_q0cos0_vsW_5gev} which shows the ratio between both. 
This ratio fluctuates around 1.0 by about $\pm 10\%$ in the chosen $W$ region, which is consistent
with the uncertainty $\Delta N(p_s,E_b)$ we assign to the normalization factor, 
see next section.
 Note that
this factor is the same for all bins in spectator angle and in ($W^*, Q^2$) for a given beam
energy setting and $p_s$ bin, allowing us to study the dependence of the data on these variables without normalization
bias. 

After applying the normalization $N(p_s,E_b)$ we 
form the ratio $R_{D/S}$
between the background and radiative tail-subtracted tagged data (integrated over a given kinematic bin)
and the normalized simulation.
This ratio can then be used to
study the kinematic dependence of any deviations between our data and our cross section model,
see Section~\ref{secPWIAcomp}. If our  spectator cross section model is valid, 
$R_{D/S}$ can be interpreted as the ratio between the  effective structure function
 $F_2^{n, \rm eff}(W^*,Q^2, p_s, \cos \theta_{pq})$ and the model input for $F_2^n(W^*, Q^2)$ for each bin:
 \begin{align}
 \label{MCratio}
R_{D/S} &=&  \frac{N_{^2{\mathrm H}(e,e'p_s)X}^{\rm data,corr}(W^*,Q^2, p_s, \cos \theta_{pq}) }
{N(p_s,E_b)  N_{^2{\mathrm H}(e,e'p_s)X}^{\rm simul}(W^*,Q^2, p_s, \cos \theta_{pq}) } \nonumber \\
&=& \left(1+
\frac{\Delta N(p_s,E_b)}{N(p_s,E_b)}\right) \frac{F_2^{n, \rm eff}(W^*,Q^2, p_s, \cos \theta_{pq})}{F_2^{n, \rm model}(W^*,Q^2)}
\end{align}
where the first factor on the second line accounts for the possible normalization uncertainty.

As a further 
result, the value of the effective structure function $F_2^{n, \rm eff}$ for a given kinematic bin in
$p_s, \cos{\theta_{qp_s}}, Q^2$ and $x^*$ or $W^*$ can be extracted from the data by
multiplying the ratio $R_{D/S}$ with the model input for the free $F_2^n$ at the center of that bin
(thus also taking bin centering into account). This method leads to an (approximate) cancellation of the model input for
$F_2^n$ since the simulated data are (roughly) proportional to it, leading to  largely unbiased results for
$F_2^{n, \rm eff}$.


\subsubsection{Systematic uncertainties}
The total systematic uncertainty on each data point consists of an overall scale uncertainty and point-to-point
uncertainties due to the various inputs and assumptions for the analysis. The scale uncertainty,
$\Delta N(p_s,E_b)$, is due
to our RTPC normalization method (Section~\ref{dataextract}) which relies on the assumption that
our model describes the data accurately for the kinematic bin chosen to normalize the simulated to the 
measured tagged inelastic data. 
 We estimate
this uncertainty by varying the $W^*$ range over which we compare data and simulation, which
 yields a scale uncertainty
of $\Delta N(p_s,E_b) = \pm 0.1 N(p_s,E_b)$. This includes an uncertainty of $5\%$ for the model value for $F_2^n$ in the 
chosen kinematic range.
 This scale uncertainty is not shown on plots, since it affects all the bins in a given distribution uniformly.
The remaining point-to-point systematic uncertainties are discussed below and summarized in Table~\ref{t_slava_syserr}.
\begin{itemize}

\begin{table*}[ht!]
\caption{\label{t_slava_syserr}Point-to-point systematic uncertanties on the extracted structure function 
$F_2^{n, \rm eff}(W^*,Q^2, p_s, \cos \theta_{pq})$ from the ``Monte Carlo method''. Each uncertainty is shown as a percentage of the structure function.}
\begin{ruledtabular}
\begin{tabular}{lll}
Source & Systematic uncertainty($\%$) & Explanation \\
\hline
$e^+$ &              1.0               & Effect of pair-symmetric contamination         \\
$\pi^-$ &             1.0           & Effect of pion contamination \\
$\Delta z$  &             1.0           & Accidental background subtraction  \\ 
$Eff(E',\theta)$  &             8.5           & Uncertainty of $E'$- and $\theta_e$-dependent CLAS efficiency  \\ 
MC  &             9.0           & Combined uncertainty due to Monte Carlo statistics and systematics  \\ 
\hline
Total &                 12.5           & Added in quadrature   
\end{tabular}
\end{ruledtabular}
\end{table*}

 \item \textbf{Accidental background subtraction.} 
Our background subtraction method (see Section \ref{sec_acc_bg}) depends somewhat on the limits chosen for
the ``wings'' in the $\Delta z$ distribution that are used to estimate the number of background events between
the cut limits of $-1.5$ cm$ < \Delta z < 1.5$ cm. We vary the  $\Delta z$  ``wings'' from the standard range  (2 -- 16 cm) to a smaller range of 2 -- 9 cm, and estimate the systematic uncertainty as the resulting change in accidental counts subtracted.
This leads to an average systematic uncertainty of the order of 1\% relative to the corrected data, with most bins having uncertainty under 1\%.
Uncertainties on the subtraction of other backgrounds ($\pi^-$ and pair-symmetric contamination)
 are of the order of 1\%, as well.

  \item \textbf{$\boldsymbol{E'}-\boldsymbol{\theta}$ dependent acceptance and efficiency uncertainty.} 
  This is the uncertainty on the estimate of the detection efficiency of the CLAS trigger electrons, calculated using the ratio of measured
  and simulated inclusive D$(e,e')$ event rates (see Section~\ref{SimulRate}) as a function of $E'$ and $\theta_e$. The uncertainty on this efficiency
  stems mostly from bin-to-bin fluctuations of the counting statistics and the uncertainty in the model 
  used for the simulation. It was estimated by using the standard deviation of these (nearly random) fluctuations.
  This yields a kinematics-dependent systematic uncertainty of 8.5$\%$ (see Section~\ref{dataextract}). 
  (An overall scale uncertainty is already accounted for, as mentioned above).
  
    \item \textbf{$F_2^n$ model dependence.} 
    An overall scale uncertainty in our model of $F_2^n$ of about $5\%$ is included in the scale factor (see above).
    Any remaining deviation of the model from the ``true'' neutron structure function is part of the information to
    be extracted from the ratio $R_{D/S}$ and cancels largely in the extracted values for 
 $F_2^{n, \rm eff}(W^*,Q^2, p_s, \cos \theta_{pq}) = R_{D/S} F_2^{n, \rm model}$ since the denominator of $R_{D/S}$ is approximately
  proportional to $F_2^{n, \rm model}$. A small residual uncertainty stems from smearing and radiative effects (that
  depend weakly on $F_2^{n, \rm model}$) and the structure function $R$ used for the simulation. It is subsumed in
  the uncertainty due to the Monte Carlo simulation.
   
  \item \textbf{Monte Carlo simulations.} 
 Besides determining the detection efficiency via inclusive count rates, the Monte Carlo simulation is used in two different steps during the data analysis:
     \begin{enumerate}
\item to determine the quasi-elastic radiative tail that is subtracted from the data in the inelastic region, and
\item to calculate the ratio $R_{D/S}$ between experimental and simulated inelastic data.
\end{enumerate}
Both steps entail uncertainties due to Monte Carlo statistics and possible deviations between
the simulated detector response and the real performance of CLAS and the RTPC. 
(The separate uncertainty due to the simulation of inclusive D$(e,e')$ rates has been discussed above).

The statistical Monte Carlo errors are calculated using simple counting statistics (Poisson distribution) and straightforward 
error propagation. Systematic point-to-point uncertainties are due to possible inaccuracies in our 
GEANT detector model and residual dependencies on the structure function models and radiative
corrections (see previous bullet). We kept the Monte Carlo statistical errors below the systematic uncertainties
in all cases. To estimate the systematic uncertainty due to the 
subtraction of the radiative quasi-elastic tail, we compared the simulated spectra in the
quasi-elastic region 0.9~GeV$/c^2 < W^* < $~1 GeV/$c^2$ with the measured one (see, \textit{e.g.}, Fig.~\ref{fig_simW5gev}). We concluded that the
normalization of the tail has an uncertainty of about 10\%, due to the slightly different shapes of
these two spectra.
\end{itemize}

The systematic uncertainties due to $E'-\theta$ efficiency, background subtraction, and Monte Carlo simulation (both parts) are added in quadrature yielding a total point-to-point uncertainty of the ratio $R_{D/S}$ of about 12.5\%. 
To convert these values to systematic uncertainties of the $F_2^n$ structure function, they are multiplied by the value of the model $F_2^n$ in the corresponding bin. These systematic uncertainties are shown as shaded bands
in all plots in Section~\ref{results} that are based on the Monte Carlo method. While they depend on kinematics, this dependence is seen to be
a relatively smooth function of the kinematic variables across the various spectra shown in Section~\ref{results}.

\subsection{Ratio method of extracting free neutron results}
\label{altanal}
\subsubsection{Overview of the Ratio Method}
The analysis method described up to this point has the advantage of using the complete available information from
all detector elements of CLAS and the RTPC to correct the raw data for acceptance, efficiency,
radiative effects and backgrounds  bin by bin over the full kinematic domain covered by our experiment. This is essential when studying the dependence of the
extracted effective structure function $F_2^{n, \rm eff}$  on all relevant kinematic variables.
In contrast, for the purpose of extracting
the (nearly) free neutron structure function $F_2^n(x,Q^2)$ from our data in the
 ``VIP'' (Very Important Proton) region
($p_s < 100$ MeV/$c$, $\theta_{pq} > 100^\circ$)
we used the alternative ``ratio method'' that is less dependent on accurate knowledge of
detector efficiencies and acceptance.
The first publication of BONuS results~\cite{prl} is based on this approach. In this section, we give a somewhat expanded explanation of the ratio method (more details can be found in~\cite{natethesis}). In 
Section~\ref{compareNS} we compare the results for $F_2^n(x,Q^2)$  from these two different analyses, which have partially
independent systematic uncertainties. As can be seen from Fig. \ref{slavanate}, the overall agreement is good and increases our confidence that all systematic experimental uncertainties of our final result have been properly accounted for.

The ratio method relies on the fact that the acceptance of the RTPC, after integration over the VIP region,
is nearly independent of $W^*$ and $Q^2$ (since it depends only on the proton kinematics which are weakly
correlated with these variables). Furthermore, the acceptance of CLAS for electrons
within a given bin of $W^*$ and $Q^2$ for tagged events is very close to that for inclusive
electrons from D$(e,e')$X events in the equivalent $W, Q^2$ bin, where $W^2 = M_p^2 + 2 M_p \nu - Q^2$ is
the usual electron missing-mass variable  (uncorrected for initial nucleon kinematics). 
We can therefore form the ratio of tagged over inclusive events, $N_{d(e,e'p_s)}(W^*,Q^2) / N_{d(e,e')}(W,Q^2)$ for each bin in $W^*$ and $Q^2$ (and the same bin in $W$). 
This ratio can be related to the ratio of structure 
functions $F_2^n(W,Q^2)/F_2^d(W,Q^2)$ via
\begin{eqnarray}
  R_{exp} = \frac{N_{d(e,e'p_s)}(W^*,Q^2) }{N_{d(e,e')}(W,Q^2)} C(E_b,W^*,W, Q^2) = \nonumber \\
   \frac{F_2^n(W^*,Q^2)}{F_2^d(W,Q^2)} \int_{VIP} d\alpha_sdp_s^{\perp}A_p(\alpha_s,p_s^{\perp})S(\alpha_s,p_s^{\perp}) .
  \label{eq_ratio}
\end{eqnarray}
Here, $C(E_b,W^*,W, Q^2)$ is a correction factor (close to 1) that accounts for the slightly
different acceptance (due to slightly different ranges in $E', \theta_e$) for inclusive
electrons belonging to the bin $(W, Q^2)$ and tagged events belonging to the
bin $(W^*, Q^2)$, as well as different radiative corrections and  background
contributions (see below). 

The integral in Eq.~\eqref{eq_ratio}
over the spectral function $S(\alpha_s,p_s^{\perp})$ times the acceptance-efficiency product 
$A_p(\alpha_s,p_s^{\perp})$ for the RTPC is largely independent of kinematics as stated before, and
taken as a normalization constant for each data taking period (corresponding to one of the beam energy settings). 
It was determined by matching
the extracted $F_2^n/F_2^d$ to a new fit to the world data on protons and deuterons~\cite{ERIC}, see 
Section~\ref{f2ndata} .
This normalization leads
to an overall scale uncertainty of 5-10\% (mostly due to the uncertainty on the fit).
$F_2^n$ can, in principle, be obtained from the ratio 
$F_2^n/F_2^d$
by multiplying it with the parametrization of $F_2^d$ from \cite{ERIC}, while the ratio
$F_2^n/F_2^p$ can be calculated by multiplying with $F_2^d/F_2^p$, again from that same
parametrization.

\subsubsection{Analysis Details}

\begin{table*}[ht!]
\caption{\label{t_nate_syserr}Point-to-point systematic uncertainties on the extracted structure function ratio $F_2^n(W,Q^2)/F_2^d(W,Q^2)$ and the structure function $F_2^n(W,Q^2)$ derived from it,
with the ratio method. Each uncertainty is shown as a percentage of the value of the result.
 An overall normalization uncertainty of about 7-10\% applies uniformly
to the complete data set for each beam energy.}
\begin{ruledtabular}
\begin{tabular}{lll}
Source & Syst. uncertainty($\%$) & Explanation \\
\hline
FSI    & 5.0                          & Effect of final state interactions \cite{cda4} \\
Target fragmentation & 1.0            & Effect of target fragmentation \cite{cda2_1}   \\
Off-shellness        & 1.0            & Effect of nucleon off-shellness \cite{meln_schr_thom}\\
$C_e^+$ &              1.0               & Effect of pair-symmetric contamination         \\
$C_{\pi}$ &             1.0           & Effect of pion contamination \\
$r_{rc}$  &             2.0           & Each value of Born and radiated cross-sections has an uncertainty of 1$\%$, \\ 
 & &
leading to a 2$\%$ overall uncertainty \\
$Int$       &             5.0           & Possible deviation from the assumption that the integral in Eq.~\eqref{eq_ratio} is constant. \\
$F_2^d/F_2^p$ &         4.2           & Fits to structure functions have point-to-point uncertainties of 3$\%$ 
\cite{bosted_christy,Christy:2007ve}, \\
&&leading to a 4.2$\%$ overall uncertainty (on extracted $F_2^n$ and $F_2^n/F_2^p$ values only)\\
\hline
Total &                 8.7           & Added in quadrature   
\end{tabular}
\end{ruledtabular}
\end{table*}

The ratio method used the same data set as described before, with the same corrections 
for  RTPC and CLAS momenta, and the same kinematic cuts. 
The treatment of accidental background events was somewhat simplified by assuming a triangular
shape for their distribution as a function of the proton-electron vertex difference $\Delta z$. 
This assumption is a natural consequence of the convolution of two flat distributions in $z$ and is 
born out by the observed shape of ``truly'' accidental coincidences, see Fig.~\ref{fig_randz}. 
We then extrapolate this background from the ``wings'' 
(outside $\pm$ 2  cm) of the distribution in  $\Delta z$
into the ``signal'' region,  $|\Delta z| \le 1.5$ cm.
This method gives essentially
the same corrections for accidental backgrounds as the one described earlier.

The correction factor $C(E_b,W^*,W, Q^2)$ in Eq.~\eqref{eq_ratio} is composed of several 
contributions, accounting for the (small) difference in electron acceptance for tagged and inclusive
events ($R_{acc}$), pair symmetric ($C_e^+ $) and pion contamination ($C_{\pi}$) and
differences in radiative corrections $ r_{rc}$
\begin{equation}
\label{rmcorr}
C(E_b,W^*,W, Q^2) = R_{acc}  C_e^+ C_{\pi} r_{rc} .
\end{equation} 
The correction factor $R_{acc}$ 
is calculated by comparing the measured inclusive rate $N_{d(e,e')}$ to
the rate predicted by the well-known cross section for inclusive scattering off deuteron, as
a function of $(E', \theta_e)$, yielding an efficiency function $\epsilon(E',\theta_e)$. This function
is integrated (weighted by the data) over the range of $(E',\theta_e)$  belonging to either
the bin $(W^*,Q^2)$ for tagged events or the bin $(W,Q^2)$ for inclusive events, and
the ratio yields $R_{acc}$. Note that the overall luminosity and average event reconstruction efficiency of
CLAS drop out in this ratio.

Radiated and Born cross-section models, $\sigma_{r}$ and $\sigma_{Born}$, for both
electron-neutron and electron-deuteron scattering
were generated by the code of P.~Bosted and E.~Christy \cite{bosted_christy,Christy:2007ve} in each 
 $(W^*/W,Q^2)$ bin. Radiative effects were again treated following Mo and Tsai~\cite{motsai}. In our final data sample, we avoided regions where the elastic tail contribution is larger than 10$\%$. The radiative correction is the ``super-ratio''
 \begin{equation}
  r_{rc} = \frac{\sigma_{Born}^n/\sigma_{r}^n}{\sigma_{Born}^d/\sigma_{r}^d},
\end{equation}
where indices $n$ and $d$ denote the neutron and the deuteron respectively. Again, this factor
is usually very close to 1.

Finally, the relative contaminations of tagged and inclusive events from pair-symmetric $e^+e^-$ decays
and misidentified pions were estimated as described in Section~\ref{bgnd} and the ratios $C_e^+$ and
$ C_{\pi}$ calculated, together with their systematic uncertainties.

All statistical errors were properly propagated from the tagged and inclusive number of counts in each
bin. The systematic uncertainties of each correction factor in Eq.~\eqref{rmcorr} were estimated
and are listed in
Table~\ref{t_nate_syserr}, together with systematic uncertainties due to other sources. 
Even after including theoretical uncertainties (first three lines in Table~\ref{t_nate_syserr}), the overall
point-to-point systematic uncertainty of the extracted $F_2^n/F_2^d$ (about 7.5\%) -- 
as well as the derived value of $F_2^n$ (about 8.7\%) -- is smaller than the 
corresponding uncertainty of the Monte Carlo method. An overall scale uncertainty due to our cross normalization to existing fits amounts to at most 10\% for each beam energy. This scale uncertainty is common to both methods (since they are both normalized to
an existing parametrization of $F_{2}^{n}/F_{2}^{d}$) and is not included in the systematic uncertainty bands
shown in the figures in the next section.

\section{Results}
\label{results}

In the following, we present the results from our analysis of the 
BONuS data. We use the results derived from the Monte Carlo based 
analysis to study deviations from spectator model expectations,
and the ratio method results for final values of the
ratios $F_2^n/F_2^d$ and $F_2^n/F_2^p$ as well as the neutron 
structure function $F_2^n$ in the region where the spectator model is
expected to work well.

\begin{figure*}[htb!]
\includegraphics[width=0.750\textwidth]{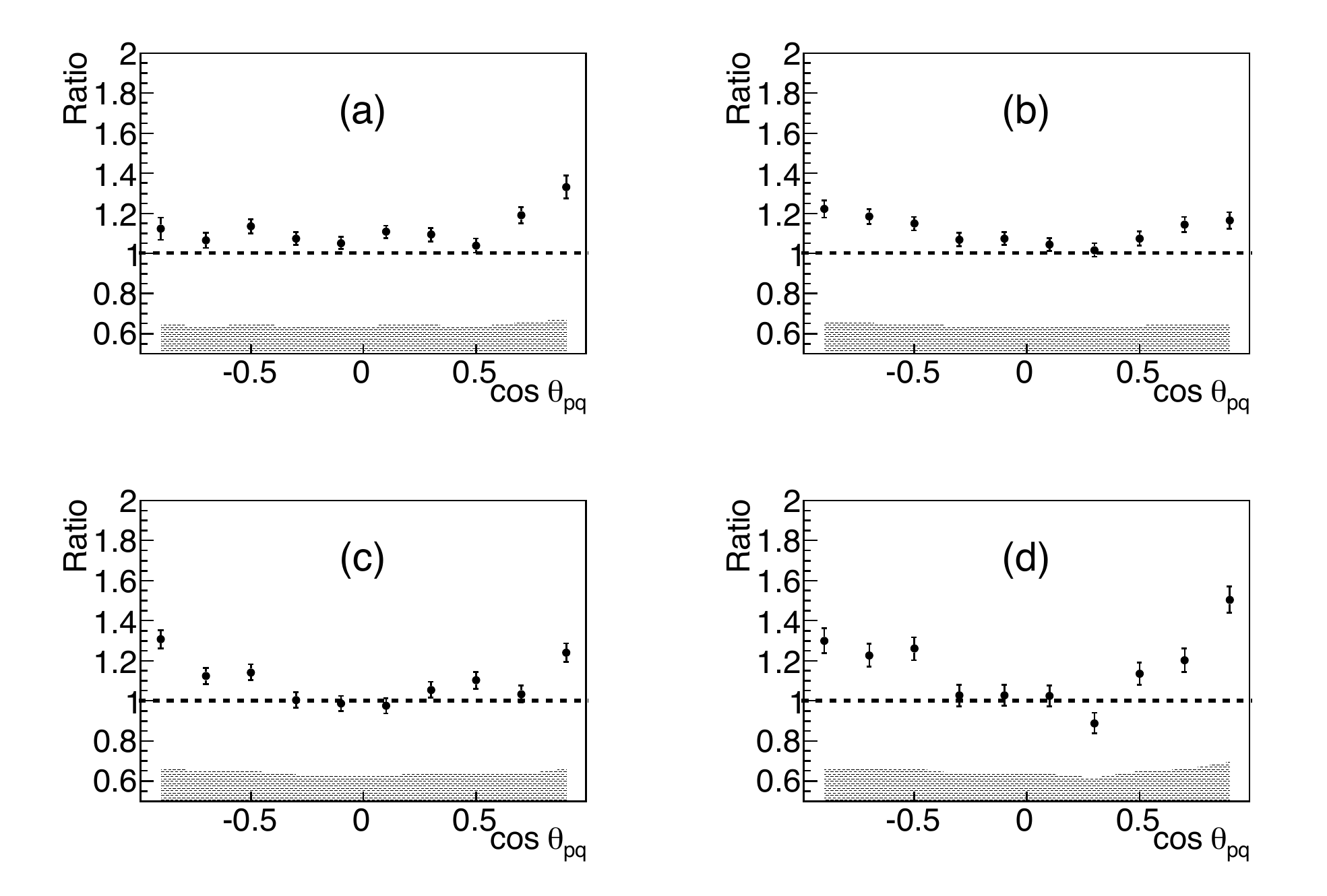}
\caption{Ratio $R_{D/S}$ of experimental data (with subtracted
	background and elastic tail) to the simulation as a function
	of $\cos\theta_{pq}$ for a selected bin in $Q^2$
	(from 1.10 to 2.23~GeV$^2/c^2$) and $W^*$  
	(from 1.35 to 1.6~GeV/$c^2$), for a beam energy of 5.25~GeV.
	Each panel corresponds to one of our 4 bins in the spectator 
	momentum $p_s$: 
	(a) 70 MeV/$c$ $-$ 85 MeV/$c$,
	(b) 85 MeV/$c$ $-$ 100 MeV/$c$,
  (c)  100 MeV/$c$ $-$ 120 MeV/$c$, (d)  120 MeV/$c$ $-$ 150 MeV/$c$.
	The error bars are statistical only, with systematic
	uncertainties shown as a band at the bottom.
\label{fig_q0W1_vscos_5gev}}
\end{figure*}

\subsection{Comparison with spectator model predictions} 
\label{secPWIAcomp}

The goal of this section is to assess in which kinematic region the 
proton spectator model describes the BONuS data, and to gain a      
quantitative understanding of the deviations from this spectator picture.
To this end, we study the dependence of the ratio of data to simulation 
on the kinematics of the spectator proton for different regions in $W^*$ and 
$Q^2$.
Any {\em systematic} dependence on spectator kinematic variables would 
indicate deviations from the spectator model, arising for instance     
from nuclear binding modifications of the effective structure function 
$F_2^{n, \rm eff}$, deviations from the input spectral function        
$S(\alpha_s, p_s^{\perp})$, and effects from FSIs.                     

As outlined in the previous section, the Monte Carlo based analysis     
leads to extracted values for the ratio $R_{D/S}$ (Eq.~\eqref{MCratio}) 
and the effective neutron structure function  			
$F_2^{n, \rm eff}(x^*,Q^2, p_s, \cos\theta_{pq})$
for a grid of values in $(x^*,Q^2)$ or $(W^*,Q^2)$ and averaged
over bins in $(p_s, \cos\theta_{pq})$.
We used 5 bins in $Q^2$ with central values 0.34, 0.61, 0.93, 1.66 and   
3.38~GeV$^2/c^2$, and 4 bins in spectator momentum: 70 -- 85, 85 -- 100, 
100 -- 120, and 120 -- 150~MeV/$c$.                                      
The dependence of $R_{D/S}$ on the angle between the  		
spectator momentum and the direction of momentum transfer is
averaged over 10 evenly spaced fine bins over the range  	
$-1.0 \le \cos\theta_{pq} \le 1.0$ or, for studies of the $W^*$
or $x^*$ dependence, in 3 coarser bins:
backward ($-1.0 \le \cos\theta_{pq} \le -0.2$),			
sideways ($-0.2 \le \cos\theta_{pq} \le  0.2$), and		
forward  ($ 0.2 \le \cos\theta_{pq} \le -1.0$).			
Similarly, $W^*$ is either binned finely in			
90 bins of 0.03-GeV/$c^2$ width or more coarsely in 6 broad regions 
covering the quasi-elastic peak (0.88 -- 1.0~GeV/$c^2$),          
the $\Delta$ resonance region (1.0 -- 1.35~GeV/$c^2$),            
the second resonance (1.35 -- 1.6~GeV/$c^2$), and third resonance 
(1.6 -- 1.85~GeV/$c^2$) regions, and two higher-$W$ regions       
(1.85 -- 2.2~GeV/$c^2$ and 2.2 -- 2.68~GeV/$c^2$).		  

\subsubsection{$\theta_{pq}$ dependence}

\begin{figure}[htb!]
\includegraphics[width=0.45\textwidth]{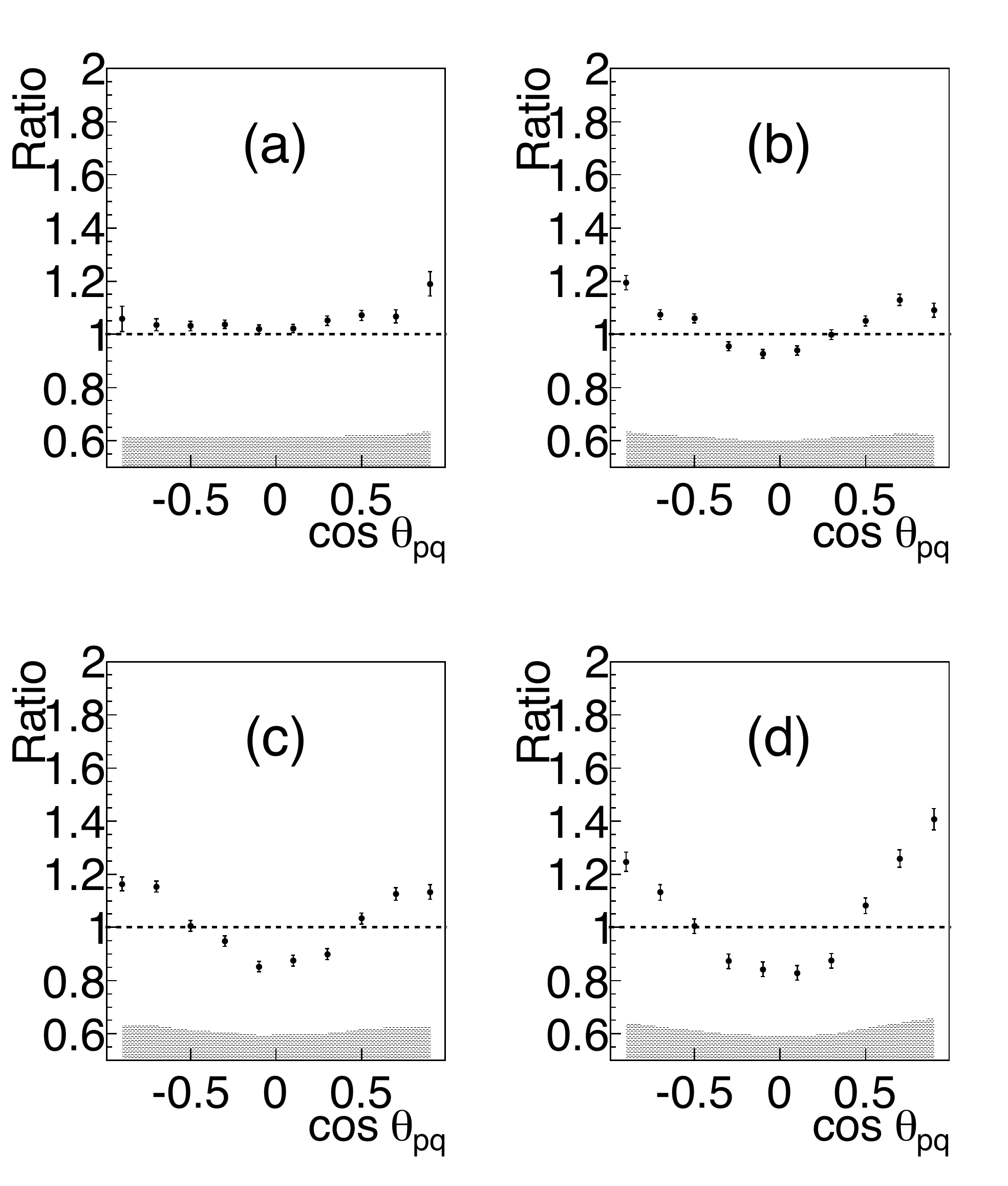}
\caption{The same ratio as in Fig.~\ref{fig_q0W1_vscos_5gev},        
	but for a higher bin in $W^*$ (from 1.85 to 2.20~GeV/$c^2$). 
\label{fig_q0W4_vscos_5gev}}
\end{figure}

\begin{figure}[htb!]
  \includegraphics[width=0.45\textwidth]{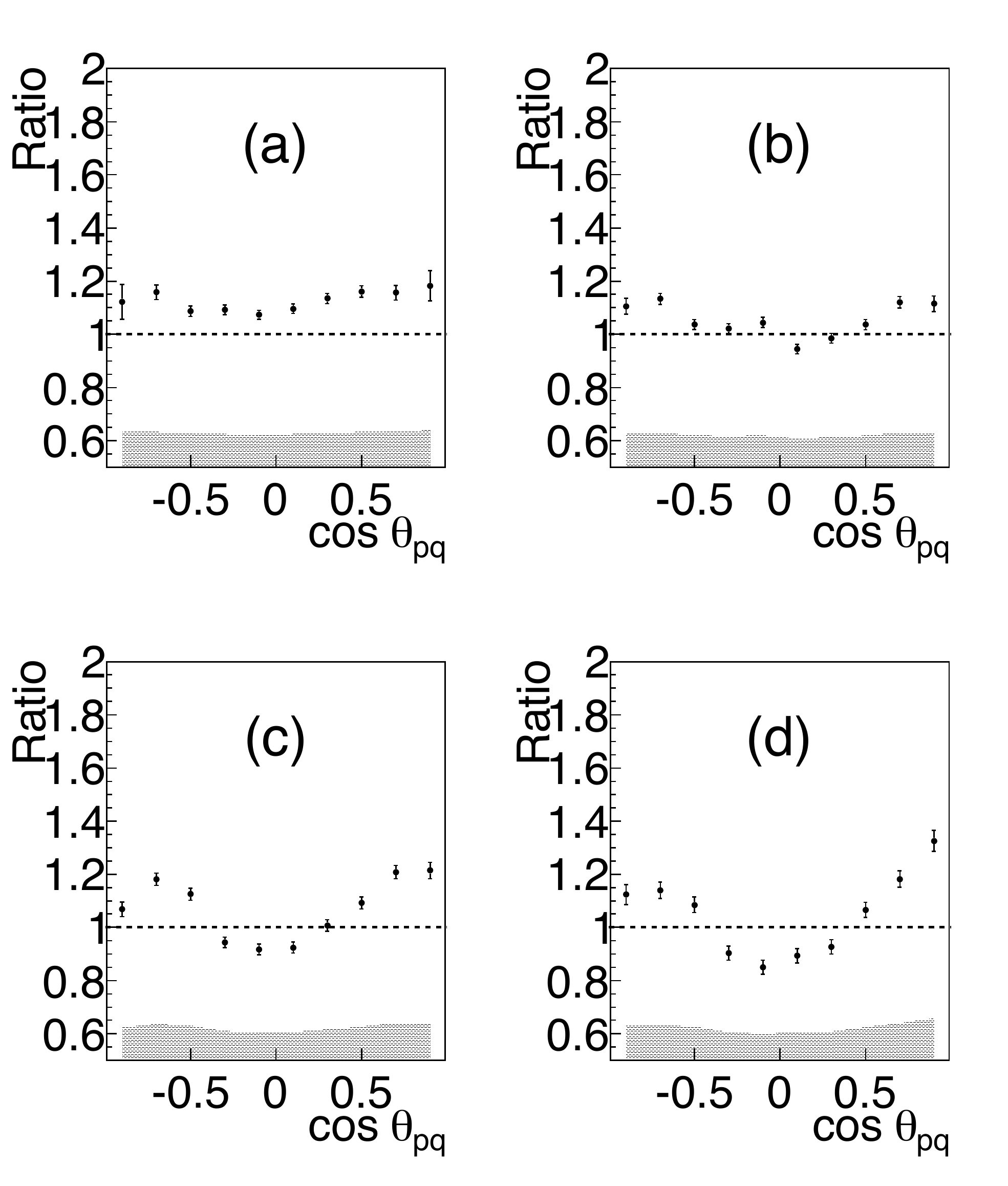}
\caption{The same ratio as in Fig.~\protect\ref{fig_q0W4_vscos_5gev} for 
	a beam energy of 4.23~GeV.
\label{fig_q0W4_vscos_4gev}}
\end{figure}

The dependence of the data-to-simulation ratio on the cosine of the
angle $\theta_{pq}$ gives us the most direct information on the validity
of the spectator picture in different kinematic domains.
In the spectator model this ratio is expected to be constant       
(equal to 1, modulo an overall normalization factor). Any overall trend, 
such as a monotonic increase or decrease with $\cos\theta_{pq}$,   
would indicate a shortcoming of the deuteron wave function model,  
while FSI effects are expected to give rise to more complicated    
structures in this ratio (see Section~\ref{sec_physics}).          
Our data on the $\cos\theta_{pq}$ spectrum for 6 bins in $W^*$, 5 
bins in $Q^2$, and 4 bins in $p_s$ are included in the supplemental material for this publication~\cite{SupMat}.
Here, we discuss a few 
representative plots (Figs.~\ref{fig_q0W1_vscos_5gev} -- 
\ref{fig_q0W4_vscos_4gev}) of this spectrum, for $Q^2$ between 1.10 and 
2.23 GeV$^2/c^2$.

Figure~\ref{fig_q0W1_vscos_5gev} shows the $\cos\theta_{pq}$ spectrum 
for a $W^*$ bin covering the second resonance region and four $p_s$ bins 
from the 5.25 GeV data set. One observes first that the data lie on 
average about 10\% higher than unity, which could be attributed to 
either an overall normalization error or a greater strength of the 
neutron structure function in this resonance region than anticipated by 
our $F_2^n$ model. Beyond that, it is clear that the data for the lowest 
$p_s$ bin fluctuate very little around this average (most points are 
less than one standard deviation away), with the possible exception of a 
slight increase at very forward angles (where target remnants from the 
struck nucleon might conceivably contribute).
The fact that the $\cos\theta_{pq}$ spectrum is flat at backward angles 
is a clear confirmation of the spectator picture for the ``VIP region"  
selected to extract the free neutron structure function.                
A slightly more pronounced $\cos\theta_{pq}$ dependence is seen in the 
next $p_s$ bin, 
and this structure becomes even more prominent for the highest two $p_s$ 
bins.
This indicates that the spectator mechanism may not be as ``pure'' at  
increasing spectator momentum, as is indeed expected from FSI models.  

These features become even more evident for the higher $W^*$ bin
(at the edge of the DIS region) shown in Fig.~\ref{fig_q0W4_vscos_5gev}.
Here, the overall normalization yields an average ratio close to 1,
due to the fact that we used part of this kinematic region for our
cross normalization.  The structure that develops as $p_s$ increases
shows a clear trend that is statistically significant, due to the
much higher count rate in this bin.  While the ratio is still mostly
flat (at least within the systematic uncertainty) for backward angles
and the lower two momentum bins, a significant depression at angles
around $90^\circ$ develops at higher $p_s$.
This is consistent with expectations from some FSI models
\cite{cda4, Cosyn11}, 
in which strength in this region is shifted to even higher momenta
through re-interaction between the struck nucleon and the spectator.
Comparison with Fig.~\ref{fig_q0W4_vscos_4gev} shows that the beam
energy (4.23~GeV in this case) has only a minor impact on the
observed pattern.

Overall we find that the $\cos\theta_{pq}$ dependence is very close  
to flat in the backward angle region for the two lowest $p_s$ bins   
(the region in which the spectator model should work well), for nearly 
all $Q^2 - W^*$ bins. (Some structure visible in the second lowest $p_s$ 
bin may in fact be ``leakage'' from  higher spectator momenta, due 
to kinematic smearing.)  This confirms that this kinematic region is  
described well by the spectator picture and therefore well-suited to 
extract (nearly) free neutron structure functions. On the other hand, 
significant deviations from this picture emerge at higher spectator 
momentum, consistent with contributions from FSI and perhaps target 
nucleon fragmentation. These data will enable tests and refinements of 
theoretical models that parametrize deviations from the spectator 
model \cite{cda4, Cosyn11, Frankfurt94}, which in turn would allow us 
to correct our $F_2^n$ data for any residual effects of this kind.    

\begin{figure*}[htb!]
  \includegraphics[width=0.75\textwidth]{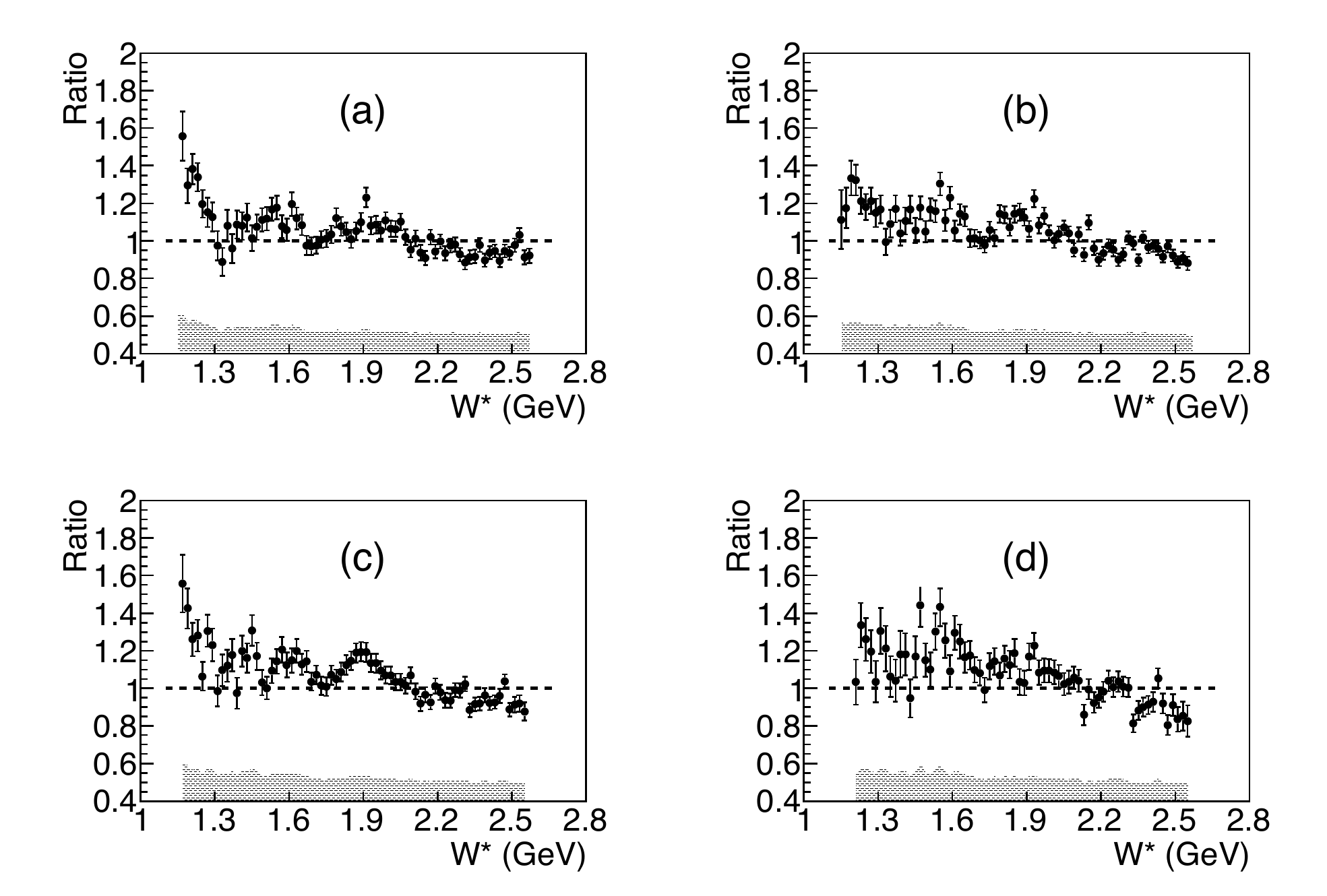}
\caption{Ratio of experimental data (after subtraction of accidental 
    background and elastic tail) to the full simulation with the        
    spectator model as a function of $W^*$. The data are for $Q^2$ from 1.10 to 
    2.23~GeV$^2/c^2$ and $\cos\theta_{pq}$ from $-1.0$ to $-0.2$.   
    Again, they are shown in four bins in the spectator 
	momentum $p_s$: 
	(a) 70 MeV/$c$ $-$ 85 MeV/$c$,
	(b) 85 MeV/$c$ $-$ 100 MeV/$c$,
  (c)  100 MeV/$c$ $-$ 120 MeV/$c$, (d)  120 MeV/$c$ $-$ 150 MeV/$c$.
    The beam energy is 5.3~GeV.
    Error bars are statistical only, with systematic uncertainties  
    shown as bands.\label{fig_q0cos0_vsW_5gev}}			    
\end{figure*}

\begin{figure*}[htb!]
  \includegraphics[width=0.75\textwidth]{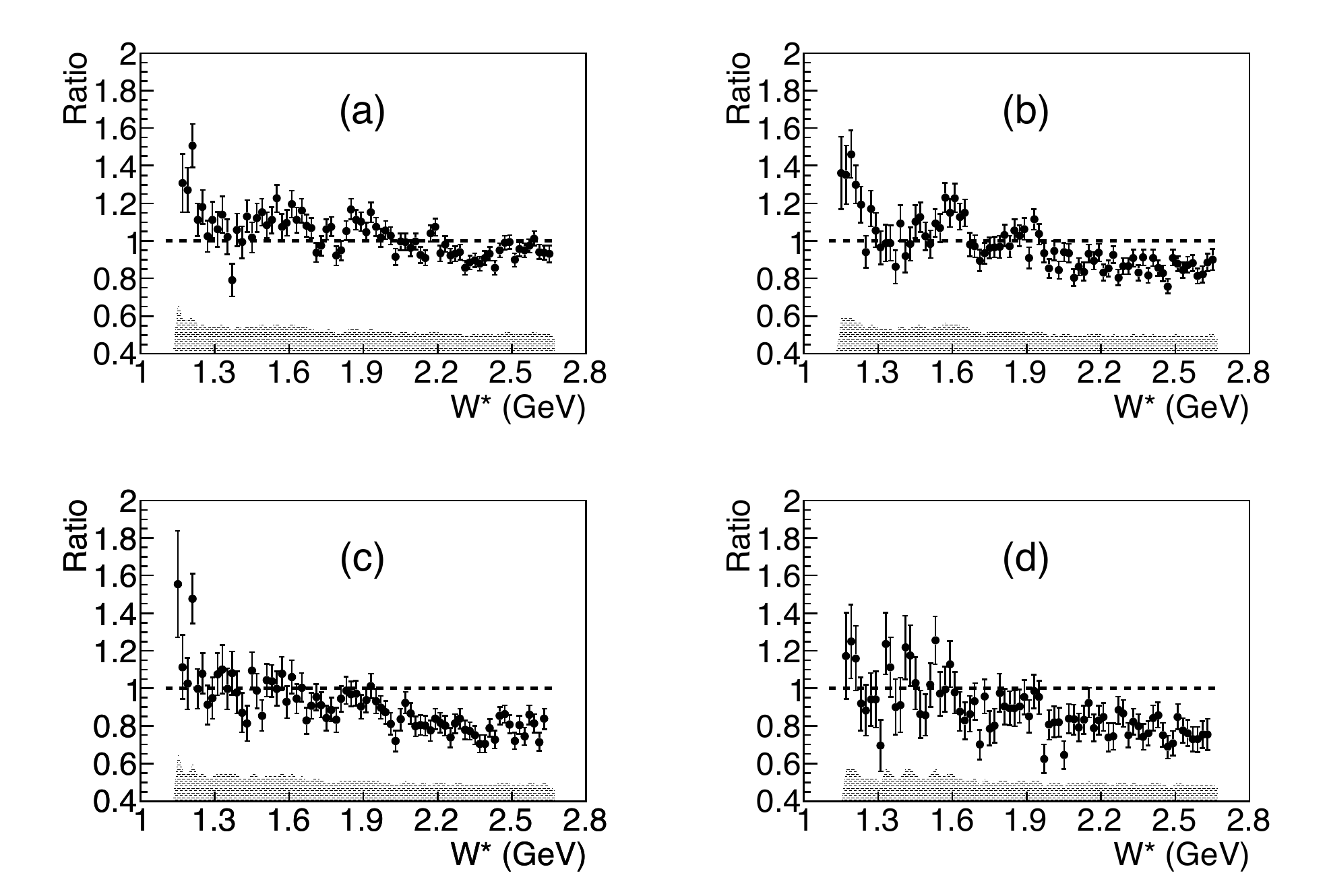}
\caption{Same ratio as in Fig.~\protect\ref{fig_q0cos0_vsW_5gev} except 
  for sideways spectator angles, $-0.2 \le \cos\theta_{pq} \le 0.2$. 
  \label{fig_q0cos1_vsW_5gev}}
\end{figure*}

\subsubsection{$W^*$ dependence}

To explore the deviations of the data from our model as a function of 
the invariant final-state mass, we show in 
Fig.~\ref{fig_q0cos0_vsW_5gev} the ratio $R_{S/D}(W^*)$ for the same bin 
in $Q^2$ as before and the highest beam energy, selecting only events in 
which the spectator proton moves backwards relative to the momentum  
transfer ($\cos\theta_{pq} \le -0.2$). The four panels again show our 
four $p_s$ bins.

We note first that there appears to be an excess of events in the region 
below and around $W^*=1.2$~GeV/$c^2$, above the model expectations. Some 
of this excess may be due to incomplete subtraction of the quasi-elastic 
radiative tail -- our systematic uncertainty (shaded band) covers nearly half of the 
statistically significant difference. However, it is possible that our 
model (which is based on inclusive deuteron data) is indeed too low in 
this region, where $F_2^n$ varies rapidly and therefore Fermi smearing 
plays an important role.
Similar, if somewhat smaller, enhancements are also visible in the second 
resonance region (around $W^* = 1.5$ and 1.6~GeV/$c^2$) and between     
$W^* = 1.8$ and 2.0~GeV/$c^2$.						
Since these features appear in most $p_s$ bins, it is unlikely that they 
are due to a breakdown of the spectator picture. A more recent fit to 
the world inclusive structure function data~\cite{ERIC} shows better
agreement with our data (see Section~\ref{f2ndata}).		      

At $W^* > 2$~GeV, the data (which have been normalized to the model in 
the region $2.0 \le W^* \le 2.2$~GeV) rarely differ more from our  
model than the combined statistical and systematic uncertainty, although 
one might discern a downward sloping trend with the higher $p_s$ bins.
Looking at the same spectra for {\em sideways}--moving spectators  
(see Fig.~\ref{fig_q0cos1_vsW_5gev}) we note a more pronounced 
depletion relative to the model at  $W^* > 2$ GeV, especially for the higher $p_s$ bins.
This could be an indication that strength in the region of higher $W^*$ 
is predominantly shifted to other kinematics ({\it e.g.}, higher proton recoil 
momenta), due to FSI between the hadronic debris from the primary 
reaction and the spectator proton. Again, this is consistent with some 
of the existing models for FSI~\cite{cda4, Cosyn11}.

Overall, our results exhibit a generally good agreement of
$F_2^{n, \rm eff}(W^*)$ with the model for all but the lowest
$W^*$ within the ``VIP'' (spectator) region of low $p_s$ bin and
backward $\theta_{pq}$. Any observed structures in this region are more likely
compatible with deficiencies in our $F_2^n$ model and the Monte Carlo
simulation of the experiment 
than with a breakdown of the spectator picture.

\begin{figure}[htb]
  \includegraphics[width=0.5\textwidth]{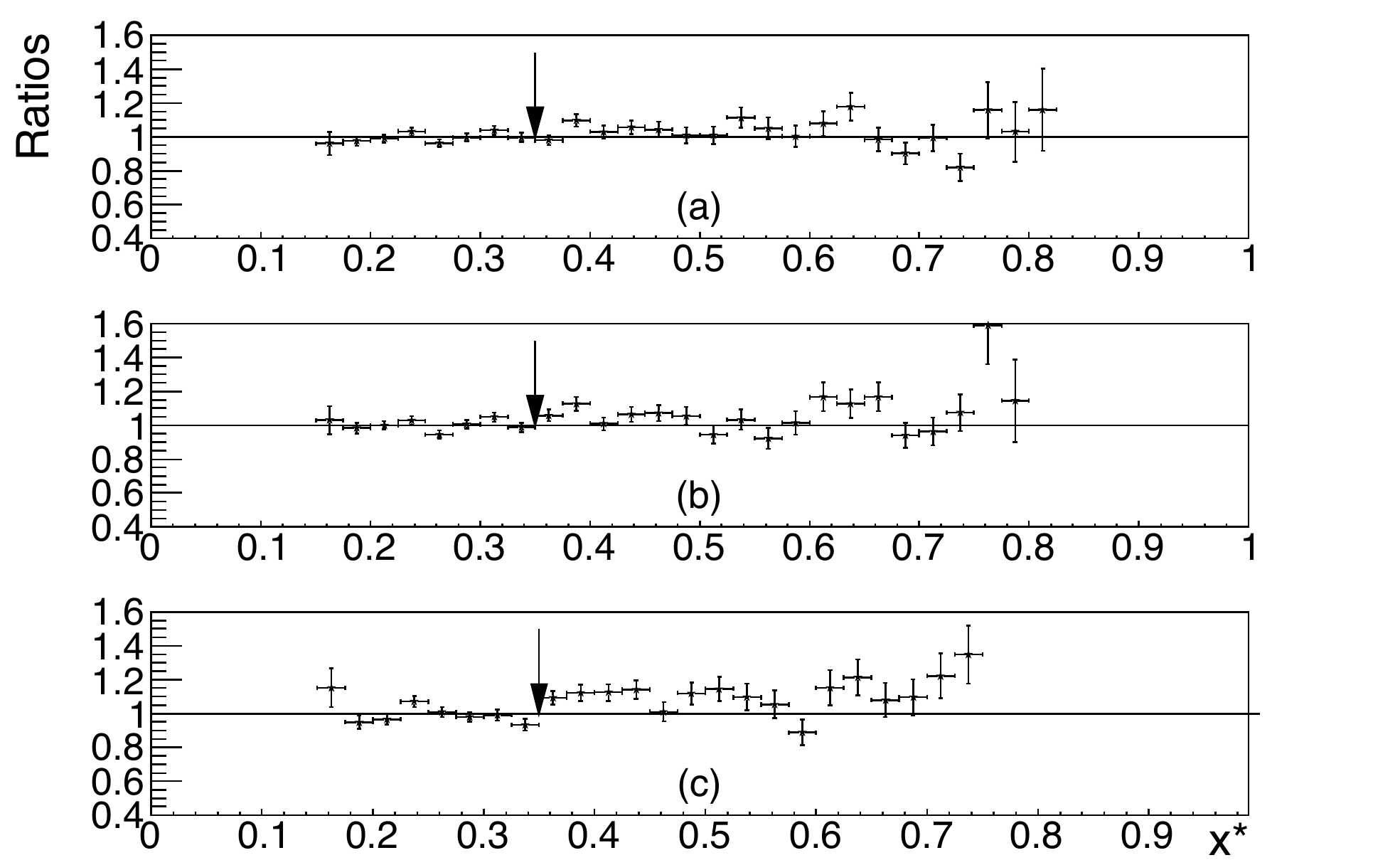}
  \caption{Ratios of $F_2^{n, \rm eff}(x^*,Q^2,p_s)$ for backward spectator momenta in each of the 
  three higher $p_s$ bins ((a) = 85 MeV/$c$ $-$ 100 MeV/$c$,
  (b) =  100 MeV/$c$ $-$ 120 MeV/$c$, (c) =  120 MeV/$c$ $-$ 150 MeV/$c$)
  to $F_2^{n, \rm eff}(x^*,Q^2,p_s )$
   for the lowest $p_s$ bin.
  Data are for $Q^2= 1.1$ -- 2.2 GeV$^2/c^2$, $\cos\theta_{pq}$ from $-$1.0 to $-$0.2, and 5.3 GeV beam energy. Error bars are statistical only. The arrows indicate the approximate location of the edge of the DIS region, $W^* = 2$ GeV.
  \label{fig_emcQ0cos0pass5}
  }
\end{figure}
 
\begin{figure}[htb]
  \includegraphics[width=0.5\textwidth]{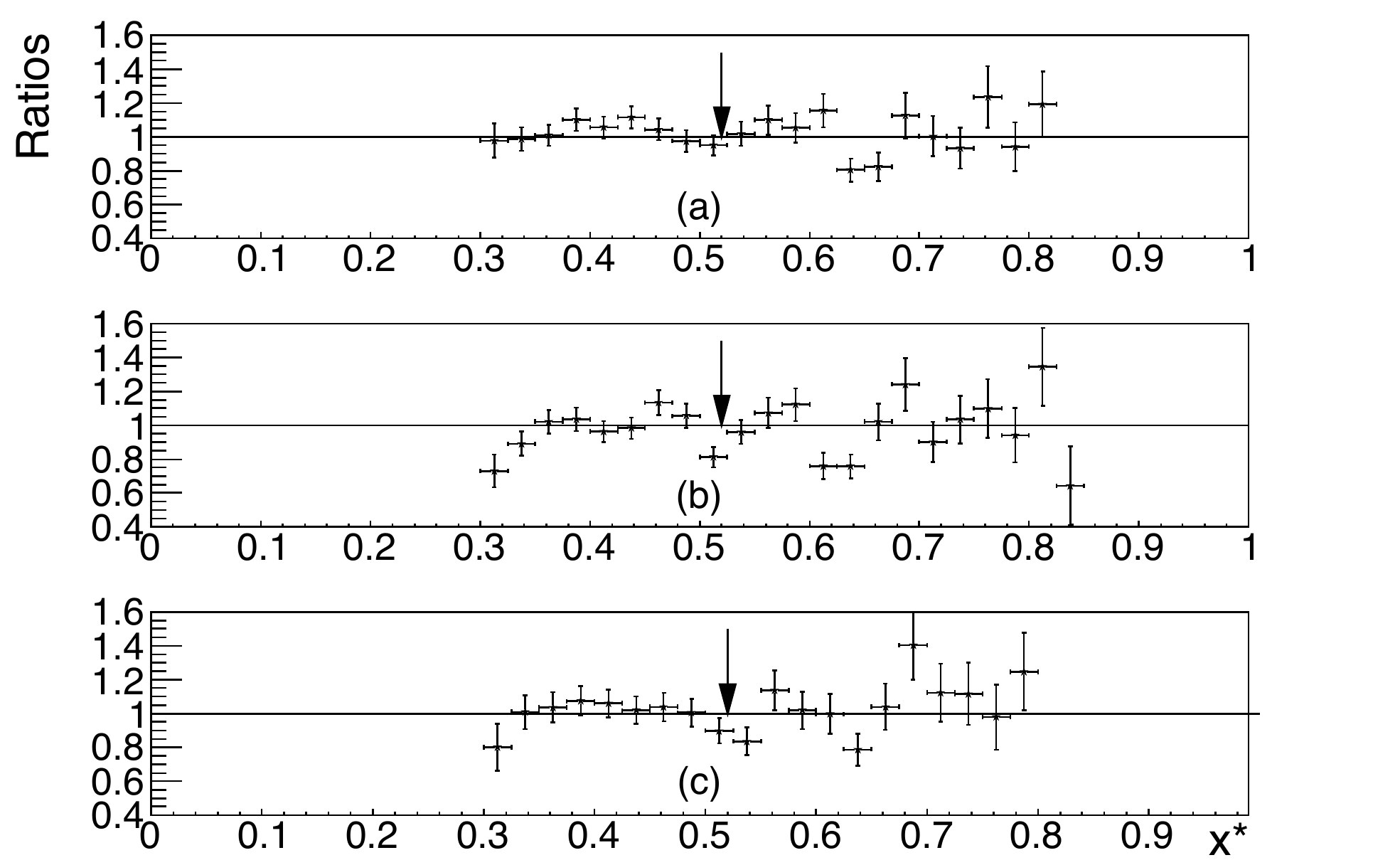}
  \caption{Same as Fig.~\protect\ref{fig_emcQ0cos0pass5}, but for a higher $Q^2$ bin, $2.2 - 4.5$~GeV$^2/c^2$. 
  \label{fig_emcQ1cos0pass5}
  }
\end{figure}

\subsubsection{Binding effects}  

We can sharpen the search for possible indications of binding and 
off-shell effects in our data by comparing the $x^*$ dependence of the 
effective neutron structure function for different spectator momenta. 
Several models of the EMC effect (see Sec.~\ref{beyond}) suggest that 
the effect can be (partially) explained by a reduction of $F_2^n$
if the struck nucleon is far from its on-shell energy
$E=\sqrt{M^2 + {\bm p}^2}$, which is equivalent within the       
spectator picture to a high-momentum backward-moving spectator.

We therefore plot ratios of our extracted structure functions
$F_2^{n, \rm eff}$ as a function of $x^*$ for different bins in
$p_s$ and our usual range of backward spectator angles, see 
Figs.~\ref{fig_emcQ0cos0pass5}--\ref{fig_emcQ1cos0pass5}.
The first figure is for a lower $Q^2$ bin where the DIS region
ends already around $x^* = 0.35$ (indicated by arrows).
It is quite apparent that the ratios are rather flat, within
the statistical uncertainties, even beyond the DIS region.
Systematic uncertainties largely cancel in this ratio.
In particular, there is no indication of a negative slope
as seen in the ratio between nuclear and nucleon structure
functions (as in the EMC effect). The same behavior repeats
itself for a higher $Q^2$ bin (Fig.~\ref{fig_emcQ1cos0pass5})
albeit with significantly larger statistical errors.
(Here, the DIS region extends to about $x^* = 0.52$.)
While our statistical precision is not sufficient to rule out
a small $p_s$ dependence of the structure function ratio,  
it appears that binding effects are still rather small for
spectator momenta up to about 150~MeV/$c$. The future BONuS       
measurement with 12~GeV beam~\cite{bonus12} will check this
conclusion with much improved precision.

\begin{figure*}[htb]
\begin{center}
  \includegraphics[width=0.85\textwidth]{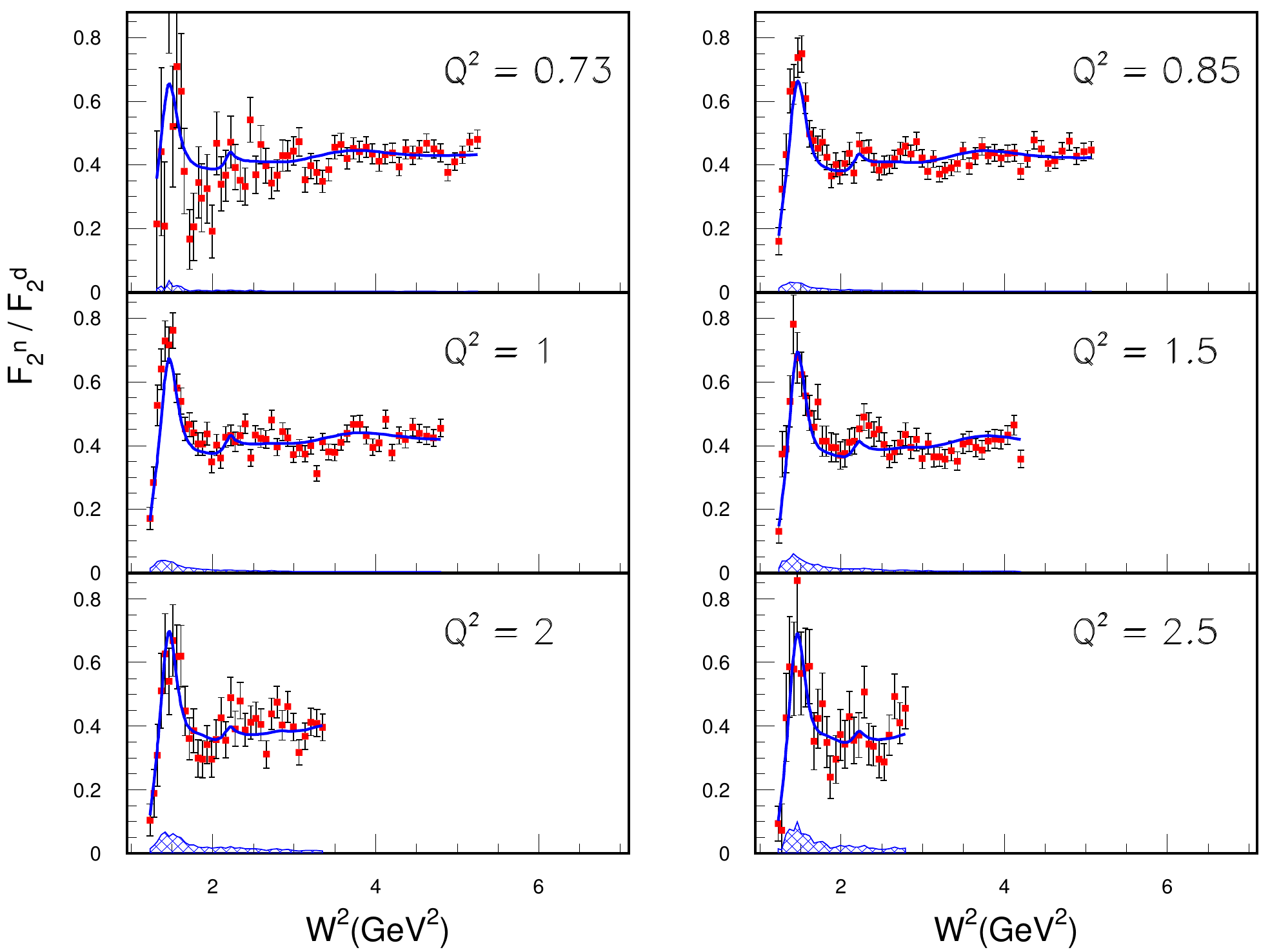}
  \end{center}
\caption{(Color online) $F_2^n/F_2^d$ {\it vs.} $W^{*2}$ 
for 4.2 GeV beam energy.
  The data are from the VIP region 
  $p_s \le 100$~MeV/$c$ and 
  $\theta_{pq} \ge 100^\circ$. Error bars indicate 
  statistical uncertainties while the bands show the point-to-point 
  systematic uncertainties. The data have been normalized to the preliminary
  new fit of the world data by Christy et al.~\cite{ERIC} (see text) which is shown as a
   solid line. We estimate that there is an overall normalization uncertainty
  of up to 10\%.
 \label{fig_q2_F2nvsW_24gev}}
\end{figure*}

\begin{figure*}[htb]
\begin{center}
  \includegraphics[width=0.85\textwidth]{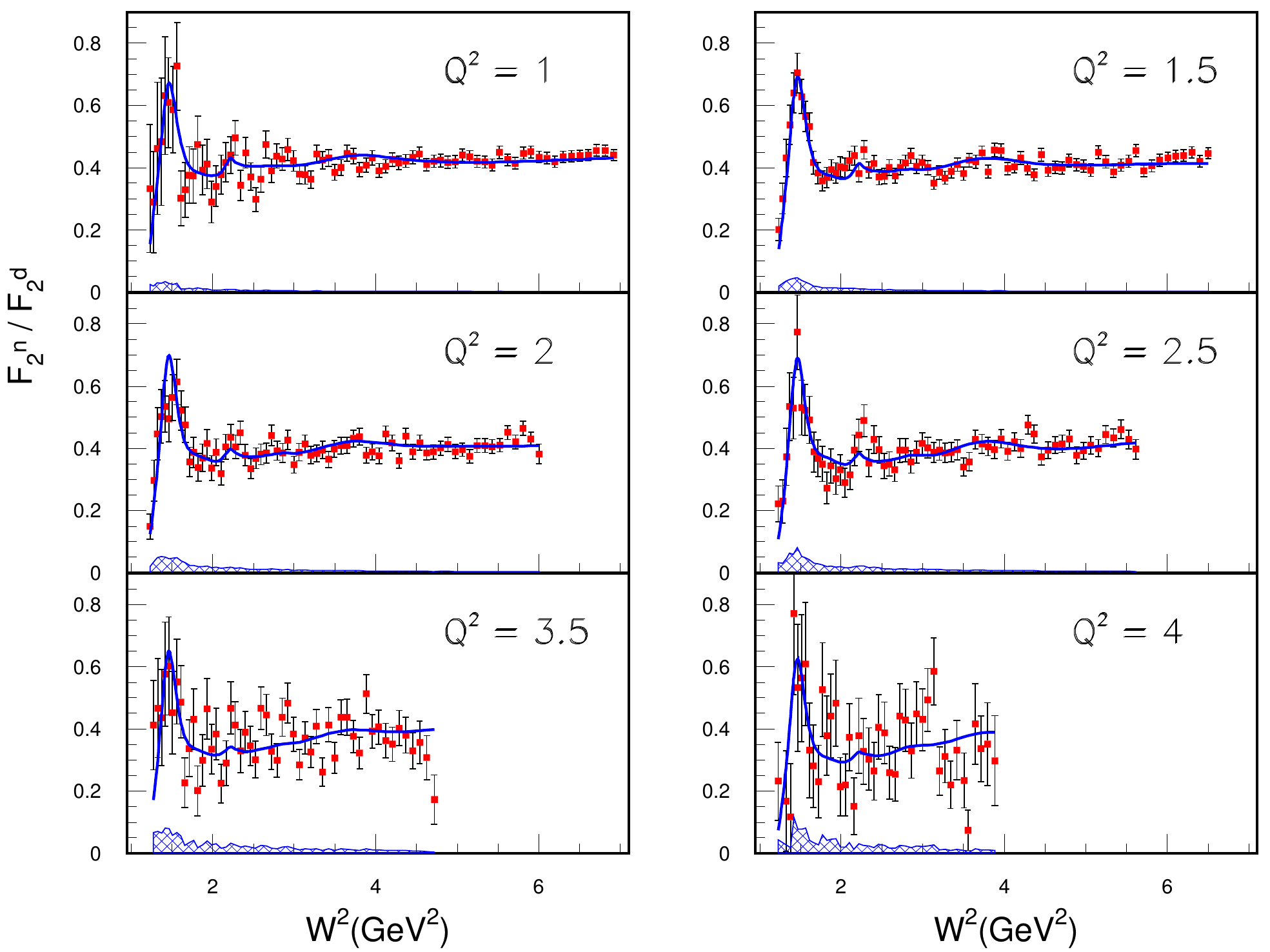}
  \end{center}
\caption{(Color online) Same as Fig.~\ref{fig_q2_F2nvsW_24gev}, for 5 GeV beam energy.
 \label{fig_q2_F2nvsW_25gev}}
\end{figure*}

\subsection{The free neutron structure function}
\label{f2ndata}

After establishing that the spectator picture is indeed a reasonably
good approximation within our VIP region ($p_s \le 100$~MeV/$c$,         
$\theta_{pq} \ge 100^\circ$), we proceed to extract results for the (nearly)
free neutron for all kinematic bins in
$W^*$ and $Q^2$, within the VIP region, using our ratio method.
Since this method determines  the ratio of
$F_2^n/F_2^d$, we show the results for this (nearly model-independent)
quantity in Figs.~\ref{fig_q2_F2nvsW_24gev}-\ref{fig_q2_F2nvsW_25gev}, separately
for our two highest beam energies. 
The error bars indicate statistical errors, while the point-to-point 
systematic uncertainties are indicated by the band at the bottom. 
As explained earlier,
there is an overall normalization uncertainty which means that the data
must be multiplied by a factor determined from other information. 
For this purpose, we used a 
recent update of the Bosted-Christy fit~\cite{bosted_christy} of the 
world data on protons and deuterons~\cite{ERIC}. This new fit uses a convolution
model~\cite{Kahn:2008nq} to combine parametrizations of proton and neutron
structure functions to model the deuteron. From this  fit 
(which does not yet include the BONuS data), the ratio $F_2^n/F_2^d$ 
can be extracted in a model-dependent way and we use the result
to determine the overall normalization constants for both beam energies,
by minimizing the $\chi^2$ of the normalized data versus the fit.
The data shown in Figs.~\ref{fig_q2_F2nvsW_24gev}-\ref{fig_q2_F2nvsW_25gev} are the main result of the
BONuS experiment -- they are available in tabular form in the supplemental material of
this publication~\cite{SupMat}
and in the CLAS 
experimental database~\cite{DataBase}. We estimate that the normalization uncertainty could be as
large as 10\%, by comparing our present result to earlier fits of $F_2^n/F_2^d$~\cite{bosted_christy}.

\begin{figure*}[htb]
  \includegraphics[width=0.85\textwidth]{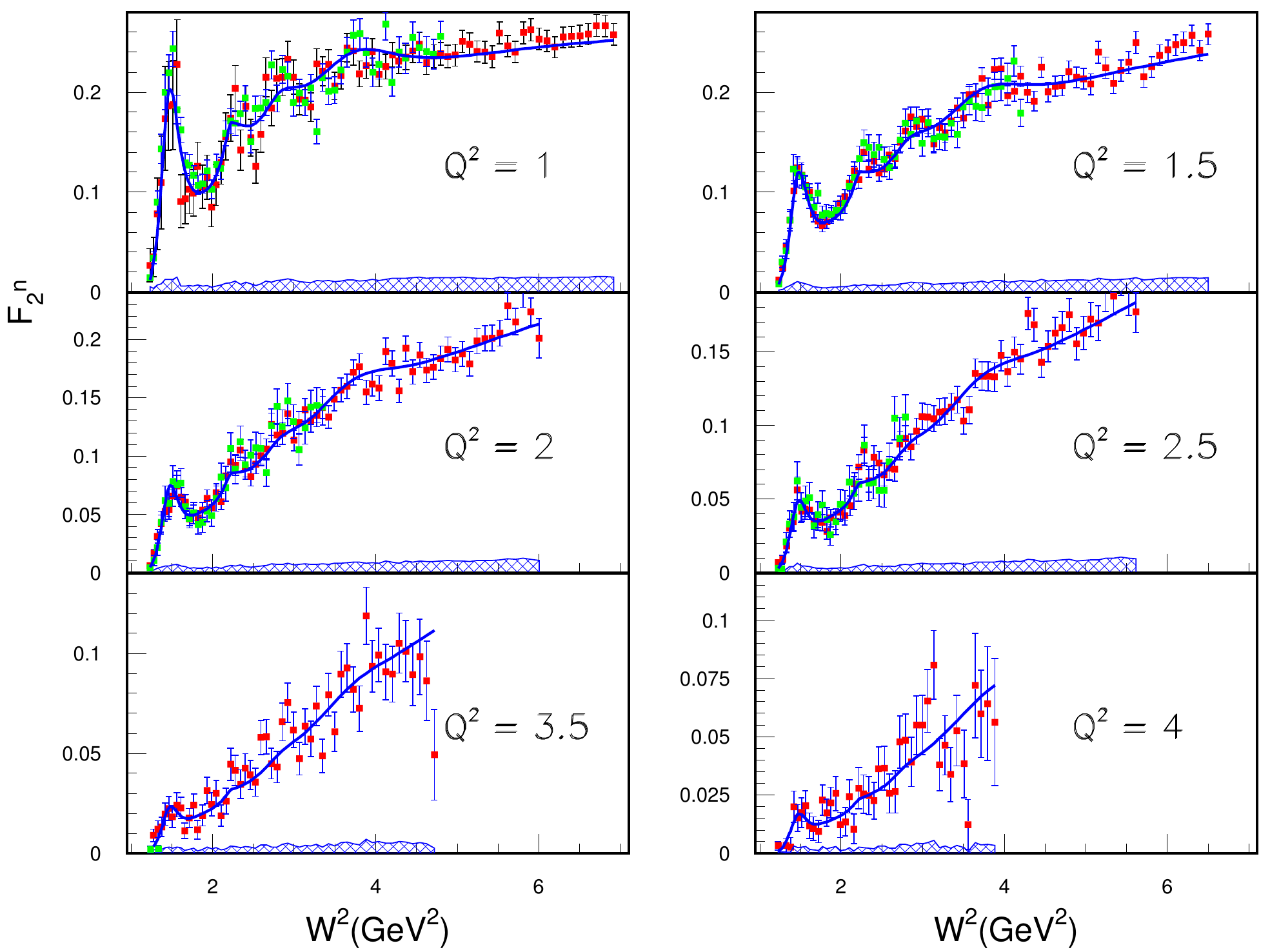}
\caption{(Color online) Extracted structure function $F_2^n$ as a 
  function of squared final-state invariant mass $W^{*2}$ for 6 bins in 
  $Q^2$. Results are shown for two beam energies, 
  4.2~GeV (green circles) and 5.3~GeV (red squares).
  Error bars are statistical only, with systematic uncertainties for the 
  5.3~GeV data shown as a band.
  (4.2 GeV systematic uncertainties are very similar in magnitude
  to the shown ones). The solid line indicates a new fit to the
  world data on deuteron and proton targets, that does not include
  BONuS data (see text).
\label{fig_F2nvsW_comp}}
\end{figure*}

Within the assumptions of the new fit, we can also extract $F_2^n$ from the ratio
by multiplying it with the fit result for $F_2^d$.
Our corresponding results  for $F_2^n$ as a function of $W^{*2}$ are shown in
Fig.~\ref{fig_F2nvsW_comp} in six different slices of $Q^2$, with both
 beam energies combined.
We point out that these results depend on the exact functional form used for
$F_2^d$ and could change if other models are used. The underlying parametrization for
$F_2^n$ from this new fit is also shown as a solid line.
We note that the agreement between $F_2^n$ obtained 
from the two energies (after cross normalization) is quite good, 
increasing our confidence that smaller corrections, such as those       
due to radiative effects, detector acceptance and the contribution      
from $R = \sigma_L / \sigma_T$, are quite small and well under control. 

We also observe a generally good agreement between the data and the new fit,
but with some indications for room to improve the latter.
In particular, the ratio between the strength at the top of the
three resonance ``peaks'' and the valleys in between appears larger 
in some of our data than in the fit.
%
%
Such a deviation from the fit (which is based on inclusive deuteron 
world data) is understandable, keeping in mind that our experiment is the
first one that does not have to rely on an unfolding prescription.
The Fermi smearing for inclusive scattering off the deuteron tends 
to wash out strong resonance features.				   
Ultimately, BONuS data will be incorporated into this new fit
to further improve its precision in describing the neutron.

\subsection{Results in the DIS region}\label{compareNS}

\begin{figure}[htb]
  \includegraphics[width=0.5\textwidth]{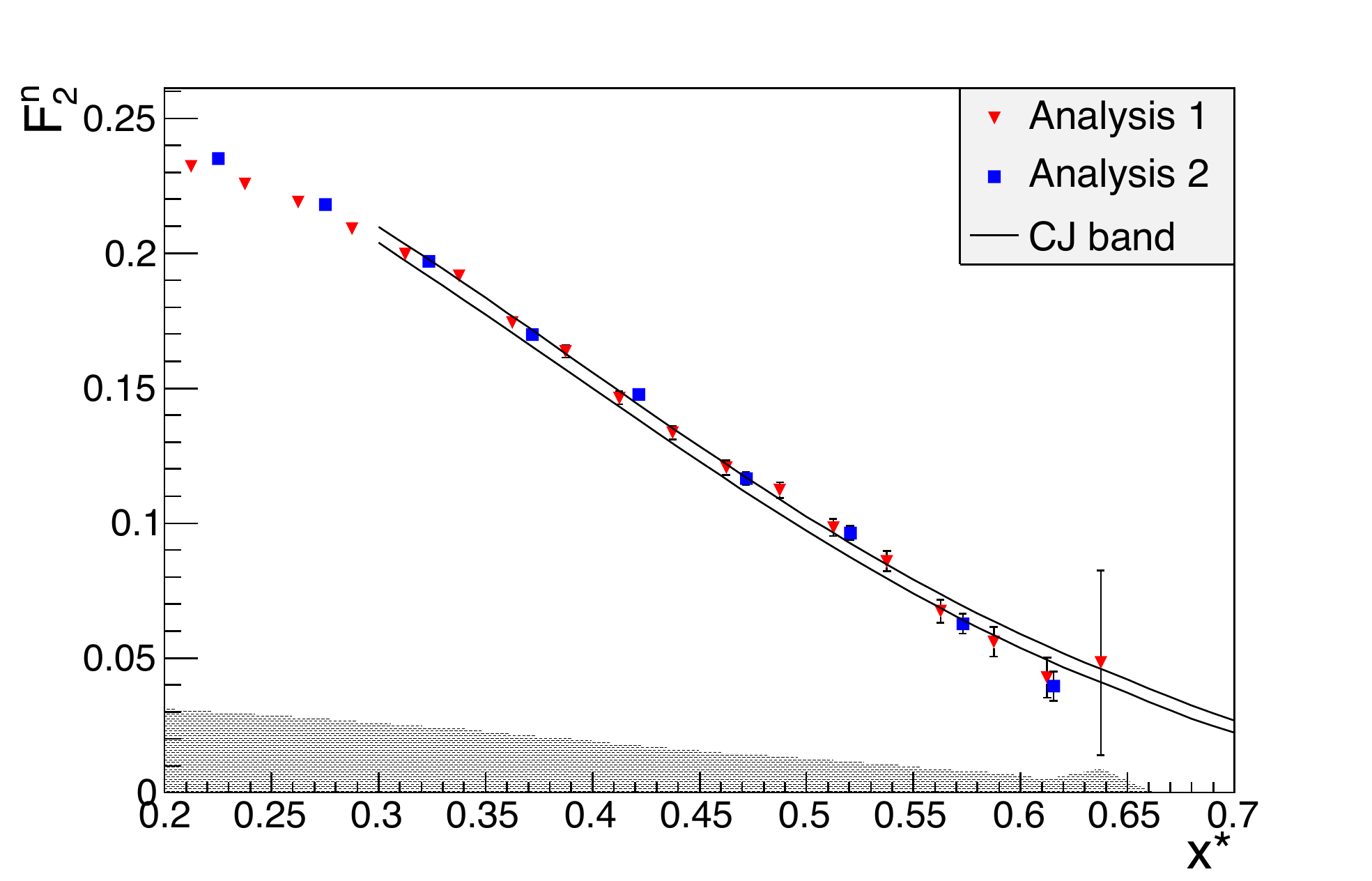}
  \caption{(Color online) 
  Results for the neutron structure function $F_2^n(x)$ (integrated over $Q^2 > 1$ GeV$^2/c^2$ while
 requiring  $W^* > 1.8$ GeV/$c^2$) from the
  Monte Carlo method  (labeled ``Analysis 1'') and  ratio method (labeled ``Analysis 2''). The
  range of $F_2^n$ from the CJ fit \cite{CJ_accardi} is shown by the two solid lines. Systematic uncertainties  for the Monte Carlo method are  shown as the shaded band. The two analysis results are cross-normalized to the average of the CJ fit 
  at $x = 0.32$.}
  \label{slavanate}
\end{figure}

Our second main goal is to pin down the behavior of $F_2^n$ at large 
$x$ but in the DIS region (see Sec.~\ref{sec_physics}).              
Unfortunately, the kinematic reach of the present BONuS experiment was 
restricted by the maximum available beam energy (5.25~GeV), which limits us 
to $x < 0.55$ if we require $W^* > 2$~GeV/$c^2$.	
Even pushing down to $W^* > 1.8$~GeV/$c^2$ 		
does not extend the $x$ range much beyond $x = 0.6$, which is the region 
where presently the uncertainty on the down quark distribution function 
becomes large. Still, we can compare our results over the measured range 
($0.2 < x < 0.65$) with existing NLO fits based on world 
data~\cite{CJ_accardi}. In Fig.~\ref{slavanate}, we show our results for $F_2^n$
using both analysis methods.		   

For both analyses, we select events in the VIP region ($p_s \le 100$~MeV/$c$,
and $\theta_{pq} \gtrsim 100^\circ$) from the highest beam energy.
We require $W^*>1.8$~GeV/$c^2$ and integrate over all $Q^2>1$~GeV$^2/c^2$
within a given $x$ bin.
We convert the values for $F_2^n/F_2^d$ from the ratio method once again using the new fit
for $F_2^d$, and for the Monte Carlo Method we multiply the ratio $R_{S/D}$ with the
model for $F_2^n$ used for the generated events in our simulation. 
Both results are normalized at $x = 0.32$
to the middle of the uncertainty band of the CJ fit~\cite{CJ_accardi} 
(given by the two solid lines in Fig.~\ref{slavanate}).
In spite of significant differences between the two approaches,
the results of the Monte Carlo method (``Analysis 1'', inverted   
triangles) and the ratio method (``Analysis 2'', squares) agree   very
well within their systematic uncertainties (given for Analysis 1 by
the shaded band). We reiterate that, apart from overall normalization 
factors ({\em not} included in the shaded band), the systematic uncertainties 
of the two methods are largely uncorrelated. 
Most of the data are within 
or close to
the uncertainty range of the CJ fit,
although some fluctuations (most likely due to remaining resonant contributions)
are visible. (The CJ band does not extend below  $x = 0.3$ since the fit is restricted
to $Q^2 > 1.6$ GeV$^2$/$c^2$ and our data fall  below that value for $x \le 0.3$.)

\begin{figure}[htb]
  \includegraphics[width=0.5\textwidth]{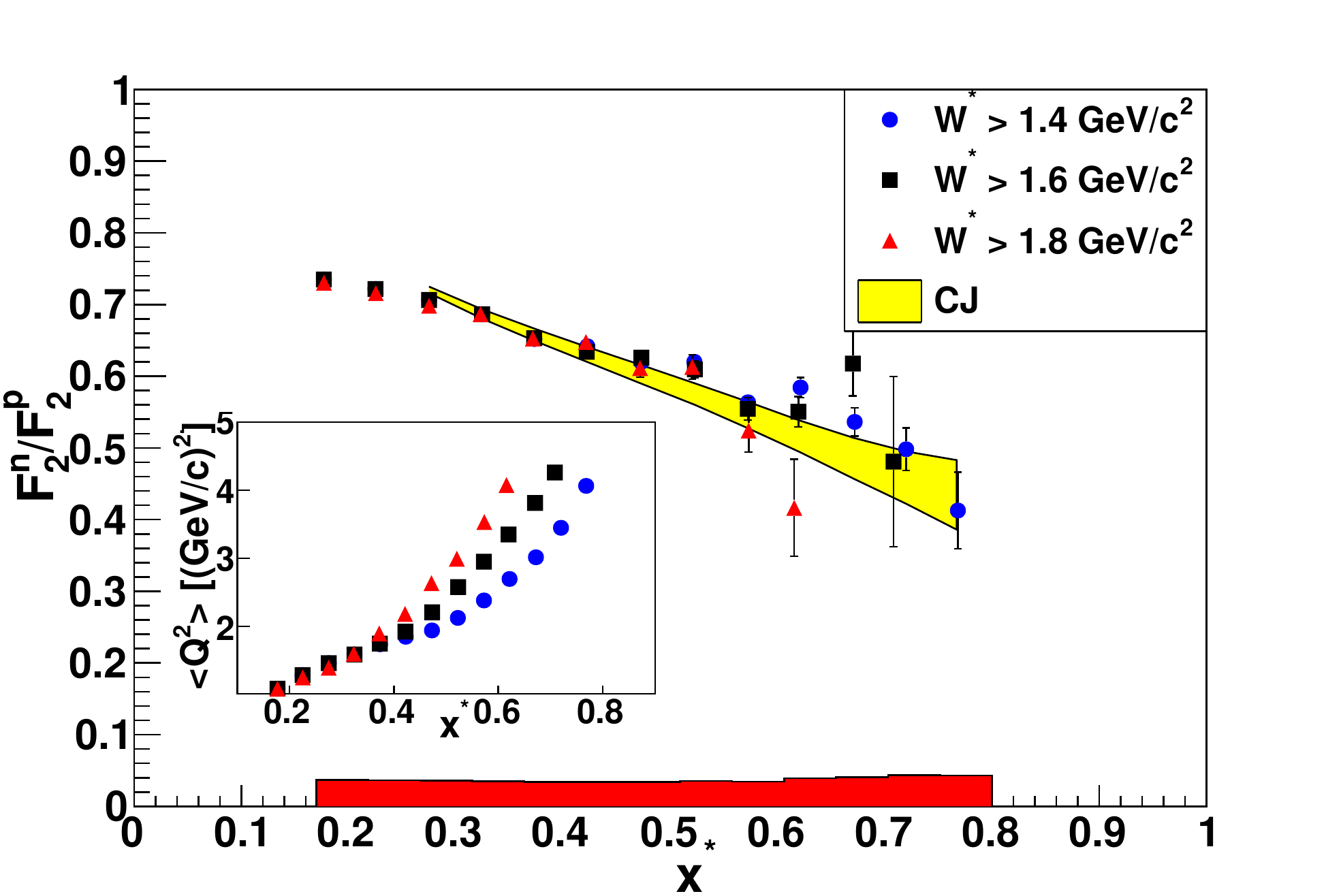}
  \caption{(Color online) 
  Results for the ratio of the neutron to proton structure functions $F_2^n/F_2^p(x)$ 
  (integrated over $Q^2 > 1$ GeV$^2/c^2$ and three different minimum values for $W^*$) from the
  ratio method. The
  uncertainty range from the CJ fit \cite{CJ_accardi} is shown by the (yellow) shaded band. Systematic uncertainties  are  shown as
  the (red) shaded band at the bottom. Our data are cross-normalized to the average of the CJ fit 
  at $x = 0.32$. The inset shows the average $Q^2$ for each data point, separately for the three lower $W^*$ limits. }
  \label{moneyplot}
\end{figure}

The ratio $F_2^n/F_2^p$, which is of high interest because of its relationship to the asymptotic
$d/u$ ratio (see Section~\ref{sec_physics}), can also be extracted from our data using a
suitable model for $F_2^p$. We showed this quantity in our
previous publication~\cite{prl}, using the ratio method. We reproduce this result here in 
Fig.~\ref{moneyplot}, updated with the new fit for $F_2^d$ and $F_2^p$.
The results are shown for three lower cuts on the range in $W^*$ over which we integrate our data. 
The red triangles are for $W^* > 1.8$ GeV, {\it i.e.}, showing the same data as in Fig.~\ref{slavanate}. They
agree reasonably well with the prediction from the CJ fit, but do not extend much beyond $x=0.6$. The
black squares ($W^* > 1.6$ GeV) and the blue circles ($W^* > 1.4$ GeV) push this limit to higher $x$, but
some clear resonant structure can be observed at large $x$. Taken at face value, the difference between these
integration regions can be interpreted as a first hint that local duality may not hold as well for the neutron as for the
proton in our kinematic region. Ultimately, only by repeating this measurement with significantly higher beam
energy can one cleanly extract the DIS limit for $F_2^n/F_2^p$ as $x \rightarrow 1$. A corresponding measurement
is planned for the CLAS12 spectrometer at Jefferson Lab after the upgrade to 11 GeV
beam energy is completed~\cite{bonus12}.

\section{Summary}
\label{sec_summary}

We have presented the full analysis and final results from the BONuS 
experiment, which accessed for the first time structure functions of
the neutron by tagging spectator protons in the reaction $^2$H$(e,e'p_s)$.
Comparison of our data to a full Monte Carlo simulation based on the
spectator model in the impulse approximation shows generally good   
agreement for the lowest spectator momenta ($p_s = 70 - 85$~MeV/$c$),
especially in the backward $\theta_{pq}$ region.  Deviations from the
spectator picture could be identified, however, at higher momenta.
The results for the dependence on the spectator proton angle tend to agree with expectations
from target fragmentation models \cite{cda2_1, cda2_2}, with the data
showing an enhancement in the region of forward $\theta_{pq}$,
as well as with final-state interaction models \cite{cda4} which
predict a dip in the vicinity of $\theta_{pq} = 90^{\circ}$.

Within the kinematic region of its applicability, the spectator model
allows us to extract the ratio $F_2^n/F_2^d$  of the free neutron structure function  
to the deuteron one
over a
wide range in $x$ or $W$ and $Q^2$.  Comparison to a new, preliminary fit for
this ratio from inclusive deuteron data using Fermi-smearing models
\cite{ERIC} shows overall good agreement, but with some room for
improvement in the detailed description of the resonance structures
present in the data.  In the DIS region, our data agree well with
existing PDF parametrizations~\cite{CJ_accardi} out to $x \approx 0.65$,
where uncertainties become large.

Structure functions extracted from the BONuS experiment using two
different analysis methods are in agreement with each other,
indicating that systematic uncertainties are under control.
The complete data set for $F_2^n/F_2^d$ over all bins in ($W^*, Q^2$) is
available from the CLAS database~\cite{DataBase} and as 
supplemental information to this paper~\cite{SupMat}.
It will aid the
improvement of existing models and parametrizations of neutron
structure functions.
These parametrizations in turn are crucial for other experimental goals,
such as the extraction of neutron spin structure functions from
polarization asymmetries, more precise studies of the nuclear EMC
effect via comparisons of nuclear cross sections with the free
proton and neutron cross sections, as well as reducing uncertainties
in PDFs used for extracting information from collider measurements.
Our data will also provide constraints on the isospin dependence of
nucleon resonant excitations and the non-resonant background,
as well as tests of quark-hadron duality.

A future experiment with the energy-upgraded accelerator at Jefferson Lab 
will allow us to improve both the statistical precision and to extend
the range in $x$ \cite{bonus12}. This experiment will finally settle the
question about the asymptotic behavior of the $d/u$ ratio in the limit
$x \to 1$.

\section*{Acknowledgements}

We thank the staff of the Jefferson Lab accelerator and Hall B for their
support on this experiment.
 This work was supported in part by 
 the Chilean Comisi\'on Nacional de Investigaci\'on Cient\'ifica y Tecnol\'ogica (CONICYT),
 the Italian Istituto Nazionale di Fisica Nucleare,
 the French Centre National de la Recherche Scientifique,
 the French Commissariat \`{a} l'Energie Atomique,
 the U.S. Department of Energy,
 the National Science Foundation,
 the Scottish Universities Physics Alliance (SUPA),
 the United Kingdom's Science and Technology Facilities Council,
 and the National Research Foundation of Korea.

 The Southeastern Universities Research Association (SURA) operates the
 Thomas Jefferson National Accelerator Facility for the United States
 Department of Energy under contract DE-AC05-84ER40150.


\begin{thebibliography}{72}%

\makeatletter
\providecommand \@ifxundefined [1]{%
 \@ifx{#1\undefined}
}%
\providecommand \@ifnum [1]{%
 \ifnum #1\expandafter \@firstoftwo
 \else \expandafter \@secondoftwo
 \fi
}%
\providecommand \@ifx [1]{%
 \ifx #1\expandafter \@firstoftwo
 \else \expandafter \@secondoftwo
 \fi
}%
\providecommand \natexlab [1]{#1}%
\providecommand \enquote  [1]{``#1''}%
\providecommand \bibnamefont  [1]{#1}%
\providecommand \bibfnamefont [1]{#1}%
\providecommand \citenamefont [1]{#1}%
\providecommand \href@noop [0]{\@secondoftwo}%
\providecommand \href [0]{\begingroup \@sanitize@url \@href}%
\providecommand \@href[1]{\@@startlink{#1}\@@href}%
\providecommand \@@href[1]{\endgroup#1\@@endlink}%
\providecommand \@sanitize@url [0]{\catcode `\\12\catcode `\$12\catcode
  `\&12\catcode `\#12\catcode `\^12\catcode `\_12\catcode `\%12\relax}%
\providecommand \@@startlink[1]{}%
\providecommand \@@endlink[0]{}%
\providecommand \url  [0]{\begingroup\@sanitize@url \@url }%
\providecommand \@url [1]{\endgroup\@href {#1}{\urlprefix }}%
\providecommand \urlprefix  [0]{URL }%
\providecommand \Eprint [0]{\href }%
\providecommand \doibase [0]{http://dx.doi.org/}%
\providecommand \selectlanguage [0]{\@gobble}%
\providecommand \bibinfo  [0]{\@secondoftwo}%
\providecommand \bibfield  [0]{\@secondoftwo}%
\providecommand \translation [1]{[#1]}%
\providecommand \BibitemOpen [0]{}%
\providecommand \bibitemStop [0]{}%
\providecommand \bibitemNoStop [0]{.\EOS\space}%
\providecommand \EOS [0]{\spacefactor3000\relax}%
\providecommand \BibitemShut  [1]{\csname bibitem#1\endcsname}%
\let\auto@bib@innerbib\@empty
\bibitem [{\citenamefont {Christy}\ and\ \citenamefont
  {Melnitchouk}(2011)}]{Christy}%
  \BibitemOpen
  \bibfield  {author} {\bibinfo {author} {\bibfnamefont {M.~E.}\ \bibnamefont
  {Christy}}\ and\ \bibinfo {author} {\bibfnamefont {W.}~\bibnamefont
  {Melnitchouk}},\ }\href@noop {} {\bibfield  {journal} {\bibinfo  {journal}
  {J. Phys. Conf. Ser.}\ }\textbf {\bibinfo {volume} {299}},\ \bibinfo {pages}
  {012004} (\bibinfo {year} {2011})}\BibitemShut {NoStop}%
\bibitem [{\citenamefont {Chen}\ \emph {et~al.}(2011)\citenamefont {Chen},
  \citenamefont {Deur}, \citenamefont {Kuhn},\ and\ \citenamefont
  {Meziani}}]{Chen:2011zzp}%
  \BibitemOpen
  \bibfield  {author} {\bibinfo {author} {\bibfnamefont {J.}~\bibnamefont
  {Chen}}, \bibinfo {author} {\bibfnamefont {A.}~\bibnamefont {Deur}}, \bibinfo
  {author} {\bibfnamefont {S.}~\bibnamefont {Kuhn}}, \ and\ \bibinfo {author}
  {\bibfnamefont {Z.}~\bibnamefont {Meziani}},\ }\href {\doibase
  10.1088/1742-6596/299/1/012005} {\bibfield  {journal} {\bibinfo  {journal}
  {J. Phys. Conf. Ser.}\ }\textbf {\bibinfo {volume} {299}},\ \bibinfo {pages}
  {012005} (\bibinfo {year} {2011})}\BibitemShut {NoStop}%
\bibitem [{\citenamefont {Melnitchouk}\ \emph {et~al.}(2005)\citenamefont
  {Melnitchouk}, \citenamefont {Ent},\ and\ \citenamefont
  {Keppel}}]{Melnitchouk05}%
  \BibitemOpen
  \bibfield  {author} {\bibinfo {author} {\bibfnamefont {W.}~\bibnamefont
  {Melnitchouk}}, \bibinfo {author} {\bibfnamefont {R.}~\bibnamefont {Ent}}, \
  and\ \bibinfo {author} {\bibfnamefont {C.~E.}\ \bibnamefont {Keppel}},\
  }\href@noop {} {\bibfield  {journal} {\bibinfo  {journal} {Phys. Rep.}\
  }\textbf {\bibinfo {volume} {406}},\ \bibinfo {pages} {127} (\bibinfo {year}
  {2005})}\BibitemShut {NoStop}%
\bibitem [{\citenamefont {Holt}\ and\ \citenamefont {Roberts}(2010)}]{Holt10}%
  \BibitemOpen
  \bibfield  {author} {\bibinfo {author} {\bibfnamefont {R.~J.}\ \bibnamefont
  {Holt}}\ and\ \bibinfo {author} {\bibfnamefont {C.~D.}\ \bibnamefont
  {Roberts}},\ }\href@noop {} {\bibfield  {journal} {\bibinfo  {journal} {Rev.
  Mod. Phys.}\ }\textbf {\bibinfo {volume} {82}},\ \bibinfo {pages} {2991}
  (\bibinfo {year} {2010})}\BibitemShut {NoStop}%
\bibitem [{\citenamefont {Jimenez-Delgado}\ \emph {et~al.}(2013)\citenamefont
  {Jimenez-Delgado}, \citenamefont {Melnitchouk},\ and\ \citenamefont
  {Owens}}]{JMO13}%
  \BibitemOpen
  \bibfield  {author} {\bibinfo {author} {\bibfnamefont {P.}~\bibnamefont
  {Jimenez-Delgado}}, \bibinfo {author} {\bibfnamefont {W.}~\bibnamefont
  {Melnitchouk}}, \ and\ \bibinfo {author} {\bibfnamefont {J.~F.}\ \bibnamefont
  {Owens}},\ }\href {\doibase 10.1088/0954-3899/40/9/093102} {\bibfield
  {journal} {\bibinfo  {journal} {J. Phys. G}\ }\textbf {\bibinfo {volume}
  {40}},\ \bibinfo {pages} {093102} (\bibinfo {year} {2013})}\BibitemShut
  {NoStop}%
\bibitem [{\citenamefont {Bleckwenn}\ \emph {et~al.}(1972)\citenamefont
  {Bleckwenn}, \citenamefont {Moritz}, \citenamefont {Schmidt},\ and\
  \citenamefont {Wegener}}]{Bleckwenn}%
  \BibitemOpen
  \bibfield  {author} {\bibinfo {author} {\bibfnamefont {J.}~\bibnamefont
  {Bleckwenn}}, \bibinfo {author} {\bibfnamefont {J.}~\bibnamefont {Moritz}},
  \bibinfo {author} {\bibfnamefont {K.~H.}\ \bibnamefont {Schmidt}}, \ and\
  \bibinfo {author} {\bibfnamefont {D.}~\bibnamefont {Wegener}},\ }\href
  {\doibase 10.1016/0370-2693(72)90395-4} {\bibfield  {journal} {\bibinfo
  {journal} {Phys. Lett. B}\ }\textbf {\bibinfo {volume} {38}},\ \bibinfo
  {pages} {265} (\bibinfo {year} {1972})}\BibitemShut {NoStop}%
\bibitem [{\citenamefont {Kobberling}\ \emph {et~al.}(1974)\citenamefont
  {Kobberling}, \citenamefont {Mortiz}, \citenamefont {Schmidt}, \citenamefont
  {Wegener}, \citenamefont {Zeller} \emph {et~al.}}]{Kobberling}%
  \BibitemOpen
  \bibfield  {author} {\bibinfo {author} {\bibfnamefont {M.}~\bibnamefont
  {Kobberling}}, \bibinfo {author} {\bibfnamefont {J.}~\bibnamefont {Mortiz}},
  \bibinfo {author} {\bibfnamefont {K.~H.}\ \bibnamefont {Schmidt}}, \bibinfo
  {author} {\bibfnamefont {D.}~\bibnamefont {Wegener}}, \bibinfo {author}
  {\bibfnamefont {D.}~\bibnamefont {Zeller}},  \emph {et~al.},\ }\href
  {\doibase 10.1016/0550-3213(74)90504-5} {\bibfield  {journal} {\bibinfo
  {journal} {Nucl. Phys.}\ }\textbf {\bibinfo {volume} {B82}},\ \bibinfo
  {pages} {201} (\bibinfo {year} {1974})}\BibitemShut {NoStop}%
\bibitem [{\citenamefont {Stuart}\ \emph {et~al.}(1998)\citenamefont {Stuart},
  \citenamefont {Bosted}, \citenamefont {Andivahis}, \citenamefont {Lung},
  \citenamefont {Alster} \emph {et~al.}}]{Stuart}%
  \BibitemOpen
  \bibfield  {author} {\bibinfo {author} {\bibfnamefont {L.~M.}\ \bibnamefont
  {Stuart}}, \bibinfo {author} {\bibfnamefont {P.~E.}\ \bibnamefont {Bosted}},
  \bibinfo {author} {\bibfnamefont {L.}~\bibnamefont {Andivahis}}, \bibinfo
  {author} {\bibfnamefont {A.}~\bibnamefont {Lung}}, \bibinfo {author}
  {\bibfnamefont {J.}~\bibnamefont {Alster}},  \emph {et~al.},\ }\href
  {\doibase 10.1103/PhysRevD.58.032003} {\bibfield  {journal} {\bibinfo
  {journal} {Phys. Rev. D}\ }\textbf {\bibinfo {volume} {58}},\ \bibinfo
  {pages} {032003} (\bibinfo {year} {1998})}\BibitemShut {NoStop}%
\bibitem [{\citenamefont {Bloom}\ and\ \citenamefont {Gilman}(1970)}]{Bloom69}%
  \BibitemOpen
  \bibfield  {author} {\bibinfo {author} {\bibfnamefont {E.~D.}\ \bibnamefont
  {Bloom}}\ and\ \bibinfo {author} {\bibfnamefont {F.~J.}\ \bibnamefont
  {Gilman}},\ }\href {\doibase 10.1103/PhysRevLett.25.1140} {\bibfield
  {journal} {\bibinfo  {journal} {Phys. Rev. Lett.}\ }\textbf {\bibinfo
  {volume} {25}},\ \bibinfo {pages} {1140} (\bibinfo {year}
  {1970})}\BibitemShut {NoStop}%
\bibitem [{\citenamefont {Whitlow}\ \emph {et~al.}(1992)\citenamefont {Whitlow}
  \emph {et~al.}}]{slacdata1}%
  \BibitemOpen
  \bibfield  {author} {\bibinfo {author} {\bibfnamefont {L.~W.}\ \bibnamefont
  {Whitlow}} \emph {et~al.},\ }\href@noop {} {\bibfield  {journal} {\bibinfo
  {journal} {Phys. Lett. B}\ }\textbf {\bibinfo {volume} {282}},\ \bibinfo
  {pages} {475} (\bibinfo {year} {1992})}\BibitemShut {NoStop}%
\bibitem [{\citenamefont {Melnitchouk}\ and\ \citenamefont
  {Thomas}(1996)}]{Melnitchouk96}%
  \BibitemOpen
  \bibfield  {author} {\bibinfo {author} {\bibfnamefont {W.}~\bibnamefont
  {Melnitchouk}}\ and\ \bibinfo {author} {\bibfnamefont {A.~W.}\ \bibnamefont
  {Thomas}},\ }\href@noop {} {\bibfield  {journal} {\bibinfo  {journal} {Phys.
  Lett. B}\ }\textbf {\bibinfo {volume} {377}},\ \bibinfo {pages} {11}
  (\bibinfo {year} {1996})}\BibitemShut {NoStop}%
\bibitem [{\citenamefont {Arrington}\ \emph {et~al.}(2009)\citenamefont
  {Arrington} \emph {et~al.}}]{Arrington09}%
  \BibitemOpen
  \bibfield  {author} {\bibinfo {author} {\bibfnamefont {J.}~\bibnamefont
  {Arrington}} \emph {et~al.},\ }\href@noop {} {\bibfield  {journal} {\bibinfo
  {journal} {J. Phys. G}\ }\textbf {\bibinfo {volume} {36}},\ \bibinfo {pages}
  {025005} (\bibinfo {year} {2009})}\BibitemShut {NoStop}%
\bibitem [{\citenamefont {Accardi}\ \emph {et~al.}(2011)\citenamefont
  {Accardi}, \citenamefont {Melnitchouk}, \citenamefont {Owens}, \citenamefont
  {Christy}, \citenamefont {Keppel} \emph {et~al.}}]{CJ_accardi}%
  \BibitemOpen
  \bibfield  {author} {\bibinfo {author} {\bibfnamefont {A.}~\bibnamefont
  {Accardi}}, \bibinfo {author} {\bibfnamefont {W.}~\bibnamefont
  {Melnitchouk}}, \bibinfo {author} {\bibfnamefont {J.~F.}\ \bibnamefont
  {Owens}}, \bibinfo {author} {\bibfnamefont {M.~E.}\ \bibnamefont {Christy}},
  \bibinfo {author} {\bibfnamefont {C.~E.}\ \bibnamefont {Keppel}},  \emph
  {et~al.},\ }\href {\doibase 10.1103/PhysRevD.84.014008} {\bibfield  {journal}
  {\bibinfo  {journal} {Phys. Rev. D}\ }\textbf {\bibinfo {volume} {84}},\
  \bibinfo {pages} {014008} (\bibinfo {year} {2011})}\BibitemShut {NoStop}%
\bibitem [{\citenamefont {Arrington}\ \emph {et~al.}(2012)\citenamefont
  {Arrington}, \citenamefont {Rubin},\ and\ \citenamefont
  {Melnitchouk}}]{Arrington:2011qt}%
  \BibitemOpen
  \bibfield  {author} {\bibinfo {author} {\bibfnamefont {J.}~\bibnamefont
  {Arrington}}, \bibinfo {author} {\bibfnamefont {J.~G.}\ \bibnamefont
  {Rubin}}, \ and\ \bibinfo {author} {\bibfnamefont {W.}~\bibnamefont
  {Melnitchouk}},\ }\href {\doibase 10.1103/PhysRevLett.108.252001} {\bibfield
  {journal} {\bibinfo  {journal} {Phys. Rev. Lett.}\ }\textbf {\bibinfo
  {volume} {108}},\ \bibinfo {pages} {252001} (\bibinfo {year}
  {2012})}\BibitemShut {NoStop}%
\bibitem [{\citenamefont {Owens}\ \emph {et~al.}(2013)\citenamefont {Owens},
  \citenamefont {Accardi},\ and\ \citenamefont {Melnitchouk}}]{Owens:2012bv}%
  \BibitemOpen
  \bibfield  {author} {\bibinfo {author} {\bibfnamefont {J.~F.}\ \bibnamefont
  {Owens}}, \bibinfo {author} {\bibfnamefont {A.}~\bibnamefont {Accardi}}, \
  and\ \bibinfo {author} {\bibfnamefont {W.}~\bibnamefont {Melnitchouk}},\
  }\href {\doibase 10.1103/PhysRevD.87.094012} {\bibfield  {journal} {\bibinfo
  {journal} {Phys. Rev. D}\ }\textbf {\bibinfo {volume} {87}},\ \bibinfo
  {pages} {094012} (\bibinfo {year} {2013})}\BibitemShut {NoStop}%
\bibitem [{\citenamefont {Kuhlmann}\ \emph {et~al.}(2000)\citenamefont
  {Kuhlmann} \emph {et~al.}}]{Kuhlmann00}%
  \BibitemOpen
  \bibfield  {author} {\bibinfo {author} {\bibfnamefont {S.}~\bibnamefont
  {Kuhlmann}} \emph {et~al.},\ }\href@noop {} {\bibfield  {journal} {\bibinfo
  {journal} {Phys. Lett. B}\ }\textbf {\bibinfo {volume} {476}},\ \bibinfo
  {pages} {291} (\bibinfo {year} {2000})}\BibitemShut {NoStop}%
\bibitem [{\citenamefont {Brady}\ \emph {et~al.}(2012)\citenamefont {Brady},
  \citenamefont {Accardi}, \citenamefont {Melnitchouk},\ and\ \citenamefont
  {Owens}}]{Brady:2011hb}%
  \BibitemOpen
  \bibfield  {author} {\bibinfo {author} {\bibfnamefont {L.~T.}\ \bibnamefont
  {Brady}}, \bibinfo {author} {\bibfnamefont {A.}~\bibnamefont {Accardi}},
  \bibinfo {author} {\bibfnamefont {W.}~\bibnamefont {Melnitchouk}}, \ and\
  \bibinfo {author} {\bibfnamefont {J.~F.}\ \bibnamefont {Owens}},\ }\href@noop
  {} {\bibfield  {journal} {\bibinfo  {journal} {JHEP}\ }\textbf {\bibinfo
  {volume} {1206}},\ \bibinfo {pages} {019} (\bibinfo {year}
  {2012})}\BibitemShut {NoStop}%
\bibitem [{\citenamefont {Frankfurt}\ and\ \citenamefont
  {Strikman}(1981)}]{frstr}%
  \BibitemOpen
  \bibfield  {author} {\bibinfo {author} {\bibfnamefont {L.~L.}\ \bibnamefont
  {Frankfurt}}\ and\ \bibinfo {author} {\bibfnamefont {M.~I.}\ \bibnamefont
  {Strikman}},\ }\href@noop {} {\bibfield  {journal} {\bibinfo  {journal}
  {Phys. Rep.}\ }\textbf {\bibinfo {volume} {76}},\ \bibinfo {pages} {4}
  (\bibinfo {year} {1981})}\BibitemShut {NoStop}%
\bibitem [{\citenamefont {Simula}(1996)}]{cda2_2}%
  \BibitemOpen
  \bibfield  {author} {\bibinfo {author} {\bibfnamefont {S.}~\bibnamefont
  {Simula}},\ }\href@noop {} {\bibfield  {journal} {\bibinfo  {journal} {Phys.
  Lett. B}\ }\textbf {\bibinfo {volume} {387}},\ \bibinfo {pages} {245}
  (\bibinfo {year} {1996})}\BibitemShut {NoStop}%
\bibitem [{\citenamefont {Melnitchouk}\ \emph {et~al.}(1997)\citenamefont
  {Melnitchouk}, \citenamefont {Sargsian},\ and\ \citenamefont
  {Strikman}}]{mel_sarg_strik}%
  \BibitemOpen
  \bibfield  {author} {\bibinfo {author} {\bibfnamefont {W.}~\bibnamefont
  {Melnitchouk}}, \bibinfo {author} {\bibfnamefont {M.}~\bibnamefont
  {Sargsian}}, \ and\ \bibinfo {author} {\bibfnamefont {M.~I.}\ \bibnamefont
  {Strikman}},\ }\href@noop {} {\bibfield  {journal} {\bibinfo  {journal} {Z.
  Phys. A}\ }\textbf {\bibinfo {volume} {359}},\ \bibinfo {pages} {99}
  (\bibinfo {year} {1997})}\BibitemShut {NoStop}%
\bibitem [{\citenamefont {Sargsian}\ and\ \citenamefont
  {Strikman}(2006)}]{sargsian}%
  \BibitemOpen
  \bibfield  {author} {\bibinfo {author} {\bibfnamefont {M.}~\bibnamefont
  {Sargsian}}\ and\ \bibinfo {author} {\bibfnamefont {M.}~\bibnamefont
  {Strikman}},\ }\href@noop {} {\bibfield  {journal} {\bibinfo  {journal}
  {Phys. Lett. B}\ }\textbf {\bibinfo {volume} {639}},\ \bibinfo {pages} {223}
  (\bibinfo {year} {2006})}\BibitemShut {NoStop}%
\bibitem [{\citenamefont {degli Atti}\ \emph {et~al.}(2004)\citenamefont {degli
  Atti}, \citenamefont {Kaptari},\ and\ \citenamefont {Kopeliovich}}]{cda4}%
  \BibitemOpen
  \bibfield  {author} {\bibinfo {author} {\bibfnamefont {C.~C.}\ \bibnamefont
  {degli Atti}}, \bibinfo {author} {\bibfnamefont {L.~P.}\ \bibnamefont
  {Kaptari}}, \ and\ \bibinfo {author} {\bibfnamefont {B.~Z.}\ \bibnamefont
  {Kopeliovich}},\ }\href@noop {} {\bibfield  {journal} {\bibinfo  {journal}
  {Eur. Phys. J. A}\ }\textbf {\bibinfo {volume} {19}},\ \bibinfo {pages} {145}
  (\bibinfo {year} {2004})}\BibitemShut {NoStop}%
\bibitem [{\citenamefont {Cosyn}\ and\ \citenamefont
  {Sargsian}(2011)}]{Cosyn11}%
  \BibitemOpen
  \bibfield  {author} {\bibinfo {author} {\bibfnamefont {W.}~\bibnamefont
  {Cosyn}}\ and\ \bibinfo {author} {\bibfnamefont {M.}~\bibnamefont
  {Sargsian}},\ }\href@noop {} {\bibfield  {journal} {\bibinfo  {journal}
  {Phys. Rev. C}\ }\textbf {\bibinfo {volume} {84}},\ \bibinfo {pages} {014601}
  (\bibinfo {year} {2011})}\BibitemShut {NoStop}%
\bibitem [{\citenamefont {Baillie}\ \emph {et~al.}(2012)\citenamefont {Baillie}
  \emph {et~al.}}]{prl}%
  \BibitemOpen
  \bibfield  {author} {\bibinfo {author} {\bibfnamefont {N.}~\bibnamefont
  {Baillie}} \emph {et~al.},\ }\href@noop {} {\bibfield  {journal} {\bibinfo
  {journal} {Phys. Rev. Lett.}\ }\textbf {\bibinfo {volume} {108}},\ \bibinfo
  {pages} {142001} (\bibinfo {year} {2012})}\BibitemShut {NoStop}%
\bibitem [{\citenamefont {Niculescu}\ \emph {et~al.}(2000)\citenamefont
  {Niculescu} \emph {et~al.}}]{Niculescu:2000tk}%
  \BibitemOpen
  \bibfield  {author} {\bibinfo {author} {\bibfnamefont {I.}~\bibnamefont
  {Niculescu}} \emph {et~al.},\ }\href {\doibase 10.1103/PhysRevLett.85.1186}
  {\bibfield  {journal} {\bibinfo  {journal} {Phys. Rev. Lett.}\ }\textbf
  {\bibinfo {volume} {85}},\ \bibinfo {pages} {1186} (\bibinfo {year}
  {2000})}\BibitemShut {NoStop}%
\bibitem [{\citenamefont {Arrington}\ \emph {et~al.}(2006)\citenamefont
  {Arrington}, \citenamefont {Ent}, \citenamefont {Keppel}, \citenamefont
  {Mammei},\ and\ \citenamefont {Niculescu}}]{Arrington:2003nt}%
  \BibitemOpen
  \bibfield  {author} {\bibinfo {author} {\bibfnamefont {J.}~\bibnamefont
  {Arrington}}, \bibinfo {author} {\bibfnamefont {R.}~\bibnamefont {Ent}},
  \bibinfo {author} {\bibfnamefont {C.~E.}\ \bibnamefont {Keppel}}, \bibinfo
  {author} {\bibfnamefont {J.}~\bibnamefont {Mammei}}, \ and\ \bibinfo {author}
  {\bibfnamefont {I.}~\bibnamefont {Niculescu}},\ }\href {\doibase
  10.1103/PhysRevC.73.035205} {\bibfield  {journal} {\bibinfo  {journal} {Phys.
  Rev. C}\ }\textbf {\bibinfo {volume} {73}},\ \bibinfo {pages} {035205}
  (\bibinfo {year} {2006})}\BibitemShut {NoStop}%
\bibitem [{\citenamefont {Malace}\ \emph {et~al.}(2010)\citenamefont {Malace},
  \citenamefont {Kahn}, \citenamefont {Melnitchouk},\ and\ \citenamefont
  {Keppel}}]{Malace1}%
  \BibitemOpen
  \bibfield  {author} {\bibinfo {author} {\bibfnamefont {S.~P.}\ \bibnamefont
  {Malace}}, \bibinfo {author} {\bibfnamefont {Y.}~\bibnamefont {Kahn}},
  \bibinfo {author} {\bibfnamefont {W.}~\bibnamefont {Melnitchouk}}, \ and\
  \bibinfo {author} {\bibfnamefont {C.~E.}\ \bibnamefont {Keppel}},\
  }\href@noop {} {\bibfield  {journal} {\bibinfo  {journal} {Phys. Rev. Lett.}\
  }\textbf {\bibinfo {volume} {104}},\ \bibinfo {pages} {102001} (\bibinfo
  {year} {2010})}\BibitemShut {NoStop}%
\bibitem [{\citenamefont {Kahn}\ \emph {et~al.}(2009)\citenamefont {Kahn},
  \citenamefont {Melnitchouk},\ and\ \citenamefont {Kulagin}}]{Kahn:2008nq}%
  \BibitemOpen
  \bibfield  {author} {\bibinfo {author} {\bibfnamefont {Y.}~\bibnamefont
  {Kahn}}, \bibinfo {author} {\bibfnamefont {W.}~\bibnamefont {Melnitchouk}}, \
  and\ \bibinfo {author} {\bibfnamefont {S.~A.}\ \bibnamefont {Kulagin}},\
  }\href {\doibase 10.1103/PhysRevC.79.035205} {\bibfield  {journal} {\bibinfo
  {journal} {Phys. Rev. C}\ }\textbf {\bibinfo {volume} {79}},\ \bibinfo
  {pages} {035205} (\bibinfo {year} {2009})}\BibitemShut {NoStop}%
\bibitem [{\citenamefont {Melnitchouk}\ \emph {et~al.}(1999)\citenamefont
  {Melnitchouk}, \citenamefont {Schreiber},\ and\ \citenamefont
  {Thomas}}]{meln_schr_thom}%
  \BibitemOpen
  \bibfield  {author} {\bibinfo {author} {\bibfnamefont {W.}~\bibnamefont
  {Melnitchouk}}, \bibinfo {author} {\bibfnamefont {A.~W.}\ \bibnamefont
  {Schreiber}}, \ and\ \bibinfo {author} {\bibfnamefont {A.~W.}\ \bibnamefont
  {Thomas}},\ }\href@noop {} {\bibfield  {journal} {\bibinfo  {journal} {Phys.
  Lett. B}\ }\textbf {\bibinfo {volume} {335}},\ \bibinfo {pages} {11}
  (\bibinfo {year} {1999})}\BibitemShut {NoStop}%
\bibitem [{\citenamefont {Kulagin}\ and\ \citenamefont
  {Petti}(2006)}]{Kulagin06}%
  \BibitemOpen
  \bibfield  {author} {\bibinfo {author} {\bibfnamefont {S.~A.}\ \bibnamefont
  {Kulagin}}\ and\ \bibinfo {author} {\bibfnamefont {R.}~\bibnamefont
  {Petti}},\ }\href {\doibase 10.1016/j.nuclphysa.2005.10.011} {\bibfield
  {journal} {\bibinfo  {journal} {Nucl. Phys.}\ }\textbf {\bibinfo {volume}
  {A765}},\ \bibinfo {pages} {126} (\bibinfo {year} {2006})}\BibitemShut
  {NoStop}%
\bibitem [{\citenamefont {West}(1971)}]{West:1972qj}%
  \BibitemOpen
  \bibfield  {author} {\bibinfo {author} {\bibfnamefont {G.~B.}\ \bibnamefont
  {West}},\ }\href {\doibase 10.1016/0370-2693(71)90358-3} {\bibfield
  {journal} {\bibinfo  {journal} {Phys. Lett. B}\ }\textbf {\bibinfo {volume}
  {37}},\ \bibinfo {pages} {509} (\bibinfo {year} {1971})}\BibitemShut
  {NoStop}%
\bibitem [{\citenamefont {Frankfurt}\ and\ \citenamefont
  {Strikman}(1988)}]{Frankfurt:1988nt}%
  \BibitemOpen
  \bibfield  {author} {\bibinfo {author} {\bibfnamefont {L.~L.}\ \bibnamefont
  {Frankfurt}}\ and\ \bibinfo {author} {\bibfnamefont {M.~I.}\ \bibnamefont
  {Strikman}},\ }\href {\doibase 10.1016/0370-1573(88)90179-2} {\bibfield
  {journal} {\bibinfo  {journal} {Phys. Rep.}\ }\textbf {\bibinfo {volume}
  {160}},\ \bibinfo {pages} {235} (\bibinfo {year} {1988})}\BibitemShut
  {NoStop}%
\bibitem [{\citenamefont {Kaptari}\ and\ \citenamefont
  {Umnikov}(1991)}]{Kaptar:1991hx}%
  \BibitemOpen
  \bibfield  {author} {\bibinfo {author} {\bibfnamefont {L.~P.}\ \bibnamefont
  {Kaptari}}\ and\ \bibinfo {author} {\bibfnamefont {A.~Y.}\ \bibnamefont
  {Umnikov}},\ }\href {\doibase 10.1016/0370-2693(91)90151-F} {\bibfield
  {journal} {\bibinfo  {journal} {Phys. Lett. B}\ }\textbf {\bibinfo {volume}
  {259}},\ \bibinfo {pages} {155} (\bibinfo {year} {1991})}\BibitemShut
  {NoStop}%
\bibitem [{\citenamefont {Frankfurt}\ \emph {et~al.}(1995)\citenamefont
  {Frankfurt}, \citenamefont {Greenberg}, \citenamefont {Miller}, \citenamefont
  {Sargsian},\ and\ \citenamefont {Strikman}}]{Frankfurt94}%
  \BibitemOpen
  \bibfield  {author} {\bibinfo {author} {\bibfnamefont {L.~L.}\ \bibnamefont
  {Frankfurt}}, \bibinfo {author} {\bibfnamefont {W.~R.}\ \bibnamefont
  {Greenberg}}, \bibinfo {author} {\bibfnamefont {G.~A.}\ \bibnamefont
  {Miller}}, \bibinfo {author} {\bibfnamefont {M.~M.}\ \bibnamefont
  {Sargsian}}, \ and\ \bibinfo {author} {\bibfnamefont {M.~I.}\ \bibnamefont
  {Strikman}},\ }\href {\doibase 10.1007/BF01292764} {\bibfield  {journal}
  {\bibinfo  {journal} {Z. Phys. A}\ }\textbf {\bibinfo {volume} {352}},\
  \bibinfo {pages} {97} (\bibinfo {year} {1995})}\BibitemShut {NoStop}%
\bibitem [{\citenamefont {Adams}\ \emph {et~al.}(1995)\citenamefont {Adams}
  \emph {et~al.}}]{Adams95}%
  \BibitemOpen
  \bibfield  {author} {\bibinfo {author} {\bibfnamefont {M.~R.}\ \bibnamefont
  {Adams}} \emph {et~al.},\ }\href {\doibase 10.1103/PhysRevLett.74.5198}
  {\bibfield  {journal} {\bibinfo  {journal} {Phys. Rev. Lett.}\ }\textbf
  {\bibinfo {volume} {74}},\ \bibinfo {pages} {5198} (\bibinfo {year}
  {1995})}\BibitemShut {NoStop}%
\bibitem [{\citenamefont {degli Atti}\ and\ \citenamefont
  {Simula}(1993)}]{cda2_1}%
  \BibitemOpen
  \bibfield  {author} {\bibinfo {author} {\bibfnamefont {C.~C.}\ \bibnamefont
  {degli Atti}}\ and\ \bibinfo {author} {\bibfnamefont {S.}~\bibnamefont
  {Simula}},\ }\href@noop {} {\bibfield  {journal} {\bibinfo  {journal} {Phys.
  Lett. B}\ }\textbf {\bibinfo {volume} {319}},\ \bibinfo {pages} {23}
  (\bibinfo {year} {1993})}\BibitemShut {NoStop}%
\bibitem [{\citenamefont {Bosveld}\ \emph {et~al.}(1994)\citenamefont
  {Bosveld}, \citenamefont {Dieperink},\ and\ \citenamefont
  {Tenner}}]{Bosveld94}%
  \BibitemOpen
  \bibfield  {author} {\bibinfo {author} {\bibfnamefont {G.~D.}\ \bibnamefont
  {Bosveld}}, \bibinfo {author} {\bibfnamefont {A.~E.~L.}\ \bibnamefont
  {Dieperink}}, \ and\ \bibinfo {author} {\bibfnamefont {A.~G.}\ \bibnamefont
  {Tenner}},\ }\href {\doibase 10.1103/PhysRevC.49.2379} {\bibfield  {journal}
  {\bibinfo  {journal} {Phys. Rev. C}\ }\textbf {\bibinfo {volume} {49}},\
  \bibinfo {pages} {2379} (\bibinfo {year} {1994})}\BibitemShut {NoStop}%
\bibitem [{\citenamefont {Heller}\ and\ \citenamefont
  {Thomas}(1990)}]{Heller90}%
  \BibitemOpen
  \bibfield  {author} {\bibinfo {author} {\bibfnamefont {L.}~\bibnamefont
  {Heller}}\ and\ \bibinfo {author} {\bibfnamefont {A.~W.}\ \bibnamefont
  {Thomas}},\ }\href {\doibase 10.1103/PhysRevC.41.2756} {\bibfield  {journal}
  {\bibinfo  {journal} {Phys. Rev. C}\ }\textbf {\bibinfo {volume} {41}},\
  \bibinfo {pages} {2756} (\bibinfo {year} {1990})}\BibitemShut {NoStop}%
\bibitem [{\citenamefont {Gross}\ and\ \citenamefont
  {Liuti}(1992)}]{gross_liuti1}%
  \BibitemOpen
  \bibfield  {author} {\bibinfo {author} {\bibfnamefont {F.}~\bibnamefont
  {Gross}}\ and\ \bibinfo {author} {\bibfnamefont {S.}~\bibnamefont {Liuti}},\
  }\href@noop {} {\bibfield  {journal} {\bibinfo  {journal} {Phys. Rev. C}\
  }\textbf {\bibinfo {volume} {45}},\ \bibinfo {pages} {1374} (\bibinfo {year}
  {1992})}\BibitemShut {NoStop}%
\bibitem [{\citenamefont {Klimenko}\ \emph {et~al.}(2006)\citenamefont
  {Klimenko} \emph {et~al.}}]{Klimenko06}%
  \BibitemOpen
  \bibfield  {author} {\bibinfo {author} {\bibfnamefont {A.~V.}\ \bibnamefont
  {Klimenko}} \emph {et~al.},\ }\href@noop {} {\bibfield  {journal} {\bibinfo
  {journal} {Phys. Rev. C}\ }\textbf {\bibinfo {volume} {73}},\ \bibinfo
  {pages} {035212} (\bibinfo {year} {2006})}\BibitemShut {NoStop}%
\bibitem [{\citenamefont {Mestayer}\ \emph {et~al.}(2000)\citenamefont
  {Mestayer} \emph {et~al.}}]{Mestayer:2000we}%
  \BibitemOpen
  \bibfield  {author} {\bibinfo {author} {\bibfnamefont {M.~D.}\ \bibnamefont
  {Mestayer}} \emph {et~al.},\ }\href {\doibase 10.1016/S0168-9002(00)00151-0}
  {\bibfield  {journal} {\bibinfo  {journal} {Nucl. Instrum. Meth. A}\ }\textbf
  {\bibinfo {volume} {449}},\ \bibinfo {pages} {81} (\bibinfo {year}
  {2000})}\BibitemShut {NoStop}%
\bibitem [{\citenamefont {Adams}\ \emph {et~al.}(2001)\citenamefont {Adams}
  \emph {et~al.}}]{Adams:2001kk}%
  \BibitemOpen
  \bibfield  {author} {\bibinfo {author} {\bibfnamefont {G.}~\bibnamefont
  {Adams}} \emph {et~al.},\ }\href {\doibase 10.1016/S0168-9002(00)01313-9}
  {\bibfield  {journal} {\bibinfo  {journal} {Nucl. Instrum. Meth. A}\ }\textbf
  {\bibinfo {volume} {465}},\ \bibinfo {pages} {414} (\bibinfo {year}
  {2001})}\BibitemShut {NoStop}%
\bibitem [{\citenamefont {Smith}\ \emph {et~al.}(1999)\citenamefont {Smith}
  \emph {et~al.}}]{essmith}%
  \BibitemOpen
  \bibfield  {author} {\bibinfo {author} {\bibfnamefont {E.~S.}\ \bibnamefont
  {Smith}} \emph {et~al.},\ }\href@noop {} {\bibfield  {journal} {\bibinfo
  {journal} {Nucl. Instrum. Meth. A}\ }\textbf {\bibinfo {volume} {432}},\
  \bibinfo {pages} {265} (\bibinfo {year} {1999})}\BibitemShut {NoStop}%
\bibitem [{\citenamefont {Amarian}\ \emph {et~al.}(2001)\citenamefont {Amarian}
  \emph {et~al.}}]{ec}%
  \BibitemOpen
  \bibfield  {author} {\bibinfo {author} {\bibfnamefont {M.}~\bibnamefont
  {Amarian}} \emph {et~al.},\ }\href@noop {} {\bibfield  {journal} {\bibinfo
  {journal} {Nucl. Instrum. Meth. A}\ }\textbf {\bibinfo {volume} {460}},\
  \bibinfo {pages} {239} (\bibinfo {year} {2001})}\BibitemShut {NoStop}%
\bibitem [{\citenamefont {Mecking}\ \emph {et~al.}(2003)\citenamefont {Mecking}
  \emph {et~al.}}]{mecking}%
  \BibitemOpen
  \bibfield  {author} {\bibinfo {author} {\bibfnamefont {B.~A.}\ \bibnamefont
  {Mecking}} \emph {et~al.},\ }\href@noop {} {\bibfield  {journal} {\bibinfo
  {journal} {Nucl. Instrum. Meth. A}\ }\textbf {\bibinfo {volume} {503}},\
  \bibinfo {pages} {513} (\bibinfo {year} {2003})}\BibitemShut {NoStop}%
\bibitem [{\citenamefont {Fenker}\ \emph {et~al.}(2008)\citenamefont {Fenker}
  \emph {et~al.}}]{bonusnim}%
  \BibitemOpen
  \bibfield  {author} {\bibinfo {author} {\bibfnamefont {H.}~\bibnamefont
  {Fenker}} \emph {et~al.},\ }\href@noop {} {\bibfield  {journal} {\bibinfo
  {journal} {Nucl. Instrum. Meth. A}\ }\textbf {\bibinfo {volume} {592}},\
  \bibinfo {pages} {273} (\bibinfo {year} {2008})}\BibitemShut {NoStop}%
\bibitem [{\citenamefont {Eckardt}\ \emph {et~al.}(2001)\citenamefont {Eckardt}
  \emph {et~al.}}]{star}%
  \BibitemOpen
  \bibfield  {author} {\bibinfo {author} {\bibfnamefont {V.}~\bibnamefont
  {Eckardt}} \emph {et~al.},\ }\href@noop {} {\  (\bibinfo {year} {2001})},\
  \Eprint {http://arxiv.org/abs/nucl-ex/0101013} {arXiv:nucl-ex/0101013}
  \BibitemShut {NoStop}%
\bibitem [{\citenamefont {Agakishiev}\ \emph {et~al.}(1999)\citenamefont
  {Agakishiev} \emph {et~al.}}]{ceres}%
  \BibitemOpen
  \bibfield  {author} {\bibinfo {author} {\bibfnamefont {G.}~\bibnamefont
  {Agakishiev}} \emph {et~al.},\ }\href {\doibase
  10.1016/S0375-9474(99)85115-X} {\bibfield  {journal} {\bibinfo  {journal}
  {Nucl. Phys.}\ }\textbf {\bibinfo {volume} {A661}},\ \bibinfo {pages} {673}
  (\bibinfo {year} {1999})}\BibitemShut {NoStop}%
\bibitem [{\citenamefont {Musa}\ \emph {et~al.}(2003)\citenamefont {Musa} \emph
  {et~al.}}]{ALTRO}%
  \BibitemOpen
  \bibfield  {author} {\bibinfo {author} {\bibfnamefont {L.}~\bibnamefont
  {Musa}} \emph {et~al.},\ }\href@noop {} {\bibfield  {journal} {\bibinfo
  {journal} {IEEE Nucl. Sci. Symp. Conf. Record}\ }\textbf {\bibinfo {volume}
  {5}},\ \bibinfo {pages} {3647} (\bibinfo {year} {2003})}\BibitemShut
  {NoStop}%
\bibitem [{\citenamefont {Musa}(2003)}]{ALICE}%
  \BibitemOpen
  \bibfield  {author} {\bibinfo {author} {\bibfnamefont {L.}~\bibnamefont
  {Musa}},\ }\href@noop {} {\bibfield  {journal} {\bibinfo  {journal} {Nucl.
  Phys.}\ }\textbf {\bibinfo {volume} {A715}},\ \bibinfo {pages} {843}
  (\bibinfo {year} {2003})}\BibitemShut {NoStop}%
\bibitem [{\citenamefont {Sauli}(1997)}]{gem_sauli}%
  \BibitemOpen
  \bibfield  {author} {\bibinfo {author} {\bibfnamefont {F.}~\bibnamefont
  {Sauli}},\ }\href@noop {} {\bibfield  {journal} {\bibinfo  {journal} {Nucl.
  Instrum. Meth. A}\ }\textbf {\bibinfo {volume} {386}},\ \bibinfo {pages}
  {531} (\bibinfo {year} {1997})}\BibitemShut {NoStop}%
\bibitem [{\citenamefont {Tkachenko}(2009)}]{mythesis}%
  \BibitemOpen
  \bibfield  {author} {\bibinfo {author} {\bibfnamefont {S.}~\bibnamefont
  {Tkachenko}},\ }\emph {\bibinfo {title} {Neutron structure functions measured
  with spectator tagging}},\ \href
  {http://www.jlab.org/Hall-B/general/thesis/Tkachenko_thesis.pdf} {Ph.D.
  thesis},\ \bibinfo  {school} {Old Dominion University} (\bibinfo {year}
  {2009}),\ \bibinfo {note}
  {http://www.jlab.org/Hall-B/general/thesis/Tkachenko\_thesis.pdf}\BibitemShut
  {NoStop}%
\bibitem [{\citenamefont {Biagi}(1999)}]{magboltz}%
  \BibitemOpen
  \bibfield  {author} {\bibinfo {author} {\bibfnamefont {S.}~\bibnamefont
  {Biagi}},\ }\href@noop {} {\bibfield  {journal} {\bibinfo  {journal} {Nucl.
  Instr. and Meth. A}\ }\textbf {\bibinfo {volume} {421}},\ \bibinfo {pages}
  {234} (\bibinfo {year} {1999})},\ \bibinfo {note} {{The MAGBOLTZ programme,
  version~2 (2005)}}\BibitemShut {NoStop}%
\bibitem [{\citenamefont {Leo}(1994)}]{leo}%
  \BibitemOpen
  \bibfield  {author} {\bibinfo {author} {\bibfnamefont {W.}~\bibnamefont
  {Leo}},\ }\href@noop {} {\emph {\bibinfo {title} {Techniques for nuclear and
  particle physics experiments}}}\ (\bibinfo  {publisher} {Springer-Verlag New
  York, LLC},\ \bibinfo {year} {1994})\BibitemShut {NoStop}%
\bibitem [{\citenamefont {Agostinelli}\ \emph {et~al.}(2003)\citenamefont
  {Agostinelli} \emph {et~al.}}]{geant4}%
  \BibitemOpen
  \bibfield  {author} {\bibinfo {author} {\bibfnamefont {S.}~\bibnamefont
  {Agostinelli}} \emph {et~al.},\ }\href@noop {} {\bibfield  {journal}
  {\bibinfo  {journal} {Nucl. Instrum. Meth. A}\ }\textbf {\bibinfo {volume}
  {506}},\ \bibinfo {pages} {250} (\bibinfo {year} {2003})}\BibitemShut
  {NoStop}%
\bibitem [{\citenamefont {Zhang}(2010)}]{jzthesis}%
  \BibitemOpen
  \bibfield  {author} {\bibinfo {author} {\bibfnamefont {J.}~\bibnamefont
  {Zhang}},\ }\emph {\bibinfo {title} {Exclusive $\pi^-$ electroproduction from
  the neutron in the resonance region}},\ \href
  {http://www.jlab.org/Hall-B/general/thesis/JixieZhang_thesis.pdf} {Ph.D.
  thesis},\ \bibinfo  {school} {Old Dominion University} (\bibinfo {year}
  {2010}),\ \bibinfo {note}
  {http://www.jlab.org/Hall-B/general/thesis/JixieZhang\_thesis.pdf}\BibitemShut
  {NoStop}%
\bibitem [{\citenamefont {Guler}(2009)}]{nguler}%
  \BibitemOpen
  \bibfield  {author} {\bibinfo {author} {\bibfnamefont {N.}~\bibnamefont
  {Guler}},\ }\emph {\bibinfo {title} {Spin structure of the deuteron}},\ \href
  {http://www.jlab.org/Hall-B/general/thesis/Guler_thesis.pdf} {Ph.D. thesis},\
  \bibinfo  {school} {Old Dominion University} (\bibinfo {year} {2009}),\
  \bibinfo {note}
  {http://www.jlab.org/Hall-B/general/thesis/Guler\_thesis.pdf}\BibitemShut
  {NoStop}%
\bibitem [{\citenamefont {Osipenko}\ \emph {et~al.}(2004)\citenamefont
  {Osipenko}, \citenamefont {Vlassov},\ and\ \citenamefont
  {Taiuti}}]{OsipenkoCut}%
  \BibitemOpen
  \bibfield  {author} {\bibinfo {author} {\bibfnamefont {M.}~\bibnamefont
  {Osipenko}}, \bibinfo {author} {\bibfnamefont {A.}~\bibnamefont {Vlassov}}, \
  and\ \bibinfo {author} {\bibfnamefont {M.}~\bibnamefont {Taiuti}},\ }\href
  {https://misportal.jlab.org/ul/Physics/Hall-B/clas/viewFile.cfm/2004-020.pdf?documentId=84}
  {\bibfield  {journal} {\bibinfo  {journal} {CLAS-NOTE 2004-020}\ } (\bibinfo
  {year} {2004})},\ \bibinfo {note}
  {https://misportal.jlab.org/ul/Physics/Hall-B/clas/viewFile.cfm/2004-020.pdf?documentId=84}\BibitemShut
  {NoStop}%
\bibitem [{\citenamefont {Abbott}\ \emph {et~al.}(2000)\citenamefont {Abbott}
  \emph {et~al.}}]{HallCdFF}%
  \BibitemOpen
  \bibfield  {author} {\bibinfo {author} {\bibfnamefont {D.}~\bibnamefont
  {Abbott}} \emph {et~al.},\ }\href@noop {} {\bibfield  {journal} {\bibinfo
  {journal} {Eur. Phys. J. A}\ }\textbf {\bibinfo {volume} {7}},\ \bibinfo
  {pages} {421} (\bibinfo {year} {2000})}\BibitemShut {NoStop}%
\bibitem [{\citenamefont {Mo}\ and\ \citenamefont {Tsai}(1969)}]{motsai}%
  \BibitemOpen
  \bibfield  {author} {\bibinfo {author} {\bibfnamefont {L.~W.}\ \bibnamefont
  {Mo}}\ and\ \bibinfo {author} {\bibfnamefont {Y.~S.}\ \bibnamefont {Tsai}},\
  }\href@noop {} {\bibfield  {journal} {\bibinfo  {journal} {Rev. Mod. Phys.}\
  }\textbf {\bibinfo {volume} {41}},\ \bibinfo {pages} {205} (\bibinfo {year}
  {1969})}\BibitemShut {NoStop}%
\bibitem [{\citenamefont {Lacombe}\ \emph {et~al.}(1981)\citenamefont {Lacombe}
  \emph {et~al.}}]{pariswf}%
  \BibitemOpen
  \bibfield  {author} {\bibinfo {author} {\bibfnamefont {M.}~\bibnamefont
  {Lacombe}} \emph {et~al.},\ }\href@noop {} {\bibfield  {journal} {\bibinfo
  {journal} {Phys. Lett. B}\ }\textbf {\bibinfo {volume} {101}},\ \bibinfo
  {pages} {139} (\bibinfo {year} {1981})}\BibitemShut {NoStop}%
\bibitem [{\citenamefont {Brodsky}(1997)}]{brodsky_lightcone}%
  \BibitemOpen
  \bibfield  {author} {\bibinfo {author} {\bibfnamefont {S.~J.}\ \bibnamefont
  {Brodsky}},\ }\href@noop {} {\  (\bibinfo {year} {1997})},\ \Eprint
  {http://arxiv.org/abs/hep-ph/9710288} {arXiv:hep-ph/9710288} \BibitemShut
  {NoStop}%
\bibitem [{\citenamefont {Kubon}\ \emph {et~al.}(2002)\citenamefont {Kubon},
  \citenamefont {Anklin}, \citenamefont {Bartsch}, \citenamefont {Baumann},
  \citenamefont {Boeglin} \emph {et~al.}}]{Kubon:2001rj}%
  \BibitemOpen
  \bibfield  {author} {\bibinfo {author} {\bibfnamefont {G.}~\bibnamefont
  {Kubon}}, \bibinfo {author} {\bibfnamefont {H.}~\bibnamefont {Anklin}},
  \bibinfo {author} {\bibfnamefont {P.}~\bibnamefont {Bartsch}}, \bibinfo
  {author} {\bibfnamefont {D.}~\bibnamefont {Baumann}}, \bibinfo {author}
  {\bibfnamefont {W.~U.}\ \bibnamefont {Boeglin}},  \emph {et~al.},\ }\href
  {\doibase 10.1016/S0370-2693(01)01386-7} {\bibfield  {journal} {\bibinfo
  {journal} {Phys. Lett. B}\ }\textbf {\bibinfo {volume} {524}},\ \bibinfo
  {pages} {26} (\bibinfo {year} {2002})}\BibitemShut {NoStop}%
\bibitem [{\citenamefont {Galster}\ \emph {et~al.}(1971)\citenamefont
  {Galster}, \citenamefont {Klein}, \citenamefont {Moritz}, \citenamefont
  {Schmidt}, \citenamefont {Wegener} \emph {et~al.}}]{Galster:1971kv}%
  \BibitemOpen
  \bibfield  {author} {\bibinfo {author} {\bibfnamefont {S.}~\bibnamefont
  {Galster}}, \bibinfo {author} {\bibfnamefont {H.}~\bibnamefont {Klein}},
  \bibinfo {author} {\bibfnamefont {J.}~\bibnamefont {Moritz}}, \bibinfo
  {author} {\bibfnamefont {K.~H.}\ \bibnamefont {Schmidt}}, \bibinfo {author}
  {\bibfnamefont {D.}~\bibnamefont {Wegener}},  \emph {et~al.},\ }\href
  {\doibase 10.1016/0550-3213(71)90068-X} {\bibfield  {journal} {\bibinfo
  {journal} {Nucl. Phys.}\ }\textbf {\bibinfo {volume} {B32}},\ \bibinfo
  {pages} {221} (\bibinfo {year} {1971})}\BibitemShut {NoStop}%
\bibitem [{\citenamefont {Bosted}\ and\ \citenamefont
  {Christy}(2008)}]{bosted_christy}%
  \BibitemOpen
  \bibfield  {author} {\bibinfo {author} {\bibfnamefont {P.~E.}\ \bibnamefont
  {Bosted}}\ and\ \bibinfo {author} {\bibfnamefont {M.~E.}\ \bibnamefont
  {Christy}},\ }\href@noop {} {\bibfield  {journal} {\bibinfo  {journal} {Phys.
  Rev. C}\ }\textbf {\bibinfo {volume} {77}},\ \bibinfo {pages} {065206}
  (\bibinfo {year} {2008})}\BibitemShut {NoStop}%
\bibitem [{\citenamefont {Abe}\ \emph {et~al.}(1998)\citenamefont {Abe} \emph
  {et~al.}}]{rcllacpol}%
  \BibitemOpen
  \bibfield  {author} {\bibinfo {author} {\bibfnamefont {K.}~\bibnamefont
  {Abe}} \emph {et~al.},\ }\href@noop {} {\bibfield  {journal} {\bibinfo
  {journal} {Phys. Rev. D}\ }\textbf {\bibinfo {volume} {58}},\ \bibinfo
  {pages} {112003} (\bibinfo {year} {1998})}\BibitemShut {NoStop}%
\bibitem [{\citenamefont {R.Brun}\ \emph {et~al.}(1982)\citenamefont {R.Brun},
  \citenamefont {M.Hansroul},\ and\ \citenamefont {J.C.Lassalle}}]{geant3}%
  \BibitemOpen
  \bibfield  {author} {\bibinfo {author} {\bibnamefont {R.Brun}}, \bibinfo
  {author} {\bibnamefont {M.Hansroul}}, \ and\ \bibinfo {author} {\bibnamefont
  {J.C.Lassalle}},\ }\href@noop {} {} (\bibinfo {year} {1982}),\ \bibinfo
  {note} {{GEANT Users Guide (CERN DD/EE/82)}}\BibitemShut {NoStop}%
\bibitem [{\citenamefont {Baillie}(2010)}]{natethesis}%
  \BibitemOpen
  \bibfield  {author} {\bibinfo {author} {\bibfnamefont {N.}~\bibnamefont
  {Baillie}},\ }\emph {\bibinfo {title} {Electron Scattering from an Almost
  Free Neutron in Deuterium}},\ \href
  {http://www.jlab.org/Hall-B/general/thesis/Baillie_thesis.pdf} {Ph.D.
  thesis},\ \bibinfo  {school} {College of William and Mary} (\bibinfo {year}
  {2010}),\ \bibinfo {note}
  {http://www.jlab.org/Hall-B/general/thesis/Baillie\_thesis.pdf}\BibitemShut
  {NoStop}%
\bibitem [{\citenamefont {Christy}\ \emph {et~al.}()\citenamefont {Christy},
  \citenamefont {Kalantarians}, \citenamefont {Ethier},\ and\ \citenamefont
  {Melnitchouk}}]{ERIC}%
  \BibitemOpen
  \bibfield  {author} {\bibinfo {author} {\bibfnamefont {M.~E.}\ \bibnamefont
  {Christy}}, \bibinfo {author} {\bibfnamefont {N.}~\bibnamefont
  {Kalantarians}}, \bibinfo {author} {\bibfnamefont {J.~J.}\ \bibnamefont
  {Ethier}}, \ and\ \bibinfo {author} {\bibfnamefont {W.}~\bibnamefont
  {Melnitchouk}},\ }\href@noop {} {}\bibinfo {note} {In
  preparation}\BibitemShut {NoStop}%
\bibitem [{\citenamefont {Christy}\ and\ \citenamefont
  {Bosted}(2010)}]{Christy:2007ve}%
  \BibitemOpen
  \bibfield  {author} {\bibinfo {author} {\bibfnamefont {M.~E.}\ \bibnamefont
  {Christy}}\ and\ \bibinfo {author} {\bibfnamefont {P.~E.}\ \bibnamefont
  {Bosted}},\ }\href {\doibase 10.1103/PhysRevC.81.055213} {\bibfield
  {journal} {\bibinfo  {journal} {Phys. Rev. C}\ }\textbf {\bibinfo {volume}
  {81}},\ \bibinfo {pages} {055213} (\bibinfo {year} {2010})}\BibitemShut
  {NoStop}%
\bibitem [{Sup()}]{SupMat}%
  \BibitemOpen
  \href@noop {} {}\bibinfo {note} {See Supplemental Material at [URL will be
  inserted by publisher] for tables of numerical results.}\BibitemShut {Stop}%
\bibitem [{\citenamefont {Bueltmann}\ \emph {et~al.}(2010)\citenamefont
  {Bueltmann} \emph {et~al.}}]{bonus12}%
  \BibitemOpen
  \bibfield  {author} {\bibinfo {author} {\bibfnamefont {S.}~\bibnamefont
  {Bueltmann}} \emph {et~al.},\ }\href@noop {} {\enquote {\bibinfo {title}
  {{The structure of the free neutron at large $x$-Bjorken}},}\ } (\bibinfo
  {year} {2010}),\ \bibinfo {note} {Jefferson Lab experiment
  E12-10-102}\BibitemShut {NoStop}%
\bibitem [{Dat()}]{DataBase}%
  \BibitemOpen
  \href@noop {} {}\bibinfo {note} {Jefferson Lab Experiment CLAS Database,
  http://clasweb.jlab.org/physicsdb}\BibitemShut {NoStop}%
\end{thebibliography}

%

\end{document}